\documentclass[11pt]{article}
\usepackage{amsmath, amssymb, amsthm,authblk}
\usepackage[usenames,dvipsnames]{color}
\usepackage{epsfig,euscript}
 \usepackage{mathabx,multirow}
\usepackage[authoryear,round]{natbib}
\usepackage{pdflscape}

\renewcommand{\baselinestretch}{1.5}
\makeatletter
\def\endp{{\hfill \vrule width 5pt height 5pt\par}}

\newcommand{\cM}{{\cal M}}

\newcommand{\cD}{\mbox{$\mathcal{D}$}}

\newtheorem{theorem}{Theorem}
\newtheorem{lemma}{Lemma}
\newtheorem{corollary}{Corollary}
\newtheorem{proposition}{Proposition}

\renewcommand{\hat}{\widehat}
\def\singlespace{\def\baselinestretch{1}\@normalsize}

\def\endp{{\hfill \vrule width 5pt height 5pt\par}}

\newcommand{\cov}{{\rm Cov}}

\newcommand{\tr}{\mbox{tr}}

\newcommand{\bA}{{\mathbf A}}
\newcommand{\bB}{{\mathbf B}}

\newcommand{\bH}{{\mathbf H}}
\newcommand{\bI}{{\mathbf I}}

\newcommand{\bM}{{\mathbf M}}
\newcommand{\bO}{{\mathbf O}}
\newcommand{\bQ}{{\mathbf Q}}

\newcommand{\bR}{{\mathbf R}}
\newcommand{\bS}{{\mathbf S}}
\newcommand{\bU}{{\mathbf U}}
\newcommand{\bV}{{\mathbf V}}

\newcommand{\bff}{{\mathbf f}}

\newcommand{\bq}{{\mathbf q}}

\newcommand{\bs}{{\mathbf s}}

\newcommand{\bx}{{\mathbf x}}
\newcommand{\by}{{\mathbf y}}
\newcommand{\bz}{{\mathbf z}}

\newcommand{\bSigma}{\boldsymbol{\Sigma}}

\newcommand{\bve}{\mbox{\boldmath$\varepsilon$}}

\newcommand{\bmu} {\boldsymbol{\mu}}

\newcommand{\bGamma} {\boldsymbol{\Gamma}}

\newcommand{\bzero}{{\mathbf 0}}

\newcommand{\calD}{{\mathcal D}}

\newcommand{\calF}{{\mathcal F}}

\newcommand{\calM}{{\mathcal M}}

\def\6bullets{\bullet\bullet\bullet\bullet\bullet\bullet}

\bibpunct{(}{)}{;}{a}{,}{,}

\title{\bf Threshold Factor Models for High-Dimensional Time Series\footnote{Xialu Liu is Assistant Professor, Management Information Systems Department, San Diego State University, San Diego, CA 92182. Email: xialu.liu@sdsu.edu.
Rong Chen is Professor, Department of Statistics, Rutgers University, Piscataway, NJ 08854. E-mail: rongchen@stat.rutgers.edu. Rong Chen is the corresponding
author. Chen's research was supported in part by National Science
Foundation grants DMS-1513409, DMS-1737857 and IIS-1741390. The authors wish to thank the editor, two anonymous referees, James Hamilton and Yixiao Sun for their insightful comments.}}
\author{Xialu Liu}
\affil{San Diego State University}
\author{Rong Chen}
\affil{Rutgers University}

\begin{document}
\maketitle

\date{}

\begin{abstract}
 \noindent
We consider a threshold factor model for high-dimensional time series
in which the dynamics of the time series is assumed to switch between
different
regimes according to the value of a threshold variable. This is an extension
of threshold modeling to a high-dimensional time series setting
under a factor structure.  Specifically, within each threshold regime,
the time series is assumed to follow a factor model. The regime switching
mechanism creates structural change in the
factor loading matrices.
It provides flexibility
in dealing with situations that the underlying states may be changing
over time, as often observed in economic time series and other applications.
We develop the procedures for the estimation of the loading spaces,
the number of factors and the threshold value, as well as
the identification of the threshold variable, which governs the regime
change mechanism.
The theoretical properties are investigated.
Simulated and real data examples are presented to illustrate the performance
of the proposed method.
 \medskip

\end{abstract}

\section{Introduction}
High-dimensional time series data analysis has drawn
attention from many researchers because of its broad range of
applications in many fields. It is a challenging problem due to
its complexity and the larger number of parameters involved.
Factor analysis is an effective approach
to alleviating the problem through effective dimension reduction.
Specifically, let $\by_t$ be an observed $p \times 1$ time series
$t=1,\ldots, n$. The general form of a factor model for time series data is
\begin{eqnarray*}
\by_t=\bA \bx_t +\bve_t,
\end{eqnarray*}
where $\bx_t=(x_{t,1}, x_{t,2}, \ldots, x_{t,k_0})'$ is a set of unobserved
(latent) factor time series with dimension $k_0$ that is much smaller than $p$,
the matrix $\bA$ is the loading matrix of the common factors, the
term $\bA \bx_t$
can be viewed as the signal component of the vector time series $\by_t$,
and $\bve_t$ is an error process or an idiosyncratic component.
The dimension reduction is achieved in the sense that, under the model,
the co-movement of the
$p$-dimensional process $\by_t$ is driven by a much lower dimensional
process $\bx_t$. The loading matrix $\bA$ reflects the impact of the common
factors $\bx_t$ on the observed process $\by_t$.

The general dynamic factor model assumes that
the latent factor process $\bx_t$ possesses certain dynamic structure such
as a vector time series structure
\citep{geweke1977, forni1998,forni2000,forni2001,bai2002, stock2002b, forni2003, forni2004, stock2005, hallin2007}. 
It is commonly assumed that the latent factors should
have an impact on most of the series (defined asymptotically).
In order to differentiate the signal component from the error process, strong cross-sectional dependence is not allowed for $\{\bve_t\}$.
As a consequence, the noise process $\{\bve_t\}$ may have weak
serial dependence, i.e.
$\frac{1}{n}\sum_{t=1}^n \sum_{s=1}^n | E(\bve_t' \bve_s)| < C$, where $C$ is a positive constant.


One disadvantage of the above assumptions is that the dynamic component and error process are not separable when the dimension is finite, since both of them have serial dependence. Another setting of factor models for time series data has become more popular in the literature.
It assumes that the error process is
white noise without serial dependence, i.e.,
$E(\bve_t' \bve_s)=0$, for $t\neq s$. Consequently the dependence of the
observed process $\by_t$ is completely driven by the common factors
\citep{pena1987,pena2006,pan2008,chang2015, liu2016}.
It ensures that the signal component is identifiable when the dimension
of the panel time
series is finite. In addition, the error process is allowed to have strong
cross-sectional correlation. \cite{pena2006} and \cite{lam2011} developed an approach that takes advantage of information from the autocovariance matrices of the observed process at nonzero leads via
eigen-decomposition to estimate the factor loading space, and they established the asymptotic properties as the dimension goes to infinity with sample size.
This method is applicable to non-stationary processes, processes with uncorrelated or endogenous regressors, and matrix-valued processes; see \cite{chang2015,wang2018}. In this paper, we adopt
these assumptions in developing the estimation procedures and
the corresponding theoretical properties.

In many applications it is often observed that the loading matrix of a
factor model may vary.
For example,
the expected market return is an
important factor of the expected return of an asset, according to
CAPM theory, and its impact
(loading) on any individual asset is often observed to change
depending on whether the stock market is volatile or stable.
In economics, risk-free rate, unemployment, and economic growth are
important to all economic activities and decisions. Again,
the behavior of these series may vary under
different fiscal policies (neutral, expansionary, or contractionary) or in
different stages of the economic cycle (expansion, peak, contraction,
or trough) \citep{kim1998}. \cite{liu2016} introduced a
Markov switching mechanism to the factor model to capture the changes of the loading matrix.
Although Markov regime-switching models are widely used in economics to describe the varying structure, it has the drawback of being less interpretable and
difficult to forecast.

To address this limitation, we propose a threshold factor model, in which a
threshold variable controls the changes of the loadings in different
regimes. Such a model
 enhances the flexibility in modeling the underlying regime
switching mechanism, and
provides a more interpretable structure and an
easier forecasting framework.
Threshold models have been extensively studied under the general framework
of autoregressive
models \citep{tong1980,chen1995a,tiao1989,tsay1998,forbes1999}, nonlinear
models \citep{petruccelli1986,gourieroux1992,tong1993}, and non-stationary
 models
 \citep{zakoian1994,li1996,balke1997}. In this paper we apply this powerful
 approach to
factor models for high-dimensional time series.

Specifically, we formally introduce a threshold factor model,
propose an estimation procedure for the
loading spaces and the number of factors
based on eigen-analysis of the cross moment matrices of the observed process,
develop an objective function for the identification of
the threshold value and the threshold variable,
and investigate their theoretical properties.
It is shown
that even when the number of factors is overestimated, our estimators are still consistent. Their asymptotic properties are the same as
 those when the number of factors is correctly specified.

 The paper focuses more on the structural change of the loading matrices and the identification of switching mechanism. We do not impose a specific dynamic structure on the factors. This makes our discussion and theory more general, but also limits the prediction ability of the model. Stationary and nonstationary structures for the factor models can be imposed, similar to those in \cite{pena1987, pena2006}, though different estimation procedures will be needed to take advantage of the structure.

The rest of the paper is organized as follows. Section 2 introduces the
detailed model setting. In Section 3, estimation procedures are developed and theoretical properties of the proposed estimators are investigated.
Section 4 proposes a three-step procedure for searching and identifying
the threshold variable. Simulation results are presented in Section 5,
and a real example is analyzed in Section 6. All detailed proofs are
contained in the Appendix.

\section{Threshold factor model}
We consider the following two-regime threshold factor model for
high-dimensional time series here.
Let $\by_t$ be an observed $p \times 1$ time series, and
$\bx_t$ be a $k_0 \times 1$ latent factor process, $t=1, \ldots, n$.
\begin{align}\label{model}
\by_t=\left\{
\begin{array}{cc}
\bA_1\bx_t + \bve_{t,1}	&z_{t}< r_0,\\
\bA_2 \bx_t +\bve_{t,2} & z_{t} \geq r_0,
\end{array}\right. \mbox{ and } \bve_{t,i} \sim N(\bzero, \bSigma_{t,i}), \quad i=1,2,
\end{align}
where $z_{t}$ is a partially known threshold variable, observable at time $t$,
possibly with a small number of unknown parameters.  The
noise process $\{\bve_{t,1},\, \bve_{t,2}\}$ is assumed to be
$p \times 1$ uncorrelated noise processes. $\{\bve_{t,1}, \bve_{t,2}\}$ and $\bx_t$ are uncorrelated with $z_{t}$ given ${\cal F}_{-\infty}^{t-1}$, where ${\cal F}_{i}^{j}$ is the $\sigma$-field generated by $\{ (\bx_t, z_{t}): i \leq t \leq j\}$.

The loading matrix is not uniquely defined, since $(\bA_i, \bx_t)$ in (\ref{model}) can be replaced by $(\bA_i \bU_i,\bU^{-1}_i \bx_t)$ for any $k_0 \times k_0$ non-singular matrix $\bU_i$, $i=1,2$. However, ${\cal M}(\bA_i)$, the space spanned by the columns of $\bA_i$, is uniquely defined under our assumptions.
To estimate the column space ${\cal M}(\bA_i)$, we will estimate
an orthonormal representative of the space, a $p\times k_0$ matrix $\bQ_i$, such
that
\begin{align}\label{gamma}
\bQ_i' \bQ_i= \bI_{k_0}, \mbox{ and } \bA_i=\bQ_i \bGamma_i, \quad i=1,2,
\end{align}
where $\bGamma_i$ is a $k_0 \times k_0$ non-singular matrix that provides the
link between $\bQ_i$ and $\bA_i$. Again, due to ambiguity, $\bGamma_i$ is not
estimable. In any case, we have
${\cal M}(\bQ_i)={\cal M} (\bA_i)$. The columns of $\bQ_i$ are $k_0$ orthonormal
vectors, and the column space spanned by $\bQ_i$ is the same as
the column space spanned by $\bA_i$. In (\ref{model}), we assume ${\cal M}(\bA_1)\neq  {\cal M}(\bA_2)$.
Let $I_{t,i}$ be the indicator function of regime $i$ at time $t$, i.e. $I_{t,1}=I(z_{t} <r_0)$ and $I_{t,2}=I(z_{t} \geq r_0)$. Let
\begin{align}\label{trans}
\bR_t= \sum_{i=1}^2 \bGamma_i \bx_t I_{t,i},
\end{align}
The threshold factor model (\ref{model}) can be written as
\begin{align}\label{r}
\by_t 
= \sum_{i=1}^2 (\bQ_i \bR_t +\bve_{t,i})I_{t,i},
\end{align}
where $\bQ_i$ are orthonormal matrices.

Our aim is to estimate the loading spaces ${\cal M}(\bA_i)$, $i=1,2$,
the number of factors $k_0$, and the threshold value $r_0$, given the
threshold variable. When the threshold variable is unknown, we also
propose a procedure for its identification.

\noindent{\bf Remark 1.} We note that \cite{pena1987,forni2000, forni2004,hallin2007} make specific assumptions about the dynamics of the common factors. With a data generating process, forecasting becomes possible. More sophisticated and more accurate estimation procedures can be constructed by fully utilizing the assumed structure. In this paper we do not specify the dynamic structure
for the factor process for several reasons. First, the procedure and theory become more general and less restrictive hence no additional model checking procedure is needed for confirming the dynamic structure. Second, we are able to use simple eigen-analysis for estimation. Third, the regimes bring more ambiguity into the model that makes the identification of the dynamic structure of the factors model difficult. Of course, without the dynamic structure, the model loses its forecasting ability, though the results can provide a good start for building a forecasting model if forecasting is the main objective.

\noindent{\bf Remark 2.} The state-space is divided into two regimes,
controlled by the threshold variable $z_t$. We assume $z_t$ is observable
at time $t$. It can be the lag variable of an observed time series.
In a more complicate setting, $z_t$ can be partially
observable with several unknown parameters. For example,
$z_t=\beta_1z_{1t}+\beta_2z_{2t}$ where $z_{1t}$ and $z_{2t}$ are
observable at time $t$ and $\beta_i$'s are unknown parameters.
Because $z_{t}$ is observable
at time $t$ given the parameters, we know precisely which regime the
process is in at time $t$, given $r_0$ and $\beta_i$'s.

\noindent{\bf Remark 3.} Constant terms can be included in model
(\ref{model}) as follows,
\begin{align}\label{model21}
\by_t=\left\{
\begin{array}{cc}
\bmu_1+ \bA_1\bx_t + \bve_{t,1}	&z_{t}< r_0\\
\bmu_2+ \bA_2 \bx_t +\bve_{t,2} & z_{t} \geq r_0
\end{array}\right., \mbox{ and } \bve_{t,i} \sim N(\bzero, \bSigma_{t,i}), \quad i=1,2.
\end{align}

If we combine these terms and loading matrices, model (\ref{model21}) can be written as a threshold factor model with $(k_0+1)$ factors. Specifically, in regime $i$, when $I_{t,i}=1$,
\begin{eqnarray*}
\by_t=
\left( \begin{array}{cc}
\bmu_i  &\bA_i
\end{array}\right)
\left( \begin{array}{c}
1\\
\bx_t
\end{array}\right)
+ \bve_{t,i},	 \mbox{ and } \bve_{t,i} \sim N(\bzero, \bSigma_{t,i}).
\end{eqnarray*}
Hence, model (\ref{model21}) is a special case of model (\ref{model}), in which one of the common factor is deterministic. In order to accommodate this
simplified setting, in the eigen-analysis when performing for loading matrix
estimation, we use cross auto-moment matrices,
instead of the traditional auto-covariance
matrices.

\noindent{\bf Remark 4.} The threshold factor model (\ref{model}) provides
a different approach from the regime switching model in \cite{liu2016}.
A typical regime switching model introduces a random switching mechanism
that is not
observed. In threshold models, regime switching is
observable, given the observable threshold variable $z_t$ and the
threshold value $r_0$. It provides easier estimation,
clearer interpretation and better
predictability. In all threshold modeling approaches, identification
of a suitable threshold variable that drives
 regime switching is the most
important modeling component and is often the most challenging one.
We will propose an identification
approach that is easy to use and can screen a large number
of potential threshold variables in Section 4.

\noindent{\bf Remark 5.} Note that the two-regime threshold factor model (\ref{model}) can be written as a one regime model as
\begin{align}\label{model2}
\by_t=\bA \bff_t +\bve_t,
\end{align}
where $\bA=[\bA_1 \:\: \bA_2]$, $\bff_t=(\bx'_t I_{t,1},\bx'_t I_{t,2})'$, and
$\bve_t=\sum_{i=1}^2 \bve_{t,i}I_{t,i}$.

Although this model seems to be structurally simpler, the switching mechanism is hidden in the factor dynamics.
Due to rotation ambiguity, detecting such changes in the factor process is much more difficult.  Although in our framework we only impose weak conditions on the factors hence we are able to estimate model (\ref{model2}) even with the hidden structural changes, the results will not be able to reveal the switching mechanism and the underlying threshold variable, which are important features of the time series one would like to know. In addition, model (\ref{model2}) uses a larger number of factors, which makes estimation less accurate and less interpretable.


Another advantage for model (\ref{model}) is that the factors are
allowed to have
different 'strengths' across the regimes,
while the estimation of model (\ref{model2}) using that in \cite{lam2011}
works better with the factors having the same strength.
Strength roughly measures the total
squared impact of a factor on the observed time series. A more formal definition is given in Section 3.

\noindent{\bf Remark 6.}
The detection of the existence of switching is an interesting but challenging problem. Possible approaches include testing the differences in the loading space, or testing the distance between the two loading spaces is zero. However, under the null hypothesis of no switching, the threshold value parameter becomes latent which poses technical challenges. A complete and rigorous study of the problem is out of scope of this paper.

\section{Estimation procedure with a given threshold variable}
In this section, we first present a procedure to estimate the loading spaces
corresponding to a partition in the form of $I_{t,1}(r_1)=I(z_t<r_1)$ and $I_{t,2}(r_2)=I(z_t \geq r_2)$ where $r_1\leq r_2$,
and show the asymptotic property of the estimator in the case of
$r_1 \leq r_0$ and $r_2 \geq r_0$, where $r_0$ is the true threshold value. Then we propose
a procedure for estimating $r_0$ using tentative threshold values to split the data, along its asymptotic properties.
The asymptotic properties of the estimated
loading spaces using the estimated threshold value are also presented.

Here are some notations. For any matrix $\bH$, let $\|\bH\|_F$ and
$\|\bH\|_2$ denote the Frobenius and L-2 norms of $\bH$, $\sigma_i(\bH)$ is the $i$-th largest singular value of $\bH$, and $\|\bH\|_{\min}$ is the square root of minimum nonzero eigenvalue of $\bH'\bH$. For a square matrix $\bH$, ${\rm tr}(\bH)$ denotes its trace. We write $a \asymp b$, if $a=O(b)$ and $b=O(a)$. We use $C$ to denote a positive constant.

\subsection{Initial estimation of the loading spaces with tentative threshold values}


Define the generalized second cross moment matrices of
 $\by_t$ of lead $h$ in different partitions as
\begin{align*}
\bSigma_{x,i,j}(h,r_1,r_2)= \frac{1}{n-h}\sum_{t=1}^{n-h}{\rm E} (\bx_t \bx_{t+h}' I_{t,i}(r_i)I_{t+h,j}(r_j)),\\
\bSigma_{y,i,j}(h,r_1,r_2)= \frac{1}{n-h}\sum_{t=1}^{n-h}{\rm E} (\by_t \by_{t+h}' I_{t,i}(r_i)I_{t+h,j}(r_j)),
\end{align*}
for $i, j=1,2$. Here $\bSigma_{y,1,1}(h,r_1,r_2)$ is the cross moment matrix of $\by_t$ and $\by_{t+h}$ when both $\by_t$ and $\by_{t+h}$ are in partition 1 with
$\{z_t<r_1\}$, and $\bSigma_{y,1,2}(h,r_1,r_2)$ is that when $\by_t$ is in partition 1 with $\{z_t<r_1\}$ and $\by_{t+h}$ is in partition 2 with
$\{z_{t+h}\geq r_2\}$.
$\bSigma_{y,2,1}$ and $\bSigma_{y,2,2}$ are similar.

Define the sum of a quadratic version of the cross moment matrices of $\by_t$,
\begin{align}\label{M}
\bM_i(r_1,r_2)=\sum_{h=1}^{h_0}\sum_{j=1}^2
\bSigma_{y,i,j}(h,r_1,r_2)\bSigma_{y,i,j}(h,r_1,r_2)',
\end{align}
for a pre-fixed maximum lead $h_0$, and $i=1,2$.

Let $\bq_{i,k}(r_1,r_2)$ and $-\bq_{i,k}(r_1,r_2)$
be the pair of unit eigenvectors of
$\bM_i(r_1,r_2)$ corresponding to its $k$-th largest eigenvalue.
In the following we
assume $\mathbf{1}'\bq_{i,k}(r_1,r_2)>0$ (which is uniquely defined)
and will use it in all our constructions.
Define
\begin{equation}
\bQ_i(r_1,r_2)=(\bq_{i,1}(r_1,r_2), \ldots, \bq_{i,k_0}(r_1,r_2)),
\label{def:QB}
\end{equation}
for $i=1,2$, where $k_0$ is the number of factors in model (\ref{model}).
For simplicity, in the rest of the paper, if $r_1=r_2$, notations $\bSigma_{y,i,j}(h,r,r)$, $\bQ_i(r,r)$
and $\bM_i(r,r)$ are simplified to $\bSigma_{y,i,j}(h,r)$, $\bQ_i(r)$, and
$\bM_i(r)$. Furthermore, we use $\bQ_i$ and $\bM_i$ to denote $\bQ_i(r_0)$ and $\bM_i(r_0)$, where $r_0$ is the true threshold value.
Using the sample version of $\bSigma_{y,i,j}(h,r_1,r_2)$ and $\bM_i(r_1,r_2)$,
we perform eigenanalysis of $\hat{\bM}_i(r_1,r_2)$ to obtain
$\hat{\bQ}_i(r_1,r_2)$.

The rationale behind the approach is the following.
Let $r_0$ be the true threshold value. Consider the case
that $r_1 \leq r_0$ and $r_2 \geq r_0$. Then partition 1 with $\{z_t<r_1\}$ is a subset of data in Regime 1 and partition 2 with $\{z_t \geq r_2\}$ is a subset of data in Regime 2.
For any integer $h>0$, it follows from model (\ref{model}) that
\begin{align*}
\bSigma_{y,1,1}(h,r_1, r_2) = \bA_1 \bSigma_{x,1,1}(h,r_1,r_2) \bA_1', \mbox{ and } \bSigma_{y,1,2} (h,r_1,r_2)= \bA_1 \bSigma_{x,1,2}(h,r_1,r_2) \bA_2',
\end{align*}
under the white noise assumption.
In this case,
\begin{equation*}
\bM_{1}(r_1,r_2)=\bA_1 \left(\sum_{h=1}^{h_0}  \sum_{j=1}^2 \bSigma_{x,1,j}(h,r_1,r_2)   \bA_j'  \bA_j \bSigma_{x,1,j}(h,r_1,r_2)' \right) \bA_1'.  \label{M1}
\end{equation*}
Then $\bM_1(r_1,r_2)$ is a positive  semi-definite matrix sandwiched by $\bA_1$. If there exists at least one $1\leq h \leq h_0$ such that
$\bSigma_{x,1,1}(h,r_1,r_2)$ or $\bSigma_{x,1,2}(h,r_1,r_2)$ is full rank,
then $\bM_1(r_1,r_2)$ has rank $k_0$.
Hence the corresponding
$\bQ_1(r_1,r_2)$ is an orthonormal representative of
${\cal M} (\bA_1)$. The results also hold for $\bM_2(r_1,r_2)$.

Let the sample versions of $\bSigma$ and $\bM$ be
\begin{align}
\hat{\bSigma}_{y,i,j}(h,r_1,r_2)= \frac{1}{n-h}\sum_{t=1}^{n-h} \by_t \by_{t+h}' I_{t,i}(r_i) I_{t+h,j}(r_j), \nonumber \\
\hat{\bM}_i(r_1,r_2)=\sum_{h=1}^{h_0}\sum_{j=1}^2
\hat{\bSigma}_{y,i,j}(h,r_1,r_2)\hat{\bSigma}_{y,i,j}(h,r_1,r_2)'.\label{hatM}
\end{align}
Let $\hat{\bq}_{i,k}(r_1,r_2)$ be the eigenvector of $\hat{\bM}_i(r_1,r_2)$ corresponding to its $k$-th largest eigenvalue. $\cM(\bA_i)$ can be estimated by
\[
\widehat{\cM(\bA_i)}= \cM(\hat{\bQ}_i(r_1,r_2)),
\]
where
$\hat{\bQ}_i(r_1,r_2)=(\hat{\bq}_{i,1}(r_1,r_2), \ldots, \hat{\bq}_{i,k_0}(r_1,r_2))$.

However, when $r_2\geq r_1>r_0$ or $r_1\leq r_2<r_0$,
the eigen-space of $\bM_i(r_1,r_2)$ may not correspond to
${\cal M} (\bA_i)$. For example, if $r_1>r_0$, the partition $I_{t,1}(r_1)=1$ contains observations in both
regimes --  $r_0 \leq z_t <r_1 $ for Regime 2 and
$z_t<r_0<r_1$ for Regime 1. Hence $\bM_{1}(r_1,r_2)$ is not
sandwiched by $\bA_1$. We will show later that, when $r_1$ and $r_2$ are sufficiently close
the $r_0$, the estimated space using the sample version of $\bM_{i}(r_1,r_2)$ is still
consistent.


\medskip
\noindent{\bf Remark 7.} It is tempting not to separate $y_{t+h}$ into
two regimes and use
\[
{\bM}_i(r_1,r_2)=\sum_{h=1}^{h_0} \bSigma_{y,i}(h,r_1,r_2)\bSigma_{y,i}(h,r_1,r_2)',
\]
where
\[
\bSigma_{y,1}(h,r_1,r_2)=\frac{1}{n-h} \sum_{t=1}^{n-h}
E(\by_t \by_{t+h}' I(z_t <r_1)), \quad \bSigma_{y,2}(h,r_1,r_2)=\frac{1}{n-h}
\sum_{t=1}^{n-h} E(\by_t \by_{t+h}' I(z_t  \geq r_2)).
\]
to replace (\ref{M}).
However, if the two loading spaces
${\cal M}(\bQ_1)$ and ${\cal M}(\bQ_2)$ are not orthogonal, cancellation may
occur, making the rank of $\bM_i(r_1,r_2)$ less than the rank of $\bA_i$.

\medskip
\noindent{\bf Remark 8.} Theoretically, ${\cal M}(\bA_i)$ can be
estimated through eigen-decomposition of one of $\{\bSigma_{y,i,j}(h,r_1,r_2) \bSigma_{y,i,j}(h,r_1,r_2)', h=1,2,\ldots\}$,
as long as it is full rank. Asymptotically
they
converge at the same rate.
The reason for using $\bM_i(r_1,r_2)$ in (\ref{M}) is that by summing over $h$,
we do not need to find a particular
$h$ to satisfy the condition. Since the strongest correlation often occurs at smaller leads,
a relatively small $h_0$ is usually adopted. The autocorrelation of each individual $\by_{t}$ often provides a good indication of the proper $h_0$ to be used.

To study the asymptotic properties of the estimator,
 we extend a distance measure of two
linear spaces used in \cite{chang2015,liu2016}.
 Let $\bS_1$ be a $p\times q_1$ matrix with rank $q_1$ and $\bS_2$ be a $p\times q_2$ matrix with rank $q_2$, where $p \geq \max\{q_1,q_2\}$.
Let the columns of $\bO_i$ be an orthonormal basis of $\cM(\bS_i)$,
for $i=1,2$. Define
\begin{equation}\label{distance_diff}
{\cal D}(\cM(\bS_1),\,\cM(\bS_2))
= \left(1 - \frac{{\rm tr}(\bO_1 \bO_1'
\bO_2 \bO_2')}{\min\{q_1,q_2\}}  \right)^{1/2},
\end{equation}
as the distance of the column spaces of $\bS_1$ and $\bS_2$. It
is a quantity between 0 and 1. It equals
to 0 if ${\cal M}(\bS_1) \subseteq {\cal M}(\bS_2)$ or
${\cal M}(\bS_2) \subseteq {\cal M}(\bS_1)$, and 1 if
$\cM(\bS_1)$ and $\cM(\bS_2)$ are orthogonal.
The distance of two linear spaces with the same dimension is defined in \cite{chang2015,liu2016}. Here (\ref{distance_diff}) is a modified version and takes into consideration the scenario that  the dimensions of two spaces may be different.

For factor models in high-dimensional cases, it is common to assume that the number of factors is fixed and the squared L-2 norm of the $p \times k_0$ loading matrix $\bA_i$ grows with the dimension $p$ \citep{bai2002,doz2011}. The growth rate is called the
strength of the factors in \cite{lam2011,lam2012,chang2015,liu2016}. Let
\begin{eqnarray*}
\|\bA_i\|_2^2 \asymp \|\bA_i\|_{\min}^2 \asymp p^{1-\delta_i}, \quad 0\leq \delta_i \leq 1.
\end{eqnarray*}
If $\delta_i=0$, regime $i$ is called a strong regime.
If $0<\delta_i <1$, the
regime is called a weak regime. If $\delta_i=1$, the regime is called an extremely weak regime. The strength of the regime measures the relative growth rate of
the amount of information about the
common factors $\bx_t$ carried by
the observed process $\by_t$, as $p$ increases, comparing
to the growth rate of the amount of noise process in regime $i$.
It is seen that in the following
theoretical development that the strength of the regime plays an important
role in estimation efficiency.

\begin{theorem}
Under Conditions 1-6 in Appendix A.1, when $r_1 \leq r_0$ and $r_2 \geq r_0$, if $p^{\delta_1/2+\delta_2/2}n^{-1/2}=o(1)$ and $\hat{\bQ}_1(r_1,r_2)$ and $\hat{\bQ}_2(r_1,r_2)$ estimated using the true $k_0$,
it holds that
\[
{\cal D}({\cM (\hat{{\bQ}}_i(r_1,r_2)), {\cM}(\bQ_i)})= O_p(p^{\delta_i/2 +\delta_{\min}/2} n^{-1/2}), \mbox{\ \  for \ \ } i=1,2,
\]
 as $n,p \to \infty$, where $\delta_{\min}=\min\{ \delta_1,\delta_2 \}$.
\end{theorem}

It is clearly not efficient to estimate the loading spaces with only partial observations. In practice, once the threshold value is estimated, we can use the full data set for the estimation of loading spaces. The asymptotic properties of the estimators based on the estimated threshold value will be discussed later in Theorems 3 and 5.


The convergence rates shown in Theorem 1 are the same
as those in \cite{liu2016}.
It is worth noting that when the two regimes have different
strengths $\delta_1$ and $\delta_2$, the convergence
rate of the estimator in the
stronger regime is the same as that in the one regime case, but the rate
of
the weaker regime is faster than that if it is the only regime.
In other words, the estimation in the stronger regime is
not hurt by the weaker regime, but the weaker regime gains efficiency from the stronger regime due to the switching of the process between the two regimes. We call it the 'helping effect'.

\subsection{Estimation of the threshold value}

Estimation of the threshold value in a threshold model
has been extensively studied in
univariate threshold models using least squares or likelihood
estimators, including those in
\cite{tong1980,tsay1989,chan1990, chan1993, caner2004,chen2006} and
\cite{wu2007}.
Here, we construct
an objective function for the estimation of
the threshold value. Since for a given finite sample, the model is not
distinguishable for all values
between two adjacent observations of $z_{t}$ as its threshold value,
we follow the
standard approach and assume that the threshold value takes on a
finite number of possible values in the set of all observed $z_t$.
Our method is to traverse all of these
possible threshold values and find the best
one that optimizes the objective function.

\medskip
When $r$ is used as the tentative threshold value to split the data into two subsets with $\{z_t<r\}$ and $\{z_t \geq r\}$, we define the objective function
\begin{align}
G(r)=\sum_{i=1}^2  \Big\|  {\bB_i}' \, \bM_i(r) \, \bB_i\Big\|_2=\sum_{i=1}^2  \Big\|\sum_{h=1}^{h_0}\sum_{j=1}^2 {\bB_i}' \bSigma_{y,i,j}(h,r)  {\bSigma_{y,i,j}(h,r) }' \bB_i\Big\|_2, \label{GG}
\end{align}
where $\bB_i$ is a $p\times (p-k_0)$ matrix for which $(\bQ_i, \bB_i)$ forms a $p\times p$ orthonormal matrix with $\bQ_i'\bB_i=\mathbf{0}$ and
$\bB_i'\bB_i=\mathbf{I}_{p-k_0}$. ${\cM}(\bB_i)$ is the orthogonal complement space of $\cM(\bQ_i)$, for $i=1,2$. Note that although $\bB_i$ is not uniquely defined and subject to any orthogonal transformation, $G(r)$ is invariant under such transformations.
The function $G(\cdot)$ measures the sum of the squared norm of the projections of the cross moment matrices
$\bSigma_{y,i,j}(h,r)$ onto $\cM(\bB_i)$ for $h=1,\ldots h_0$.

If $r=r_0$ (the true threshold value), the observations in
the two subsets identified do belong to the correct regimes. Then
$\bM_1$ is sandwiched by $\bA_1$ and $\bM_2$ is
sandwiched by $\bA_2$. Hence
\[
G(r_0)=
\sum_{i=1}^2 \Bigg\| \sum_{h=1}^{h_0}\sum_{j=1}^2   {\bB_i}' \, \bA_i  \bSigma_{x,i,j}(h,r_0) \bA_j'  \bA_j \bSigma_{x,i,j}(h,r_0)   \bA_i' \, \bB_i   \Bigg\|_2=0.
\]

If $r \neq r_0$, one of two subsets contains observations from both regimes.
Then the projection will not be zero. The following proposition formally states that, under mild conditions, we have $G(r)>0$ for
any $r\neq r_0$.

\begin{proposition}
Under Conditions 1-9 in Appendix A.1, if $r \neq r_0$, then $G(r)>0$.
\end{proposition}

The proof of the proposition is in Appendix A.2.

To obtain the sample version of $G(r)$, we assume {\it apriori}
that $r_0$ is in a known region of the support of $z_t$,
$r_0 \in (\eta_1, \eta_2)$. Such an assumption is standard in
threshold model estimation. Under this assumption, we can use data
corresponding to $z_t \leq \eta_1$ and  $z_t \geq \eta_2$ to obtain estimates for $\cM({\bB}_1)$ and
$\cM({\bB}_2)$, respectively. Specifically, $\bB_i(\eta_1,\eta_2)$ is estimated by
\[
\hat{\bB}_i(\eta_1,\eta_2)=(\hat{\bq}_{i,k_0+1}(\eta_1,\eta_2),\ldots, \hat{\bq}_{i,p}(\eta_1,\eta_2)),
\]
for $i=1,2$. By Theorem 1, both of them are consistent. Define
\begin{align*}
\hat{G}(r)=
\sum_{i=1}^2 \Big\|  \hat{\bB}_i(\eta_1,\eta_2)' \, \hat{\bM}_i(r) \, \hat{\bB}_i(\eta_1,\eta_2)\Big\|_2.
\end{align*}
We estimate $r_0$ by
\begin{align}\label{threshold}
\hat{r}=\arg \min_{ r \in \{z_1,\ldots, z_n \}\bigcap (\eta_1,\eta_2)}
\hat{G}(r).
\end{align}

\medskip
\noindent{\bf Remark 9.}
In the above procedure we require that there are sufficient samples
corresponding to
$z_{t}\leq \eta_1$ and $z_{t}\geq \eta_2$, to ensure the accuracy of the
estimated ${\cal M}(\bB_i)$, for $i=1,2$. When the values of $\eta_1$
and $\eta_2$ are not clear, it is possible to use a sequential procedure
based on the ranked sequence of $z_{t}$, similar to that in \cite{tsay1989}.
  
\medskip
\noindent{\bf Remark 10.} An alternative objective function
is the likelihood of $\bB_i\by_t$. However, we do not want to
involve the structure of the covariance matrices of the
noise process which both are $p\times p$ matrices. Here we still take
advantage of the whiteness assumption of the noise process,
and use the cross moment matrices of $\by_t$ at nonzero leads.

\begin{theorem}
Under Conditions 1-9 in Appendix A.1, if
$p^{\delta_1/2+\delta_2/2}n^{-1/2}=o(1)$, with true $k_0$, it holds that, for any $\epsilon>0$,
\begin{eqnarray*}
P(\hat{r}<r_0-\epsilon)\leq \frac{Cp^{\delta_1/2+\delta_{\min}/2}}{ \epsilon n^{1/2}}, \quad
P(\hat{r}>r_0+\epsilon)\leq \frac{Cp^{\delta_2/2+\delta_{\min}/2}}{ \epsilon n^{1/2}},
\end{eqnarray*}
as $n,p \to \infty$.
\end{theorem}

Theorem 2 shows that the estimator $\hat{r}$ in (\ref{threshold}) is consistent under some mild conditions. The estimation performance depends on the strength of both regimes. If the two regimes are both strong ($\delta_1=\delta_2=0$), the estimation is immune to the curse of dimensionality. However,
if at least one regime is weak,
the estimator becomes less efficient as $p$ increases,
and would require larger sample size $n$ for consistency.
When the two regimes have different strengths,
the probability that $\hat{r}$ falls in the stronger regime is smaller than
that in the weaker regime (the one with larger estimation error). Hence
the overall rate of convergence of $\hat{r}$ depends on the strength of the
weaker regime.

\subsection{Estimation of the loading spaces with estimated threshold value}

The final estimation of ${\cal M}(\bA_i)$ is obtained
using $\hat{r}$ as the threshold value and the procedure in Section 3.1.
Specifically, we define
\[
\hat{\bQ}_i(\hat{r})=(\hat{\bq}_{i,1}(\hat{r}), \ldots, \hat{\bq}_{i,k_0}(\hat{r})),
\]
where $\hat{\bq}_{i,k}(\hat{r})$ is the unit eigenvector of
$\hat{\bM}_i(\hat{r})$ corresponding to its $k$-th largest eigenvalue, and $\hat{\bM}_i(\hat{r})$ is defined in (\ref{hatM}).

Theorem 3 presents the asymptotics of the estimated loading spaces when the estimated threshold value is used.

\begin{theorem}
Under Conditions 1-9 in Appendix A.1, if $p^{\delta_1/2+\delta_2/2}n^{-1/2}=o(1)$ and $\hat{\bQ}_1(\hat{r})$ and $\hat{\bQ}_2(\hat{r})$ are estimated with the true $k_0$, it holds that
\begin{align*}
{\cal D}({\cal M}(\hat{\bQ}_i(\hat{r})),\,{\cal M}(\bQ_i))&=O_p(p^{\delta_i/2+\delta_{\min}/2}n^{-1/2}), \mbox{\ \  for \ \ } i=1,2,
\end{align*}
as $n$, $p\to \infty$.
\end{theorem}

Theorem 3 shows that
the rates of loading space estimators are the same as
that when the true threshold value is known.

Let $\bs_t$ be the signal (or dynamic) part of  $\by_t$, defined as
$\bs_t=\bA_1 \bx_t I_{t,1}+\bA_2 \bx_tI_{t,2}$.
Since the column space of $\bA_i$
is identifiable only up to a nonsingular transformation across regimes,
 we cannot estimate the latent process $\bx_t$ directly, but we have a natural estimator
for $\bs_t$ and
the latent process $\bR_t$ with standardized loadings in (\ref{trans}),
\begin{eqnarray}\label{trans_est}
\hat{\bs}_t(\hat{r})= \sum_{i=1}^2 \hat{\bQ}_i(\hat{r}) \hat{\bQ}_i(\hat{r})'\by_t I_{t,i}(\hat{r}), \quad
\hat{\bR}_{t}(\hat{r})= \sum_{i=1}^2 \hat{\bQ}_i(\hat{r})' \by_tI_{t,i}(\hat{r}).
\end{eqnarray}

\subsection{When the number of factors is unknown}

In practice, the number of factors $k_0$ is usually unknown.
This quantity can be estimated through a similar eigenvalue ratio estimator
used in \cite{lam2011}. Specifically, again we assume $r_0$ is in a
known interval $(\eta_1,\eta_2)$, and let
\begin{align}\label{est_num}
\hat{k}_i=\arg \min_{1 \leq k \leq R}
\frac{{\hat{\lambda}_{i,k+1}(\eta_1,\eta_2)}}{{\hat{\lambda}_{i,k}(\eta_1,\eta_2)}},
\mbox{\ \ for \ \ } i=1,2,
\end{align}
where $\hat{\lambda}_{i,k}(\eta_1,\eta_2)$ is the $k$-th largest eigenvalue
of $\hat{\bM}_i(\eta_1,\eta_2)$.
We follow \cite{lam2012} and use $R=p/2$, under the assumption that $k_0\ll p$.

\begin{corollary}
Under the Conditions 1-9 in Appendix A.1,
if $p^{\delta_1/2+\delta_2/2}n^{-1/2}=o(1)$, for $i=1,2$,
as $n,p \to \infty$, it holds that
\begin{eqnarray*}
\hat{\lambda}_{i,k+1}(\eta_1,\eta_2)/\hat{\lambda}_{i,k}(\eta_1,\eta_2)
&\asymp & 1 \mbox{\ \  for \ \ } k=1, \ldots, k_0-1,\\
\hat{\lambda}_{i,k_0+1}(\eta_1,\eta_2)/\hat{\lambda}_{i,k_0} (\eta_1,\eta_2)
& = & O_p(p^{\delta_i+\delta_{\min}}n^{-1}).
\end{eqnarray*}
\end{corollary}

Corollary 1 gives
the order of the ratios of the estimated eigenvalues, and shows that the probability of underestimating $k_0$ with (\ref{est_num}) goes to zero asymptotically.
For $k>k_0$, the eigen values $\lambda_{i,k}$ are theoretically zero hence the property of the ratio $\hat{\lambda}_{i,k+1}/\hat{\lambda}_{i,k}$ is difficult to obtain.
Though the consistency of (\ref{est_num}) cannot been confirmed theoretically \citep{lam2012}, the estimator performs well in numerical experiments \citep{chang2015,liu2016}. In Theorem 4 and Theorem 5 we will show that even when the numbers of factors are overestimated,
{ the asymptotic properties of
the threshold value is the same as that when the numbers of factors is known.
The estimated loading spaces with an overestimated $\hat{k}_0$, if restricted to the correct dimension is also consistent. }

Note that the convergence rate of
$\hat{\lambda}_{i,k_0+1}(\eta_1,\eta_2)/\hat{\lambda}_{i,k_0}(\eta_1,\eta_2)$ is
faster in the stronger regime (smaller $\delta_i$).
Since $k_0$ is common to both regimes, we choose
the one identified by the regime with a larger 'strength', reflected
by the scale of
$\|\hat{\bM}_i(\eta_1,\eta_2)\|_2$ \citep{liu2016}. Hence,
we use
\begin{align}\label{est_num2}
\hat{k}=\hat{k}_{\hat{\ell}},
\mbox{ where } \hat{\ell}=\arg \max_{\ell=1,2} \|\hat{\bM}_{\ell}(\eta_1,\eta_2)\|_2.
\end{align}

In the following we present
the asymptotic properties of the proposed estimators when
the number of factors is not correctly estimated.
We will show that if $k_0$ is overestimated,
the proposed
method
can still estimate the threshold value and loading spaces accurately.

Let
\[
\hat{G}_{{k}}(r) =\sum_{i=1}^2
\big\| {\hat{\bB}_{i,{k}}(\eta_1,\eta_2)'}  \,
\hat{\bM}_i(r)\, \hat{\bB}_{i,{k}}(\eta_1,\eta_2)\big\|_2,
\]
where $\hat{\bB}_{i,k}(\eta_1,\eta_2)= (\hat{\bq}_{i,k+1}(\eta_1,\eta_2), \ldots, \hat{\bq}_{i,p}(\eta_1,\eta_2))$, for $i=1,2$.
When $k_0$ is unknown, $r_0$ is estimated by
\[
\widetilde{r}=\arg \min_{r \in \{z_1,\ldots, z_n\} \bigcap (\eta_1,\eta_2)} \hat{G}_{\hat{k}}(r).
\]

The loading spaces are estimated using $\hat{k}$ as the number of factors and $\tilde{r}$ as the threshold value. Specifically,
\[
\widetilde{\bQ}_{i}(\hat{k}, \widetilde{r})= (\hat{\bq}_{i,1}(\widetilde{r}), \ldots, \hat{\bq}_{i,\hat{k}}(\widetilde{r})),
\]
where $\hat{\bq}_{i,k}(\widetilde{r})$ is the unit eigenvector
of $\hat{\bM}_i(\widetilde{r})$ corresponding to its $k$-th
largest eigenvalue, for $i=1,2$.

\begin{theorem}
Under Conditions 1-10 in Appendix A.1, if $p^{\delta_1/2+\delta_2/2}n^{-1/2}=o(1)$ and $k_0<\hat{k}<2k_0-\nu$, it holds that
\begin{eqnarray*}
P(\widetilde{r}<r_0-\epsilon)\leq \frac{Cp^{\delta_1/2+\delta_{\min}/2}}{ \epsilon n^{1/2}}, \quad
P(\widetilde{r}>r_0+\epsilon)\leq \frac{Cp^{\delta_2/2+\delta_{\min}/2}}{ \epsilon n^{1/2}},
\end{eqnarray*}
as $n, p \to \infty$, for $\epsilon>0$, where $\nu=\dim( {\cal M}(\bQ_1) \cap {\cal M}(\bQ_2))$.
\end{theorem}

Theorem 5 will show that the space spanned by the first $k_0$ columns of $\widetilde{\bQ}_i(\hat{k}, \widetilde{r})$ provides an estimate of ${\calM}(\bQ_i)$
which converges as fast as $\cM(\hat{\bQ}_i(\hat{r}))$ in Theorem 3. Define $\widetilde{\bQ}_{i}(\widetilde{r})=(\hat{\bq}_{i,1}(\widetilde{r}), \ldots, \hat{\bq}_{i,k_0}(\widetilde{r}))$, consisting of the first $k_0$ columns of $\widetilde{\bQ}_i(\hat{k},\widetilde{r})$.

\begin{theorem}
Under Conditions 1-10 in Appendix A.1, if $p^{\delta_1/2+\delta_2/2}n^{-1/2}=o(1)$ and $k_0<\hat{k}<2k_0-\nu$, it holds that
\begin{align*}
{\cal D}({\cal M}(\widetilde{\bQ}_i(\widetilde{r})),\,{\cal M}(\bQ_i))&=O_p(p^{\delta_i/2+\delta_{\min}/2}n^{-1/2}), \mbox{\ \  for \ \ } i=1,2,
\end{align*}
as $n$, $p\to \infty$.
\end{theorem}

Theorems 4 and 5 state that when the number of factors is overestimated, our estimators for threshold value and the loading spaces are still consistent. Their asymptotic properties are the same as those when the number of factors is correctly estimated. Of course, when $k_0$ is over estimated,
we lose some efficiency.

\section{Searching for threshold variable}

When there is no prior knowledge on the threshold variable,
a data-driven procedure is needed in order to search for a suitable
one. In standard univariate threshold models, a typical
candidate pool
is the lag variables \citep{tong1980, tong1990, chan1993,tong1993} and
identification is often done by using model comparison procedures.
However, in the high dimensional setting,
the candidate pool can very very large hence the trial-and-error approach can be
extremely time consuming, complicated more by the
multiple comparison problem at the end.
Here we propose a reverse approach that is closely related to
the procedure proposed in
\cite{wu2007}. Specifically, we propose to follow a three step procedure:
classification, screening and model selection.
First, a regime-switching factor
model of \cite{liu2016} is built to obtain an initial regime identification
for each time $t$, without engaging the threshold mechanism. Then
the estimated regime identification is screened against all threshold
variable candidates in a possibly very large
candidate pool and a small set of
candidates is selected by checking whether a candidate
variable will produce regime identifications similar to the estimated
identifications obtained in the classification step. Lastly model
comparison procedure is used to select the most suitable threshold
variable among the small subset.

\noindent
{\bf Classification:} Following \cite{liu2016}, a most likely
regime identification $\hat{I}_t\in \{1,2\}$ is obtained for $t=1,\ldots,n$,
using an iterative procedure of Vertibi algorithm and factor model estimation.
For more details, see \cite{liu2016}. Slightly different from that in
\cite{liu2016}, we assume independent switching instead of Markov switching,
with prior probability $P(I_t=1)=0.5$. Classification step generates initial regime identifications for the next steps. Although it may not be perfect, it is able to reveal the possible relationship between the observed series and threshold variable candidates, and thus provide guidelines to narrows down a large candidates pool.

\noindent
{\bf Screening:} It is noted that if a variable $z_t$ is indeed the threshold
variable and the true threshold value is $r_0$,
then the true regime identification $I_t$ satisfies
$I_t=1+I(z_t\geq r_0)$.
Let
\[
Q(\{z_t\})=\max_{r} \left|\sum_{t=1}^n 2({I}_t-1.5)(2I(z_t \geq r)-1)\right|.
\]
If the variable $z_t$ is the true threshold variable and $r=r_0$ is the
true threshold, then $Q(\{z_t\})=n$, reaching its maximum
value. In fact $Q(\{z_t\})$ is the maximum of a binary CUSUM statistic.

Using the estimated $\hat{I}_t$ from the classification step, we obtain
\[
\hat{Q}(\{z_t\})=\max_{r}\left|
\sum_{t=1}^n 2(\hat{I}_t-1.5)(2I(z_t\geq r)-1)
\right|.
\]
For a set of candidate pool $S=\{z_t^{(\ell)}\}$, we screen each of them by
calculating $\hat{Q}_\ell=\hat{Q}(\{z_t^{(\ell)}\})$, which can be done very efficiently.
The variables with the largest
$\hat{Q}$ values then form a small set of candidates for more careful
examinations.

\noindent{\bf Remark 11.}
It is also possible to identify a linear combination of several variables as a
threshold variable, using a supervised learning algorithm such as support
vector machine or classification tree, with $\hat{I}_t$ obtained from the
classification step as the
response classification.  Since $\hat{I}_t$ is an estimate with
error, the combination parameters needs to be re-estimated under the original
model, similar to our approach of estimating the threshold value $r_0$, though
much more complicated.

\noindent{\bf Remark 12.}
{Searching among a very large pool of candidates will inevitably find spurious relationships and meaningless variables that happen to match the underlying regime switching dynamics in the observed period. Hence a close inspection of the chosen variables is necessary. Prior knowledge is also important. In certain contexts there are known good candidates, such as composite indexes which summarizes contemporaneous information, and their lag variables.
For example, in modeling a panel of economic indicators,
potential threshold variable candidates can be the recession and expansion indicator
of the previous quarter, and its lag variables, since the dynamics of the economy are potentially
different in recession or expansion periods.
In modeling stock returns, the volatility
of the market index and its lag variables can potentially be good threshold variables as
stocks often
behave differently in markets with different volatility.}

\noindent
{\bf Model comparison procedure:} With a small set of possible threshold
variables, a more careful analysis can be carried out. We use the estimation methods in Section 3 to obtain the cross validated residual sum of squares for model comparison.

Specifically, for each threshold variable candidate $z_t$, we estimate
the loading matrices and the
threshold value using data $\{\by_1,\ldots, \by_{t_0}\}$. With those estimates, we calculate the residual sum of squares for the remaining data $\{\by_{t_0+1}, \ldots, \by_{n}\}$,
\begin{align}\label{E}
E=\sum_{t=t_0+1}^n \sum_{i=1}^2 \left(\hat{\bB}_i(\hat{r}) '\by_t\right)' \left(\hat{\bB}_i(\hat{r})' \, \by_t\right) {I}_{t,i}(\hat{r}).
\end{align}
If the threshold variable is correctly identified and $r_0$ is given, then $\bB_{i}'\by_tI_{t,i}(r_0)= \bB_{i}' \bve_{t,i}I_{t,i}(r_0)$. It measures the residual sum of squares after we extract the common factor process. The preferred model is the one with minimum $E$.

\noindent{\bf Remark 13.}
Although we use all samples to build a Markov regime switching model and use it for the purpose of screening, we use a cross validation type of criterion in
(\ref{E}) for model comparison among a small group of candidate threshold variables. The criterion in (\ref{E}) is used simply for computational consideration. More sophisticated approaches such as $m$-fold cross-validation can be used, with additional computational cost.


\noindent{\bf Remark 14.} When calculating $E$, the number of factors is needed. For threshold factor models, even when the number of factors is overestimated, we still can estimate the threshold value and loading spaces as shown in Theorems 4 and 5. Hence, we begin with a one-regime factor model, and estimate the largest possible value for $k_0$. This estimate can be used to compare different threshold variable candidates.

\section{Simulation}
In this section, we demonstrate the performance of the proposed estimators
with numerical experiments, and compare the estimation errors under different settings.

In all the examples, we
use $h_0=1$. The idiosyncratic noise process
$\{\bve_{t,1}, \bve_{t,2}\}$ are independent vector white noise
processes whose covariance matrix has 1 on the diagonal and 0.5 for all
off-diagonal entries.
For estimation of the threshold value, we use the
30-th and 70-th quantiles of $\{z_t\}$ as $\eta_1$ and $\eta_2$.
Estimation error of $\hat{\calM({\bA}_i)}$ is measured by $\calD(\hat{{\cal M}(\bA_i}), {\cal M}(\bA_i))$ where $\calD$ is
defined in (\ref{distance_diff}). The error of $\hat{r}$ is measured by $|\hat{r}-r_0|$. Sample sizes used are $n=200$ and $1000$, and the dimensions
considered are $p=20, 40, 100$. For each setting,
we repeated the simulation 100 times.

In Examples 1 to 3, we consider three settings with different regime strengths.
 In Setting 1, both regimes are strong, with $\delta_1=\delta_2=0$.  In
Setting 2,
one regime is strong, and the other one is weak. In Setting 3,
both regimes are weak. All $p \times k_0$
entries in $\bA_i$ were
generated independently from the uniform distribution on
$[-p^{-\delta_i/2}, p^{-\delta_i/2}]$ to ensure that the strength
of $\bA_i$ is $\delta_i$.
In Example 4, we discuss the impact of the distance $\cD(\cM(\bQ_1), \cM(\bQ_2))$ of the loading spaces of the two regimes on the estimation accuracy.

\medskip
\noindent{\bf Example 1.} In this experiment, we use one
factor process ($k_0=1$) following an AR(1) model with autoregressive
coefficient $0.9$ and N$(0,4)$ noise process. Weak regimes in Settings 2
and 3 are extremely weak with strength 1, which means that
as $p$ increases, we only collect noise and no more useful signal.
The threshold variable $z_t$ is independent of $\bx_t$ and $\by_t$,
following an AR(1) process with AR coefficient 0.3 and $N(0,1)$ noise process.
The threshold value used is $r_0=0$. We assume that $k_0=1$ is known.

\begin{figure}[h]
\centering
\includegraphics[width=2in]{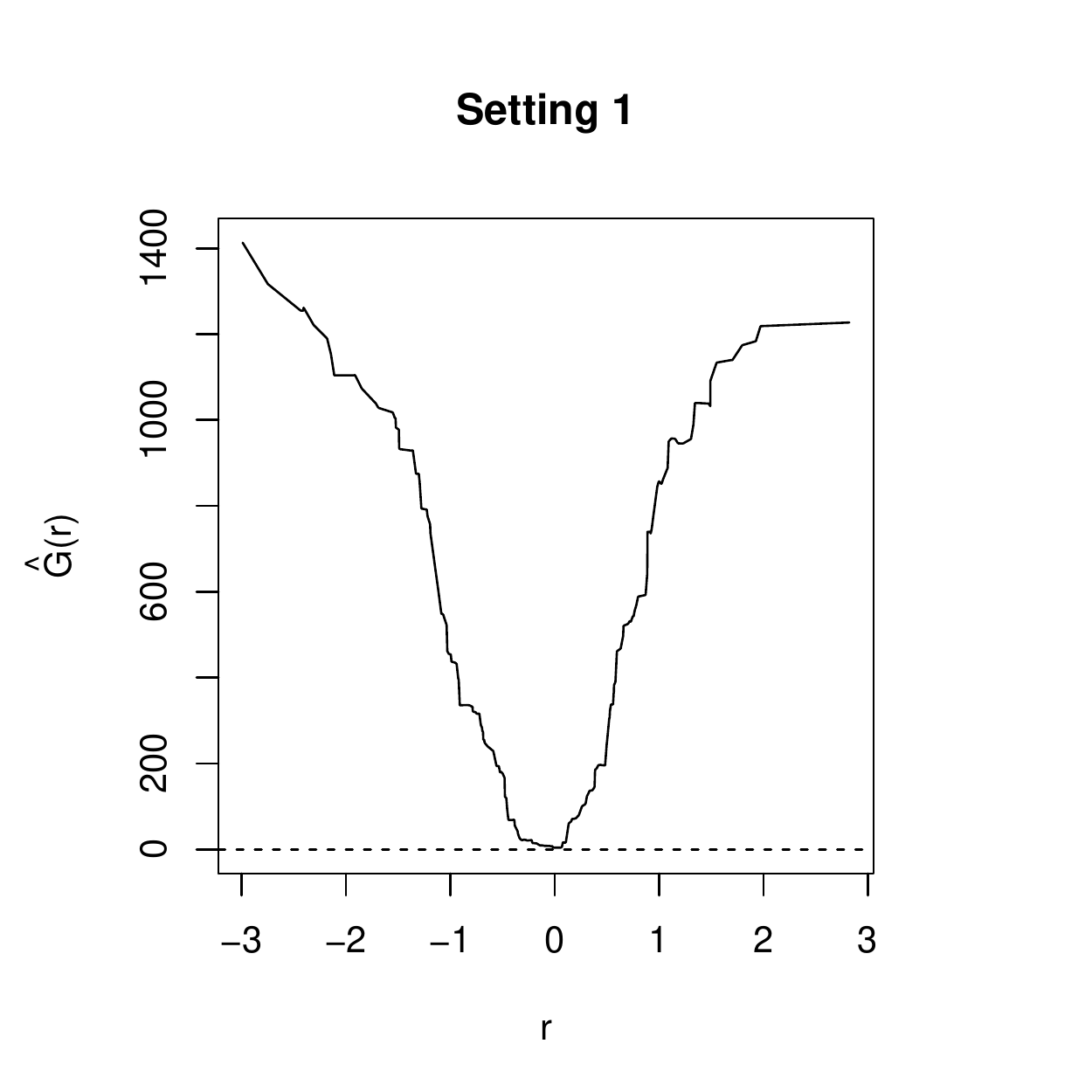}
\includegraphics[width=2in]{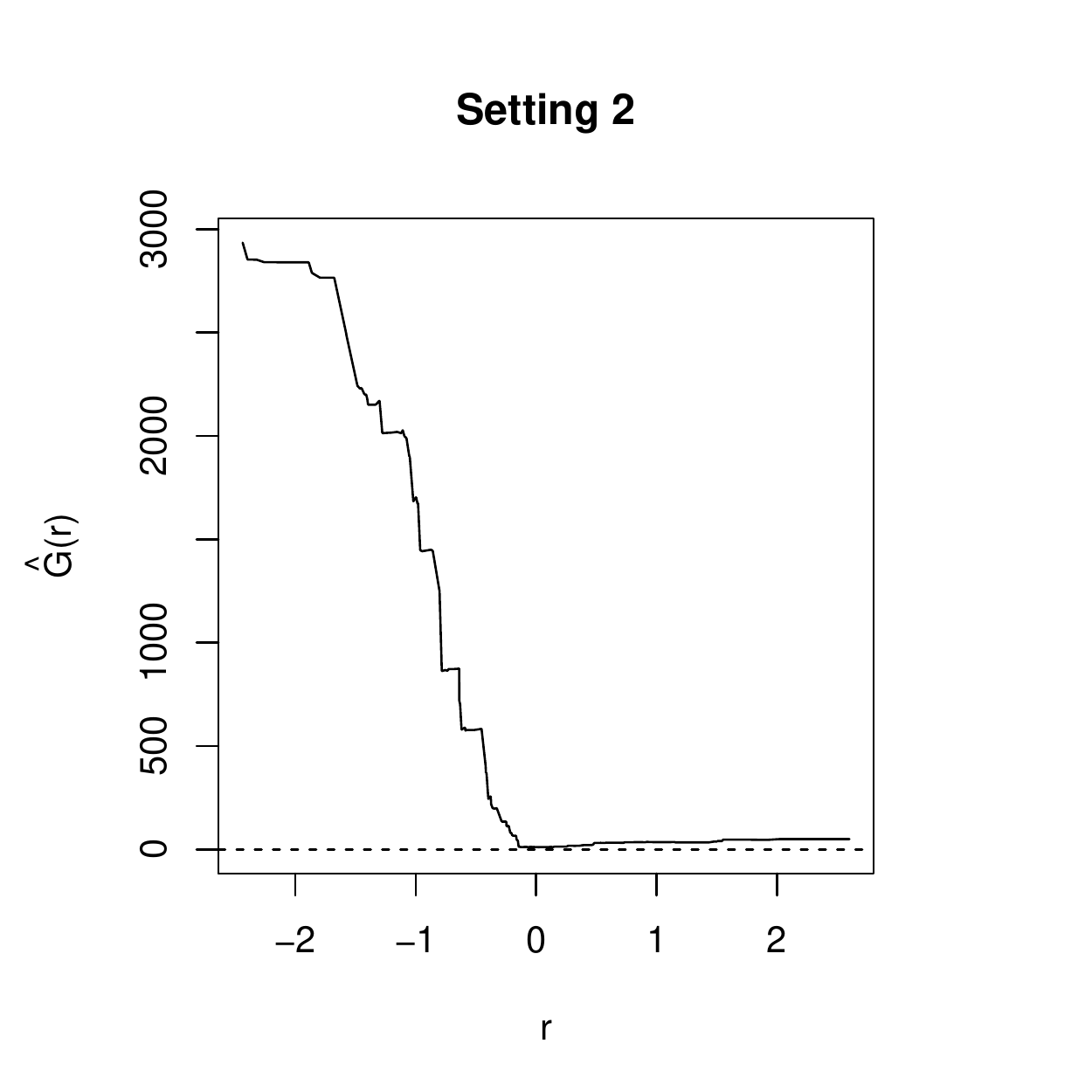}
\includegraphics[width=2in]{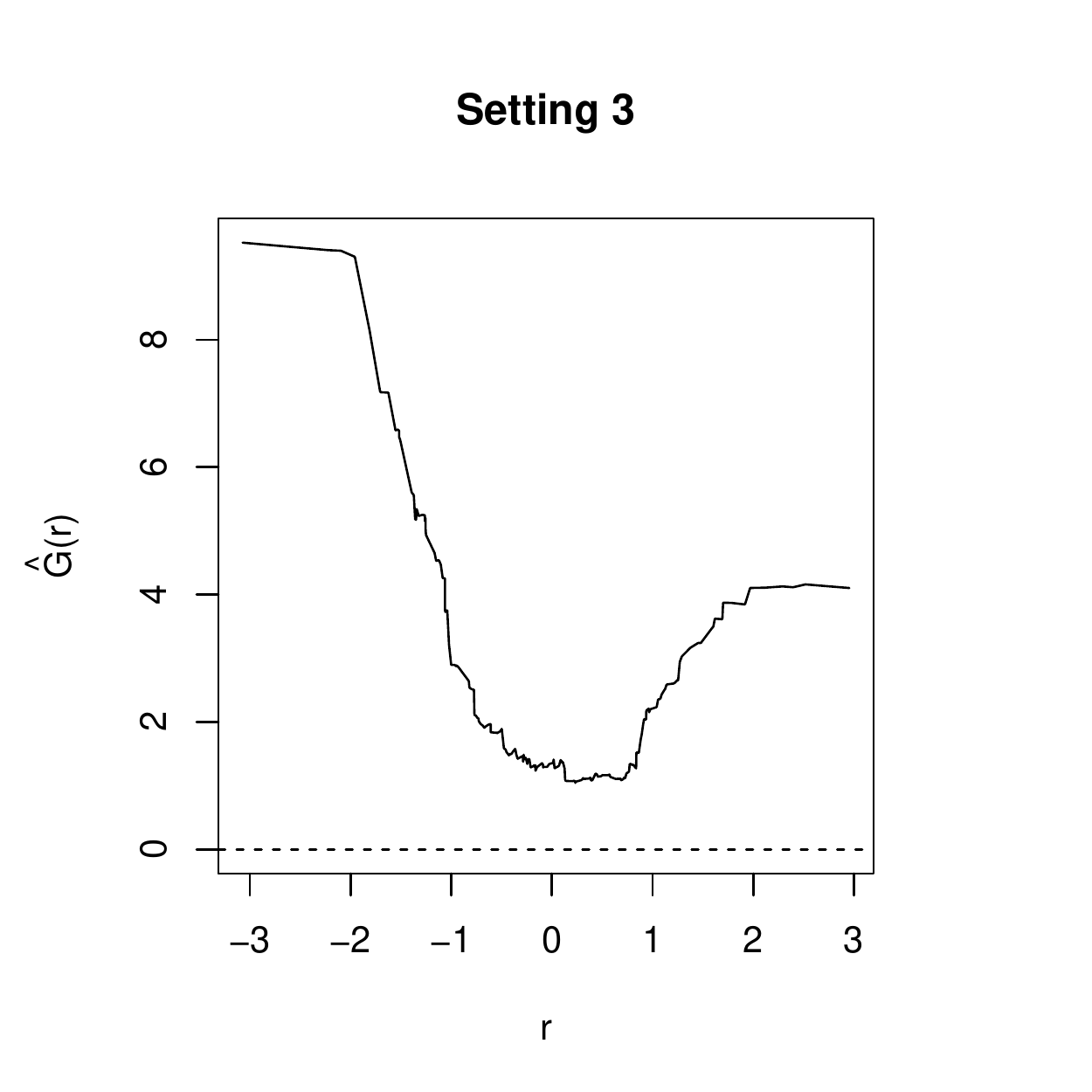}
\caption{Plots of $\hat{G}(r)$ under three settings for a typical data set of Example 1, $n=200$, $p=20$.}
\end{figure}

\begin{figure}[h]
\centering
\includegraphics[width=2in]{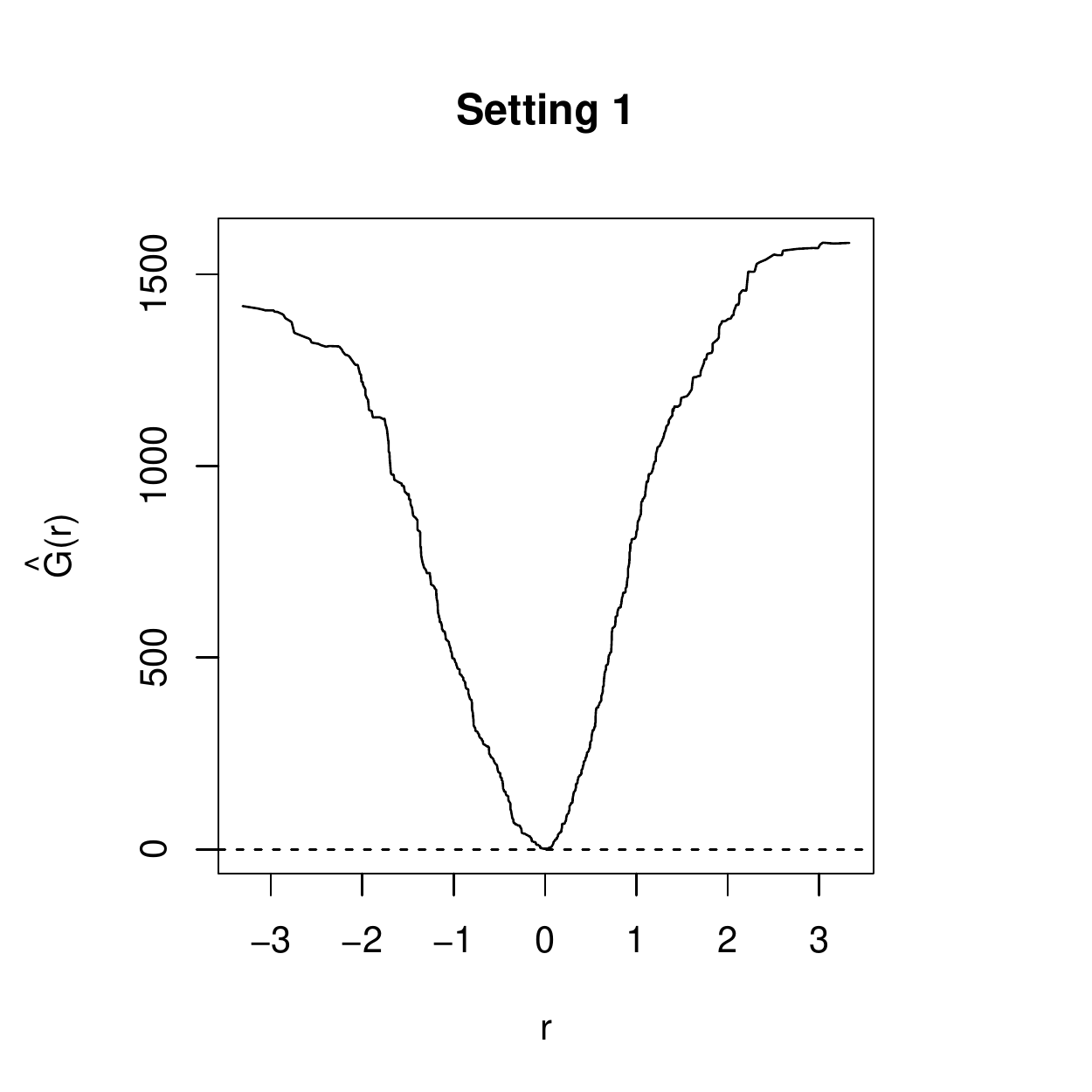}
\includegraphics[width=2in]{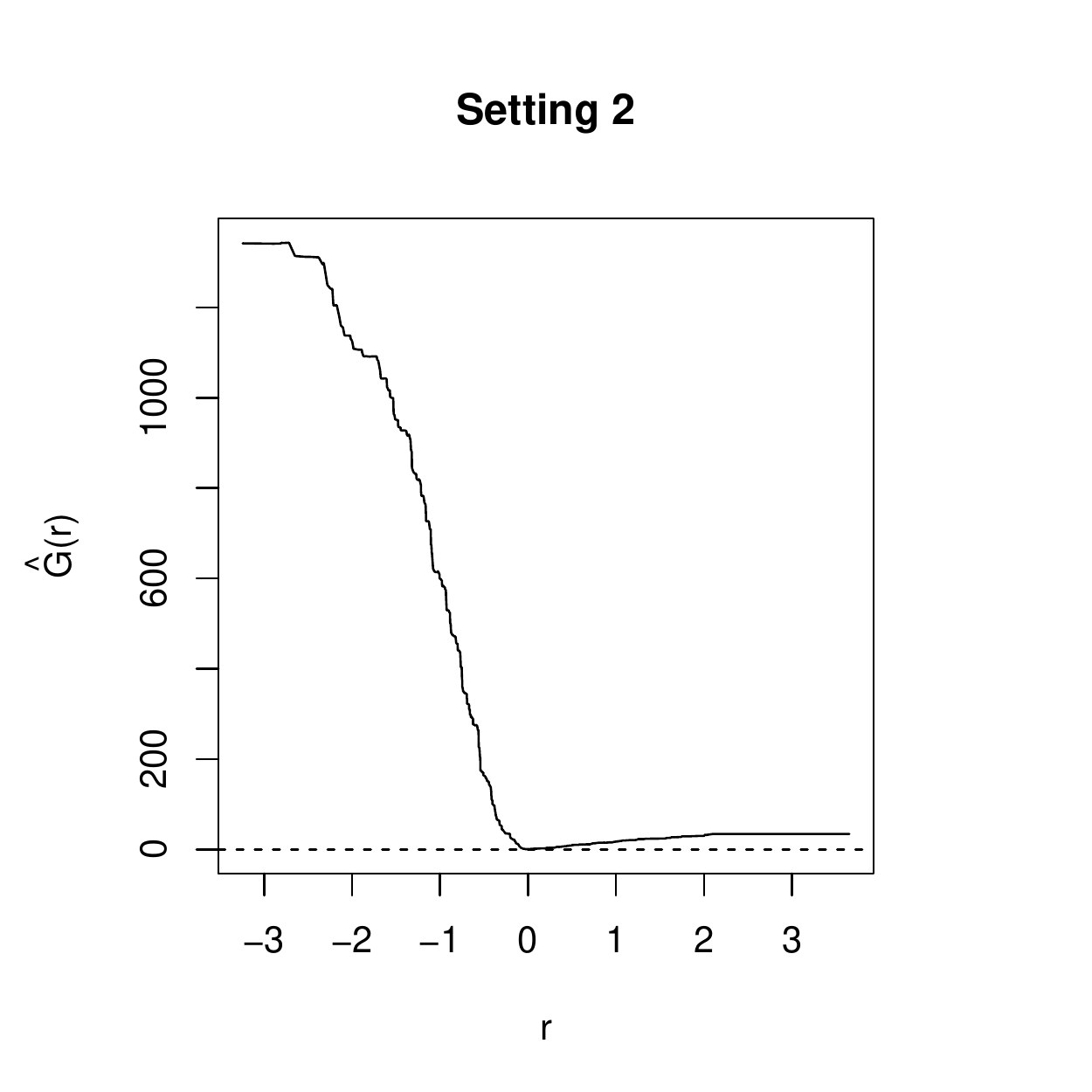}
\includegraphics[width=2in]{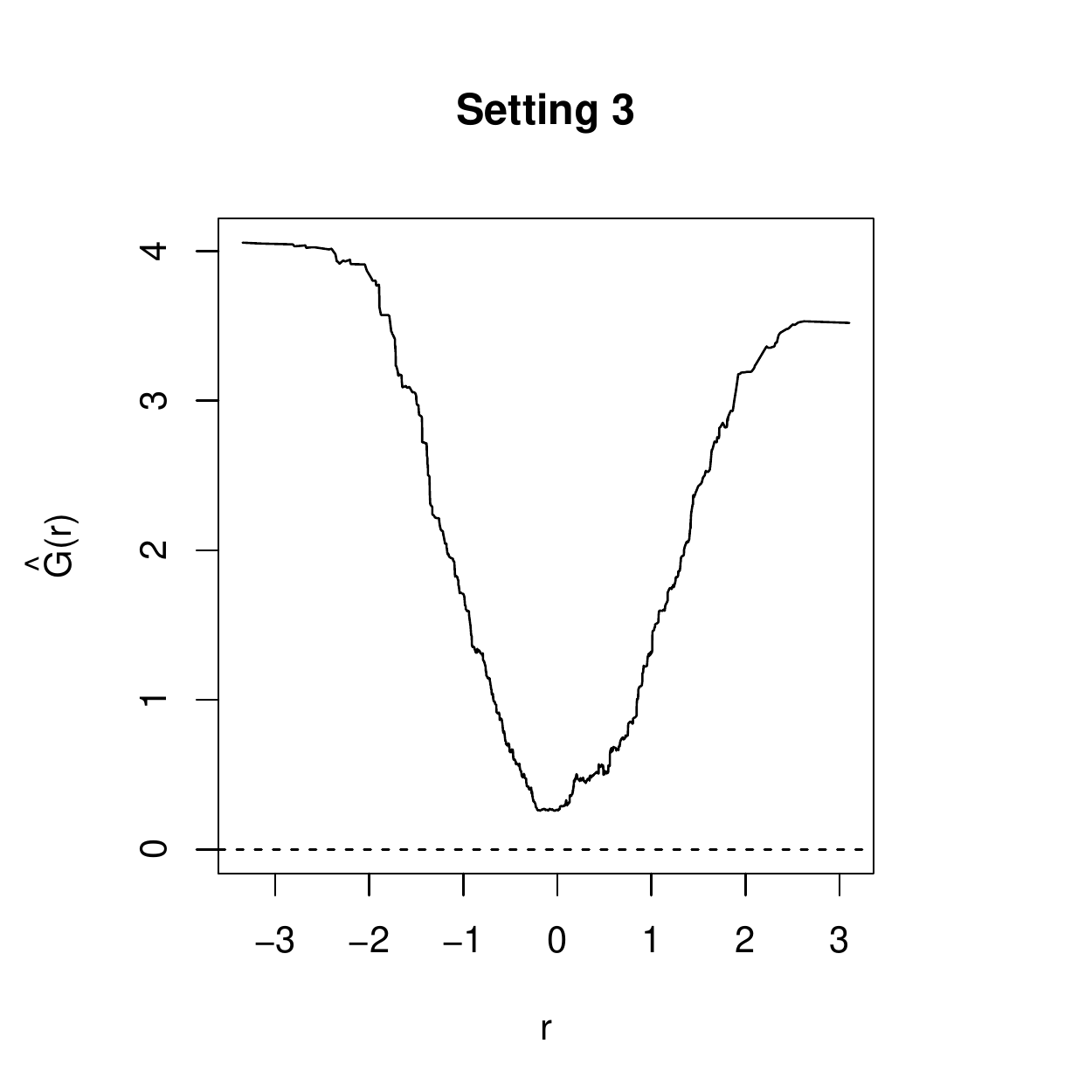}
\caption{Plots of $\hat{G}(r)$ under three settings for a typical data set of Example 1, $n=1000$, $p=20$.}
\end{figure}

Figures 1 and 2 show the function $\hat{G}(r)$ for three
 typical data sets of Example 1, one for each of the three settings,
with $n=200$ and $n=1000$, respectively. Note that the curves are not smooth
since $\hat{G}(r)$ is only evaluated at the discrete set of all observed
values of the
threshold variable.
We can see that $\hat{G}(r)$ approaches the theoretical minimum value 0 of
$G(r)$ around $r=0$ when at least one regime is strong (Settings 1 and 2).
When two regimes are both extremely weak, the range of $\hat{G}(r)$ is very small and the minimum value is above 0, but also occurs around $r=0$. In Setting 2
where two regimes have different levels of strength, $\hat{G}(r)$
is much larger in the stronger regime $r<r_0$ than
that in the weaker regime $r> r_0$, a property shown in Lemma 5 in Appendix A.2.

\begin{table}[htbp!]
\caption{The relative frequency that $\hat{r}<r_0$ when $k_0$ is known for Example 1}
\centering
\begin{tabular}{l| ccc | ccc}
\hline \hline
$n$	    &\multicolumn{3}{c|}{$200$} &\multicolumn{3}{c}{$1000$}\\ \hline
$p$	         &$20$  &$40$	&$100$     &$20$  &$40$  &$100$   \\ \hline
Setting 1    &0.48	&0.48	&0.47	&0.52	&0.49	&0.53	\\ \hline
Setting 2    &0.37	&0.38	&0.41	&0.32	&0.35	&0.30	 \\ \hline
Setting 3    &0.44	&0.50	&0.52	&0.47	&0.51	&0.54 \\ \hline \hline
\end{tabular}
\end{table}

\begin{table}[htbp!]
\caption{Average estimation errors $|\hat{r}-r_0|$ when $k_0$ is known
for Example 1}
\centering
\begin{tabular}{l|c| ccc | ccc}
\hline \hline
$n$	  &  &\multicolumn{3}{c|}{$200$} &\multicolumn{3}{c}{$1000$}\\ \hline
$p$	  &      &$20$  &$40$	&$100$    &$20$  &$40$  &$100$ \\ \hline
Setting 1 & $\hat{r}<r_0$ 	&0.046 &0.070 &0.049	&0.020	&0.017	&0.021\\
          & $\hat{r}>r_0$ 	&0.063 &0.056 &0.060	&0.020	&0.018	&0.014\\ \hline
Setting 2 & $\hat{r}<r_0$ 	&0.089 &0.099	&0.095	&0.029	&0.034	&0.035\\
          & $\hat{r}>r_0$ 	&0.223 &0.230	&0.276	&0.094	&0.117	&0.153\\ \hline
Setting 3 & $\hat{r}<r_0$ 	&0.391 &0.546	&0.537	&0.178	&0.291	&0.587\\
          & $\hat{r}>r_0$ 	&0.413 &0.507	&0.528	&0.162	&0.331	&0.681   \\ \hline \hline
\end{tabular}
\end{table}

\begin{table}[htbp!]
\caption{Average estimation errors $\cD(\hat{{\cM}(\bA_i)},
{\cM}(\bA_i))$, when $k_0$ is known and $\hat{r}<r_0$ for Example 1}
\centering
\begin{tabular}{l|l | rr r|  r rr}
\hline \hline
 $n$ &&\multicolumn{3}{c|}{$200$} &\multicolumn{3}{c}{$1000$}\\ \hline
 $p$ &        &$20$  &$40$   &$100$   	&$20$  &$40$ &$100$	\\ \hline
Setting 1  & ${\cM}(\bA_1)$($\delta_1=0$)  	&0.044	&0.046	&0.043	&0.019	&0.018	&0.019	\\
	      &${\cM}(\bA_2)$ ($\delta_2=0$)  		&0.054	&0.052	&0.050	&0.022	&0.021	&0.023	\\ \hline
Setting 2  &${\cM}(\bA_1)$ ($\delta_1=0$) 	&0.056	&0.060	&0.067	&0.022	&0.026	&0.030	\\
	      &${\cM}(\bA_2)$ ($\delta_2=1$)  		&0.302	&0.445	&0.572	&0.144	&0.193	&0.314	\\ \hline
Setting 3 &${\cM}(\bA_1)$ ($\delta_1=1$)   	&0.318	&0.637	&0.921	 &0.116	&0.165	&0.681	\\
	     &${\cM}(\bA_2)$ ($\delta_2=1$)    	&0.362	&0.557	&0.927	&0.153	&0.250	&0.603	\\ \hline \hline	
\end{tabular}
\end{table}

\begin{table}[htbp!]
\caption{Average estimation errors $\cD(\hat{{\cM}(\bA_i)},
{\cM}(\bA_i))$, when $k_0$ is known and $\hat{r}> r_0$ for Example 1}
\centering
\begin{tabular}{l|l | rrr|  rr r}
\hline \hline
 $n$ & &\multicolumn{3}{c|}{$200$} &\multicolumn{3}{c}{$1000$}\\ \hline
 $p$ &           &$20$  &$40$   &$100$ 	&$20$  &$40$ &$100$	\\ \hline
Setting 1  &${\cM}(\bA_1)$  ($\delta_1=0$)  	&0.055	&0.052	&0.054	 &0.023	&0.022	&0.021	\\
	      &${\cM}(\bA_2)$ ($\delta_2=0$)  		&0.049	&0.040	&0.046  	&0.019	&0.018	&0.018	\\ \hline
Setting 2  &${\cM}(\bA_1)$ ($\delta_1=0$)  	&0.069	&0.073	&0.071	 &0.032	&0.032	&0.033	\\
	      &${\cM}(\bA_2)$ ($\delta_2=1$)  		&0.284	&0.376	&0.527	&0.111	&0.161	&0.264	\\ \hline
Setting 3 &${\cM}(\bA_1)$ ($\delta_1=1$)   	&0.361	&0.590	&0.931	&0.151	&0.260	&0.611	\\
	     &${\cM}(\bA_2)$ ($\delta_2=1$)    	&0.342	&0.641	&0.949	 &0.093	&0.195	&0.722	\\ \hline \hline	
\end{tabular}
\end{table}

Table 1 reports the relative frequency that $\hat{r}<r_0$ for different settings when $k_0$ is known. In Setting 2, the frequencies to underestimate $r_0$ is much lower than these to overestimate $r_0$. The results are in line with our conclusions in Theorem 2.

Tables 2 to 4 show the estimation errors of the threshold value and loading spaces. We report the results under $\hat{r}<r_0$ and $\hat{r}>r_0$ cases
separately to highlight the impact of over- and under-estimate
the threshold value.
Results for threshold value estimation  and loading space estimation share many common characteristics. It is seen
that as sample size increases and as the strength increases,
estimation improves almost in
all settings. Regarding the impact of the dimension,
the estimate accuracy suffers from the curse of dimensionality
when one regime is weak (in Setting 2 and Setting 3);
when two regimes are both strong,
the accuracy does not change much with different values of $p$
(in Setting 1).

From Table 2, it is seen
that misclassification and unbalanced regime strength do
have impacts on the estimation of threshold value.
Estimates are more accurate when $\hat{r}<r_0$ than when $\hat{r}>r_0$ for Setting 2. The reason is that the order of $G(r)$ is higher in the stronger regime ($r<r_0$) than in the weaker regime ($r>r_0$), resulting in a much flatter curve in the region ${r}>r_0$ (See middle panels in Figures 1 and 2 and Lemma 5 in Appendix A.2). Hence, it is more likely to overestimate $r_0$ as shown in
Table 1, and even when the threshold value is underestimated, the error is much smaller as shown in Table 2.

The estimation results for loading spaces also show
some properties that do not apply to those for threshold value. Comparing Regime 1 in Setting 2 to both regimes in Setting 1, we see that the
estimate accuracy of ${\cM}(\bA_1)$ of the stronger regime
does not change much after introducing a weak regime in Table 3 and Table 4.
Comparing the estimation of ${\cM}(\bA_2)$ in Setting 2 to
both ${\cM}(\bA_1)$ and ${\cM}(\bA_2)$  in Setting 3, we can see
that the estimation of ${\cM}(\bA_2)$ of
the weak regime benefits from the existence of a strong regime,
especially when $p$ is large.
There is indeed a 'helping effect' for the weak regime after
adding a strong regime. These observations are in line with the
observations shown  in \cite{liu2016}.

\noindent{\bf Example 2.} In this experiment, we investigate the performance
 of the proposed estimator for the number of factors $k_0$, and
study the estimators of loading spaces and threshold value when
$k_0$ is not correctly estimated. We also
consider the case when
the threshold variable is not correctly identified.
The number of factors here is set to 3. The
factor process is set to be three independent AR(1) processes with N$(0,4)$
noises process and AR coefficients $0.9$, $-0.7$, and $0.8$.
The threshold variable $z_t$ is independent of $\bx_t$ and $\by_t$, following an AR(1) process with AR coefficient -0.7 and $N(0,1)$ noise process. The threshold
value is $r_0=0$. The strength for the weak regimes in Setting 2 and Setting 3 is
set to be $0.5$.

\begin{table}[htbp!]
\caption{The relative frequency estimates of $\hat{k}$ for  Example 2}
\centering
\begin{tabular}{l| ccc |ccc }
\hline \hline
$\hat{k}$  &\multicolumn{3}{c|}{$n=200$} &\multicolumn{3}{c}{$n=1000$}\\ \hline
$p$	            &$20$  &$40$   	&$100$ &$20$  	&$40$   	&$100$         \\ \hline
Setting 1   	&0.66	&0.75	&0.74	&0.97	&0.99	&1.00				 \\ \hline
Setting 2        	&0.77	&0.84	&0.83	&0.99	&0.99	&1.00	\\ \hline
Setting 3     	&0.33	&0.24	&0.18	&0.90	&0.82	&0.75 \\ \hline \hline
\end{tabular}
\end{table}

\begin{figure}[h]
\centering
\includegraphics[width=2in]{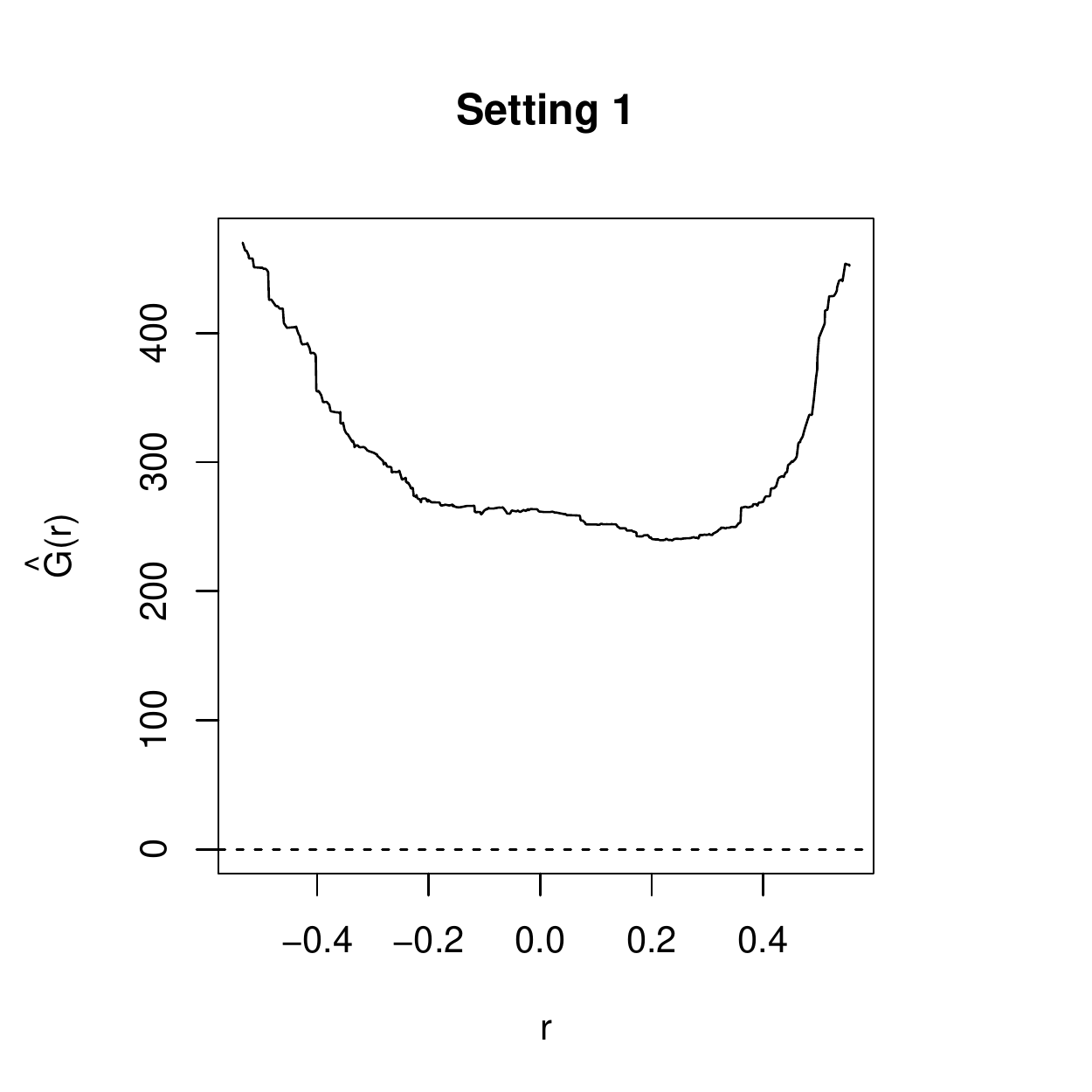}
\includegraphics[width=2in]{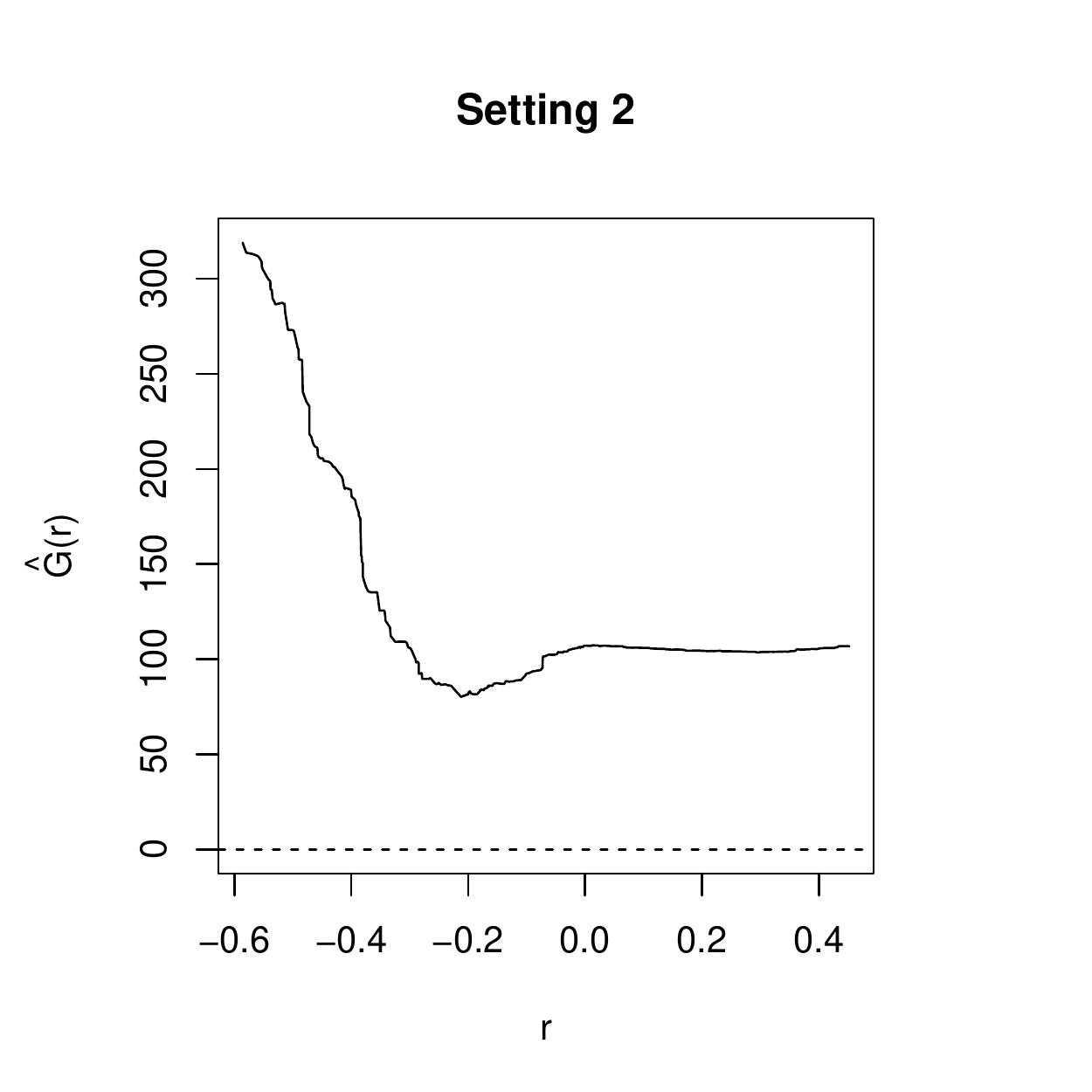}
\includegraphics[width=2in]{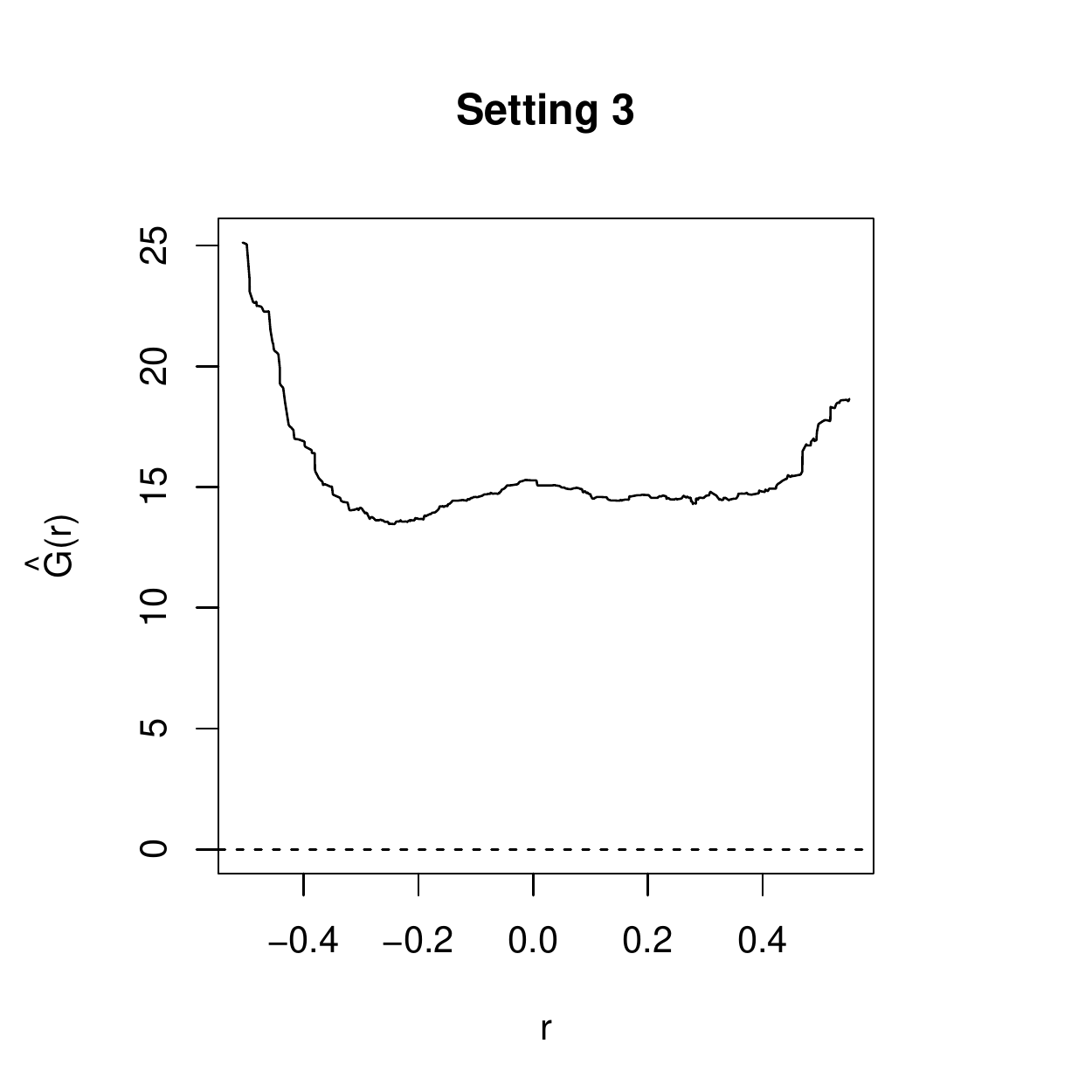}
\caption{Plots of $\hat{G}(r)$ of
a typical data set under each of the three settings for Example 2,
when $n=1000$, $p=20$, and the number of factors is underestimated
($\hat{k}=2$).}
\end{figure}

\begin{figure}[h]
\centering
\includegraphics[width=2in]{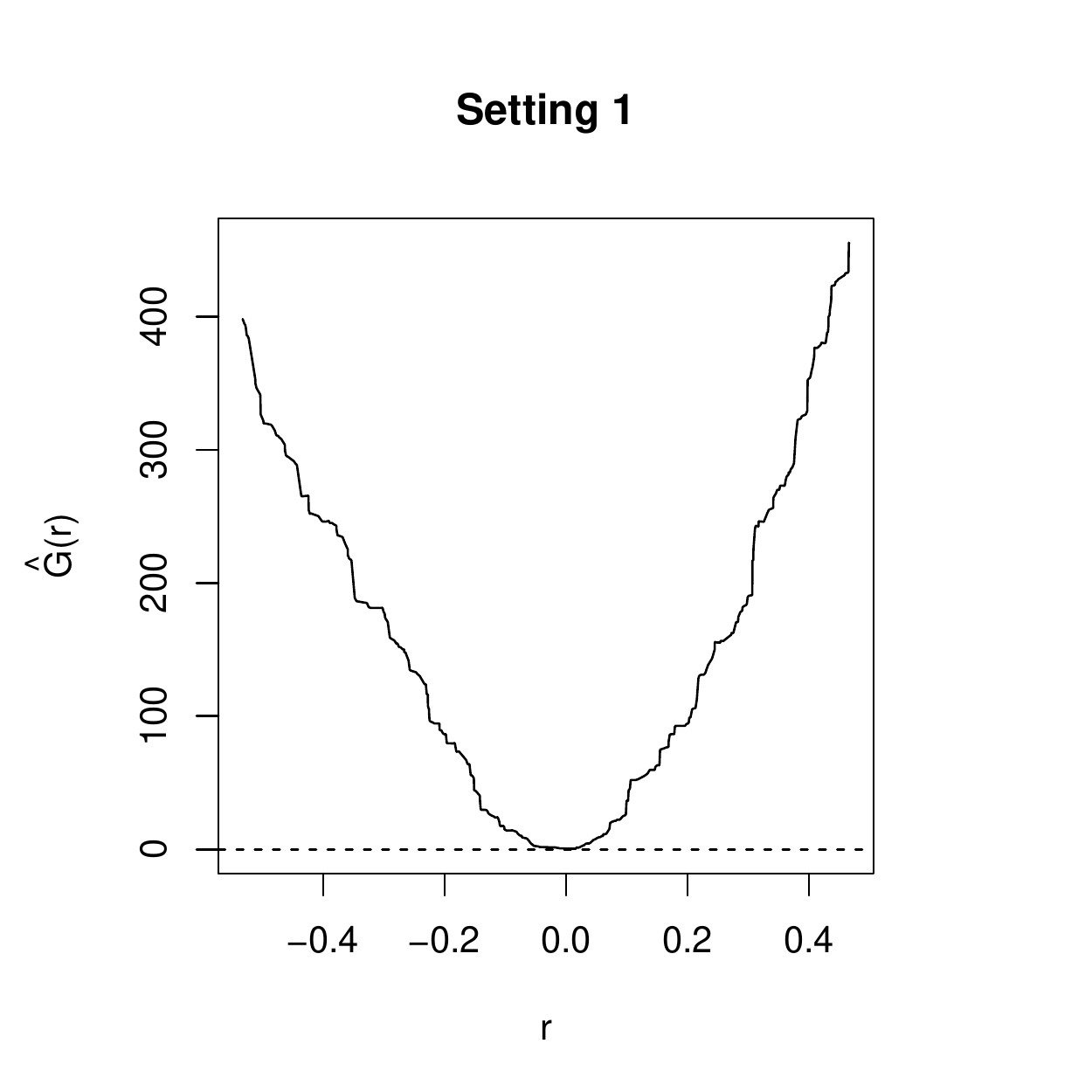}
\includegraphics[width=2in]{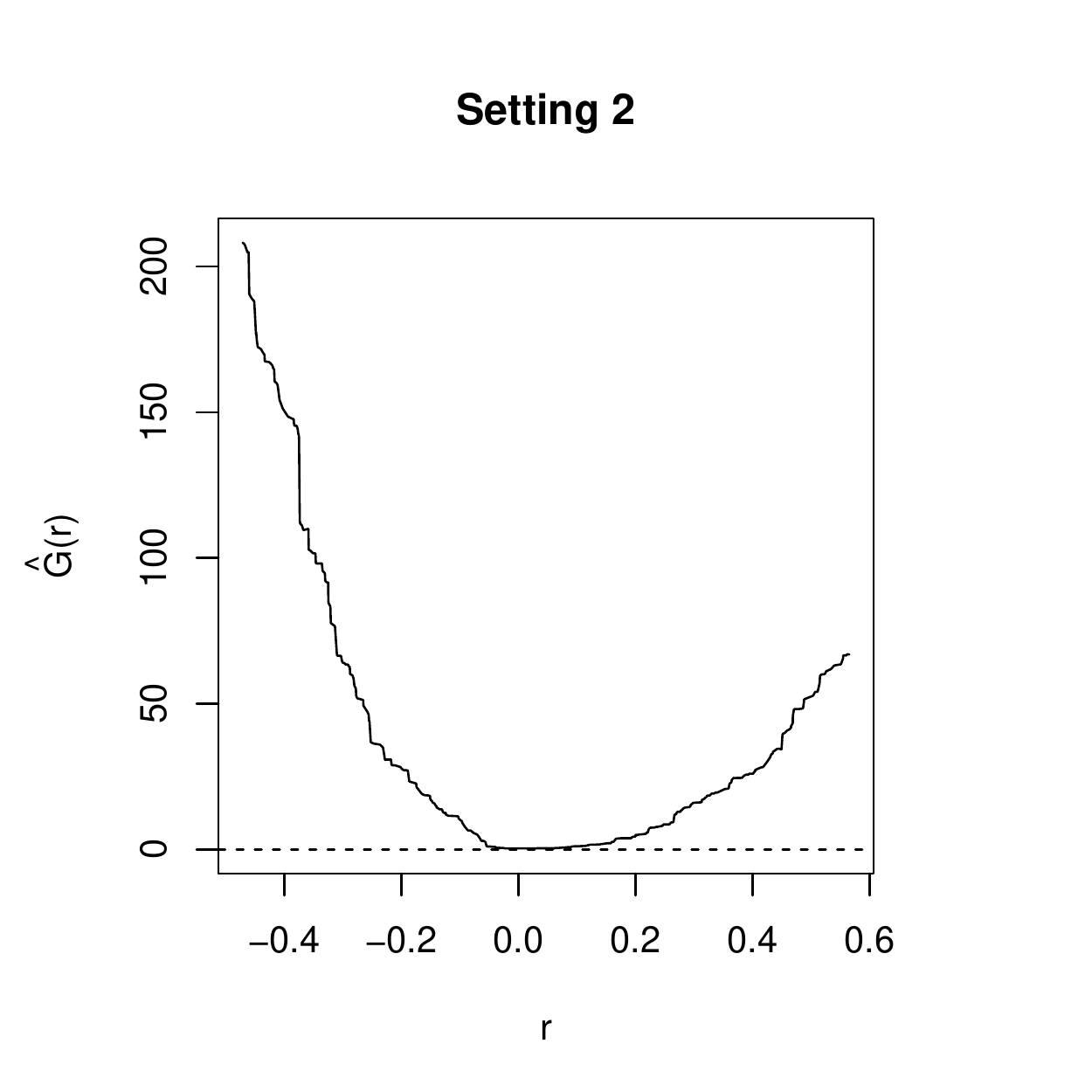}
\includegraphics[width=2in]{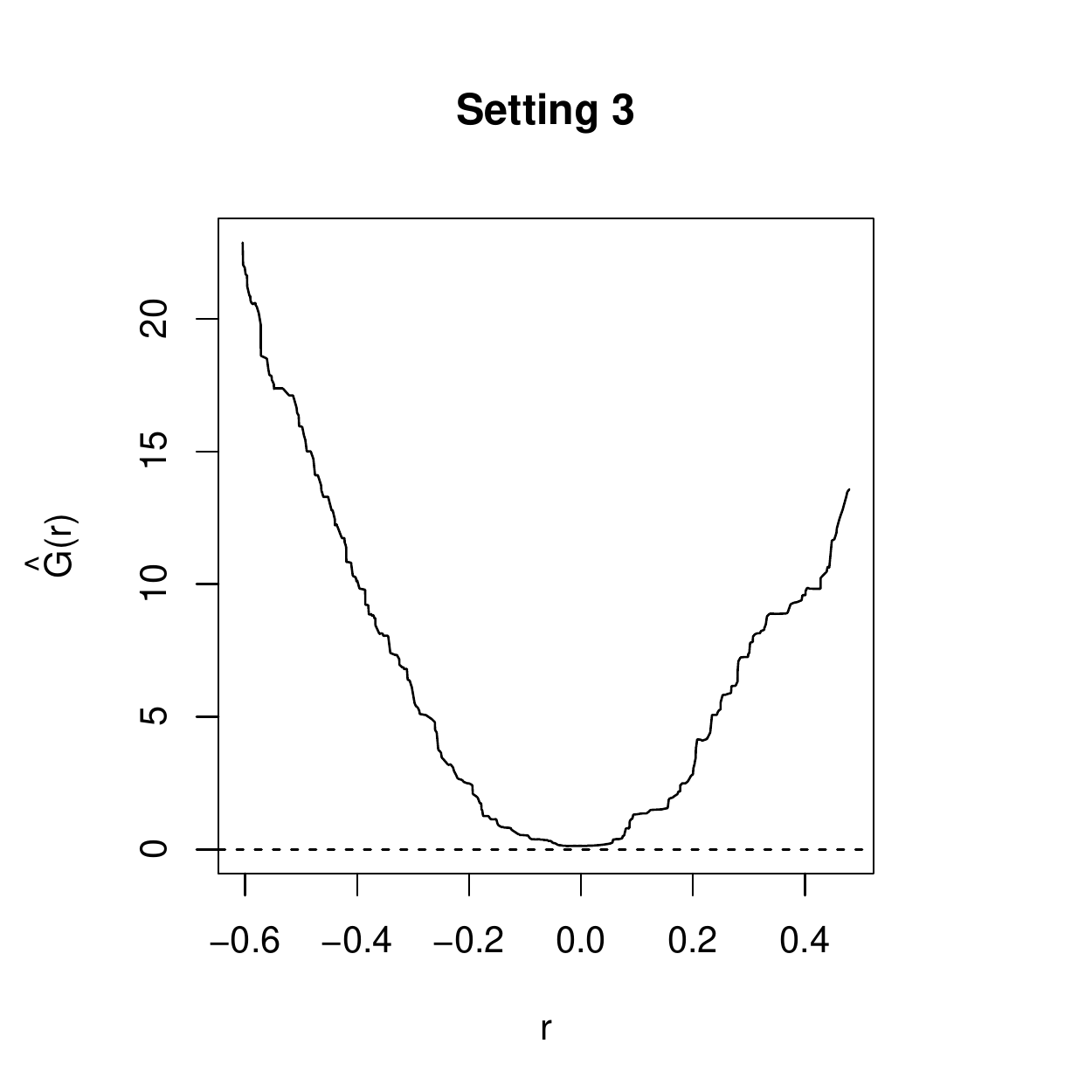}
\caption{Plots of $\hat{G}(r)$ of
a typical data set under each of the three settings for Example 2,
when $n=1000$, $p=20$, and the number of factors is overestimated
($\hat{k}=4$).}
\end{figure}

\begin{table}[htbp!]
\caption{Average estimation error  $|\hat{r}-r_0|$ when $k_0$ is
unknown for Example 2, $n=1000$}
\centering
\begin{tabular}{l| ccc | ccc|ccc}
\hline \hline
$\hat{k}$	    &\multicolumn{3}{c|}{$2$ (underestimated)} &\multicolumn{3}{c|}{$3$ (correctly specified)} &\multicolumn{3}{c}{$4$ (overestimated)}\\ \hline
$p$	        		 &$20$  &$40$	&$100$     &$20$  &$40$  &$100$   &$20$  &$40$  &$100$  \\ \hline
Setting 1    	&0.440	&0.458	&0.474 		&0.035	&0.036	&0.039			&0.012	&0.012	&0.012	 \\ \hline
Setting 2   	&0.363	&0.356	&0.362 		&0.067	&0.075	&0.104     			&0.021	&0.022	&0.030	 \\ \hline
Setting 3    	&0.446	&0.441	&0.471 		&0.090	&0.117	&0.184     			&0.026	&0.029	&0.031   \\ \hline \hline
\end{tabular}
\end{table}

\begin{table}[htbp!]
\small
\caption{Average estimation errors $\cD(\hat{{\cM}(\bA_i)},
{\cM}(\bA_i))$
when $k_0$ is unknown for Example 2, $n=1000$}
\centering
\begin{tabular}{l|l | ccc| ccc|ccc}
\hline \hline
	  &   $\hat{k}$&\multicolumn{3}{c|}{$2$ (underestimated)} &\multicolumn{3}{c|}{$3$ (correctly specified)}&\multicolumn{3}{c}{$4$ (overestimated)}\\ \hline
	  &    $p$                                     &$20$  &$40$   &$100$   	&$20$  &$40$ &$100$  &$20$  &$40$ &$100$	\\ \hline
Setting 1  &${\cM}(\bA_1)$  ($\delta_1=0$)  &0.134	&0.144	&0.148 		&0.038	&0.036	&0.036 		&0.033	&0.033	&0.032	\\
	      &${\cM}(\bA_2)$  ($\delta_1=0$)   &0.160	&0.166	&0.180 		&0.036	&0.036	&0.034		&0.032	&0.032	&0.032	\\ \hline
Setting 2  &${\cM}(\bA_1)$  ($\delta_1=0$) &0.039	&0.034	&0.034 		&0.044	&0.043	&0.043		&0.037	&0.038	&0.038	\\
	      &${\cM}(\bA_2)$  ($\delta_1=0.5$) &0.466	&0.582	&0.695 		&0.091	&0.110	&0.150		&0.083	&0.096	&0.131	\\ \hline
Setting 3 &${\cM}(\bA_1)$  ($\delta_1=0.5$)   &0.161	&0.193	&0.190 		&0.087	&0.105	&0.164		&0.074	&0.087	&0.106	\\
	     &${\cM}(\bA_2)$  ($\delta_1=0.5$)     &0.174	&0.179	&0.215 		&0.084	&0.105	&0.158 		&0.072	&0.083	&0.104	\\ \hline \hline	
\end{tabular}
\end{table}

Table 5 shows the relative frequencies that $\hat{k}=k_0$, when the true threshold variable is chosen but the threshold value $r_0$ is unknown, and only partial data with $\{z_t \leq \eta_1\}$ and $\{z_t \geq \eta_2\}$ are used. As $n$ increases from $200$ to $1000$,
the estimates improve in all settings. For the impact of regime strength, the results show that the existence of a strong regime (Setting 1 and Setting 2) results in much more accurate estimates
for the number of factors $k_0$. Regarding the impact of $p$, it is seen that the estimation performance remains about the same as
$p$ increases, when one or more strong regimes exist,
benefiting from a 'blessing of dimensionality'; see \cite{lam2011}. However, when both regimes are weak, the number of correct estimations may decrease as $p$ increases. This is because the signal to noise ratio in the system
decreases as $p$ increases.

Figures 3 and 4 plot $\hat{G}(r)$ of a typical data set
under each of the three settings for Example 2,
when $\hat{k}$ is underestimated ($\hat{k}=k_0-1$) and overestimated
($\hat{k}=k_0+1$), respectively, with sample size $n=1000$. 
It is seen that,
when $\hat{k}=2$,
$\hat{G}(r)$ does not show a sharp V-shape, and the minimum value is far above 0 in all panels of Figure 3. When $\hat{k}=4$, $\hat{G}(r)$ in Figure 4 reaches its minimum value around $r=0$ for all the settings.

Tables 6 and 7 show the estimation errors of the threshold value and
loading spaces. When the number of factors is overestimated as $\hat{k}=4$, we can obtain consistent estimators for threshold value and loading spaces. It
confirms the conclusions in Theorems 4 and 5. However, when the number of factors is underestimated, the performance of $\hat{\bQ}_i(\hat{k},\widetilde{r})$ and
$\widetilde{r}$ is rather poor. This is because when defining $G(\cdot)$ we take advantage of the complement of loading spaces, $\cM(\bB_i)$. Smaller number of factors makes the complement space to be estimated larger than it is. As a result, the estimates are much worse when $\hat{k}<k_0$ than that when $\hat{k} \geq k_0$.

\begin{table}[htbp!]
\caption{The relative frequency to correctly identify the threshold variable when $k_0$ is unknown and $n=1000$ for Example 2}
\centering
\begin{tabular}{l|ccc|ccc|ccc}
\hline \hline
$\hat{k}$	     &\multicolumn{3}{c|}{$2$ (underestimated)}  &\multicolumn{3}{c|}{$3$ (correctly specified)}&\multicolumn{3}{c}{$4$ (overestimated)}\\ \hline
$p$	         &$20$  &$40$  &$100$   &$20$  &$40$  &$100$   &$20$  &$40$  &$100$  \\ \hline
Setting 1   &0.99  		&0.99	&0.99	&1.00	&1.00	&1.00 	&1.00	&1.00	&1.00		 \\ \hline
Setting 2   &0.30		&0.17	&0.05	&1.00	&1.00	&1.00     	&1.00	&1.00	&1.00	\\ \hline
Setting 3   &0.90		&0.96	&0.85	&1.00	&1.00	&1.00   	 &1.00	&1.00	&1.00	\\ \hline \hline
\end{tabular}
\end{table}

To demonstrate the properties of the threshold variable selection
procedure, four candidates $\{ z_{t-\ell} \mid \ell=0,\ldots,3\}$
are considered in this example.
Table 8 reports the performance of threshold variable selection when $1000$ and $t_0=n/2$. The results show that our method can identify the threshold variable correctly when $k_0$ is correctly specified or overestimated. However, underestimating $k_0$ may lead to poor results, especially in
Setting 2.

Figure 5 plots $\hat{G}(r)$ of a typical data set under each of the three
settings of Example 2, when $z_{t-1}$ is used as the threshold variable.
When the threshold variable is not correctly specified, $\hat{G}(r)$ does
not show a clear V-shape curve. There may be multiple minimum points, and the
minimum value is significantly larger than 0.

\begin{figure}[h]
\centering
\includegraphics[width=2in]{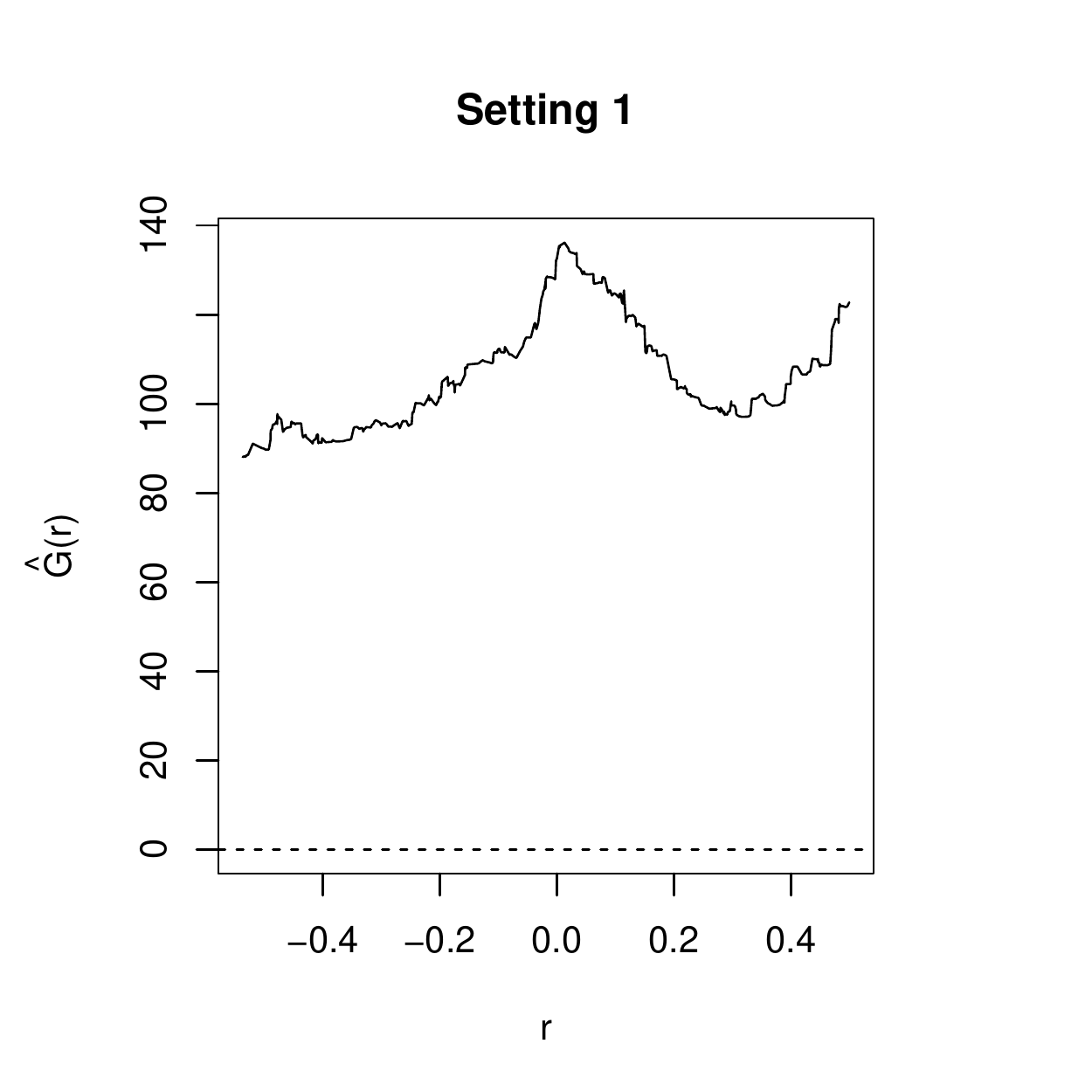}
\includegraphics[width=2in]{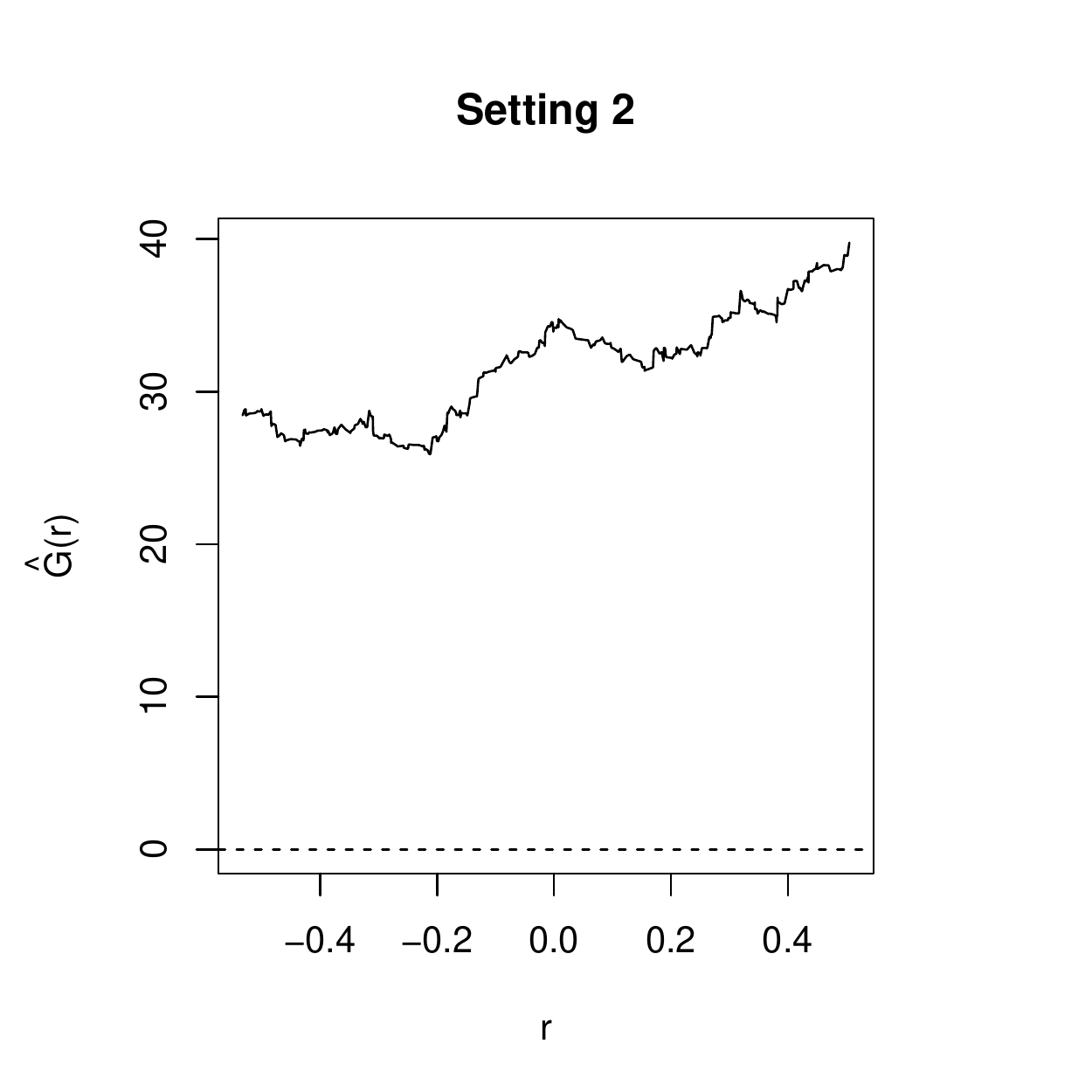}
\includegraphics[width=2in]{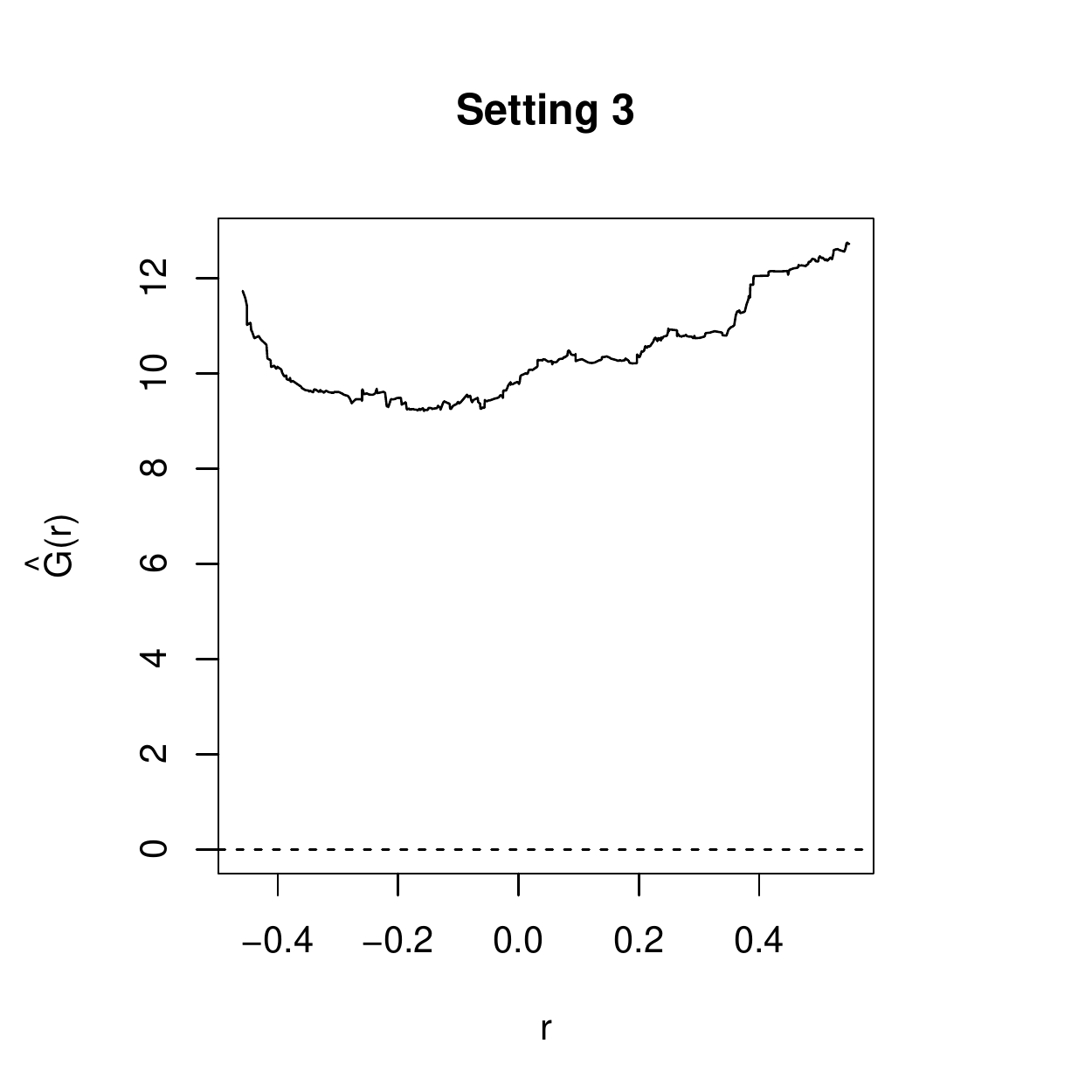}
\caption{Plots of $\hat{G}(r)$ of
a typical data set under each of the three settings
of Example 2, when $n=1000$, $p=20$, and $z_{t-1}$ is used as the threshold variable instead of true threshold variable $z_t$.}
\end{figure}

\noindent{\bf Example 3.} In this experiment, we examine the estimation
performance in a more complicated context, where the threshold variable
$\bz_t $ is correlated to the lag variable of $\by_t$.
Specifically, the threshold variable used is the cross-sectional
standard deviation of $\by_{t-1}$,
  \[
z_t=\sqrt{\frac{1}{p-1}\sum_{q=1}^p (y_{t-1,q}-\bar{y}_{t-1})^2},
\]
where $y_{t,q}$ is the $q$-th entry in $\by_t$ and $\bar{y}_{t}=\sum_{q=1}^p{y_{t,q}}/p$, for $t=1,\ldots, n$.
The factor process is generated from an AR(1) process with AR
coefficient 0.9 and N$(0,4)$ noise process. The strength of the
 weak regimes in Setting 2 and Setting 3 is set to be 0.5.
The threshold values used are $r_0=1.5, 1.2$ and $1$ for Settings 1 to 3,
respectively.

\begin{table}[htbp!]
\caption{Average estimation errors $|\hat{r}-r_0|$ when $k_0$ is known for Example 3}
\centering
\begin{tabular}{l| ccc| ccc}
\hline \hline
$n$	    &\multicolumn{3}{c|}{$200$} &\multicolumn{3}{c}{$1000$}\\ \hline
$p$	         &$20$  &$40$   &$100$   &$20$  &$40$   &$100$   \\ \hline
Setting 1    &0.174	&0.162	&0.187		&0.056	&0.052	&0.047	 \\ \hline
Setting 2    &0.085	&0.072	&0.094		&0.037	&0.034	&0.032	 \\ \hline
Setting 3    &0.093	&0.104	&0.128		&0.042	&0.057	&0.076	\\ \hline \hline
\end{tabular}
\end{table}

\begin{table}[htbp!]
\caption{Average estimation errors $\cD(\hat{{\cM}(\bA_i)},
{\cM}(\bA_i))$ when $k_0$ is known for Example 3}
\centering
\begin{tabular}{l|l | rr r|  rr r}
\hline \hline
 $n$ & &\multicolumn{3}{c|}{$200$} &\multicolumn{3}{c}{$1000$}\\ \hline
 $p$ &       &$20$  &$40$   &$100$   	&$20$  &$40$ &$100$	 \\ \hline
Setting 1  & ${\cM}(\bA_1)$ ($\delta_1=0$)  	&0.142	&0.143	&0.143	&0.053	&0.051	&0.052	\\
	      &${\cM}(\bA_2)$ ($\delta_2=0$)  		&0.032	&0.032	&0.032	&0.013	&0.013	&0.013	\\ \hline
Setting 2  &${\cM}(\bA_1)$ ($\delta_1=0$) 	&0.113	&0.096	&0.079	&0.052	&0.045	&0.037	\\
	      &${\cM}(\bA_2)$ ($\delta_2=0.5$)  	&0.007	&0.085	&0.139	&0.027	&0.033	&0.054	\\ \hline
Setting 3 &${\cM}(\bA_1)$ ($\delta_1=0.5$)   	&0.279	&0.323	&0.436	&0.101	&0.118	&0.118	\\
	     &${\cM}(\bA_2)$ ($\delta_2=0.5$)    	&0.071	&0.089	&0.142	&0.029	&0.040	&0.070\\ \hline \hline	
\end{tabular}
\end{table}

Tables 9 and 10 present the estimation errors for threshold value and loadings spaces. They show a similar pattern to the results in Example 1.
Note that the three settings are not comparable in this case because each
setting yields different sample size in each regime due to the dependency
of $z_t$ and $\by_{t-1}$.


\noindent{\bf Example 4.} We present the influence of the distance of loading spaces $\cD(\cM(\bQ_1), \cM(\bQ_2))$
on the estimation accuracy. One factor ($K=1$) is used since the distance between two one-dimensional spaces is easier to interpret. For simplicity, strong regimes $\delta_1=\delta_2=0$ are used. The latent factor $\{x_t\}$ follows an AR(1) process with AR coefficient 0.9 and $N(0,1)$ noise process.
$\{z_t\}$ follows i.i.d $N(0,1)$, and threshold value is 0. The $i$-th entries in $\bA_1$ and $\bA_2$ are generated from a bivariate normal distribution with unit variances and covariance $(1-d^2)$ so that
$\cD(\cM(\bA_1), \cM(\bA_2))\approx |d|$. Simulation is repeated 100 times for each setting of $d$.

Table \ref{dist_r0} reports the average estimation errors for threshold value.
It is not surprising that when the two loading spaces are well apart, the estimates are more accurate. For the case of $d=0$ (there is no switching mechanism), the threshold value $r_0$ is unavailable ($r_0=0$ is used in the calculation in the table). The forcedly estimated $\hat{r}$ is random in the middle of
the range of $z_t\sim N(0,1)$.

Table \ref{dist_load} reports the average estimation errors of the loading spaces. As expected, the accuracy does not change as $p$ increases since both regimes are strong. Larger sample size $n$ improves the accuracy. It is noted that the distance between the two loading spaces has almost no effect on the estimation accuracy of the loading spaces, even the estimated threshold values are not accurate when the distance is small as shown in Table \ref{dist_r0}. One possible reason may be that the distortion created by the misclassified observations (due to the misplaced threshold value) is relatively small, as the two loading spaces are close. In the case of $d=0$, there is no distortion.

Table \ref{dist_var} shows the average and sample standard deviation of the distance between the estimated loading spaces. It is seen that the distances of the estimated loading spaces are close to the true distance with small variation, since the estimated loading spaces are estimated accurately as shown in Table \ref{dist_load}. The variation of the estimated space distance is smaller for larger $p$, as in this case the distance is related to the sample correlation of two vectors ($\hat{\bQ}_1$ and $\hat{\bQ}_2$) of length $p$. The variation does not change much for larger $n$.

\begin{table}[htbp!]
\caption{Average estimation errors $\cD(\hat{{\cM}(\bA_i)},
{\cM}(\bA_i))$ when $k_0$ is known for Example 4}
\centering
\begin{tabular}{l|l | rr r|  rr r}
\hline \hline \label{dist_load}
 $n$ & &\multicolumn{3}{c|}{$200$} &\multicolumn{3}{c}{$1000$}\\ \hline
 $p$ &       &$20$  &$40$   &$100$   	&$20$  &$40$ &$100$	 \\ \hline
$d=1$  & Regime 1&0.054	&0.054	&0.054	&0.024	&0.024	&0.023	\\
	      &Regime 2						&0.052	&0.054	&0.050	&0.023	&0.023	&0.024	\\ \hline
$d=0.7$  &Regime 1 	&0.052	&0.053	&0.049	&0.023	&0.022	&0.022	\\
	      &Regime 2				            &0.052	&0.051	&0.052	&0.022	&0.021	&0.021\\ \hline
$d=0.3$  &Regime 1 	&0.050	&0.052	&0.051	&0.022	&0.021	&0.020	\\
	      &Regime 2			            	&0.052	&0.049	&0.051	&0.021	&0.022	&0.022\\ \hline
$d=0.2$  &Regime 1 	&0.051	&0.051	&0.051	&0.024	&0.021	&0.020	\\
	      &Regime 2				            &0.052	&0.051	&0.048	&0.022	&0.022	&0.022\\ \hline
$d=0.1$  &Regime 1 	&0.050	&0.048	&0.051	&0.023	&0.023	&0.021	\\
	      &Regime 2				            &0.049	&0.048	&0.052	&0.022	&0.021	&0.021\\ \hline
$d=0$  &Regime 1 	&0.049	&0.048	&0.049	&0.023	&0.021	&0.022	\\
	      &Regime 2				            &0.054	&0.049	&0.048	&0.022	&0.031	&0.021\\
	      \hline \hline	

\end{tabular}
\end{table}

 \begin{table}[htbp!]
\caption{Average and standard deviation(in brackets) of  $\cD(\hat{\cM(\bA_1)},\hat{\cM(\bA_2)})$ for Example 4}
\centering
\begin{tabular}{ l| ccc| ccc}
\hline \hline \label{dist_var}
$n$	  &\multicolumn{3}{c|}{$n=200$} &\multicolumn{3}{c}{$n=1000$}\\ \hline
$p$	         &$20$  &$40$   &$100$   &$20$  &$40$   &$100$   \\ \hline
$d=1$       &0.972(0.041)	&0.986(0.018)	&0.993(0.007)   &0.9723(0.036)	&0.990(0.013)	&0.995(0.007)		 \\ \hline
$d=0.7$     &0.685(0.117)	&0.696(0.083)	&0.688(0.051)    &0.701(0.100)	&0.696(0.083)	&0.697(0.042)		 \\ \hline
$d=0.3$     &0.290(0.066)	&0.291(0.055)	&0.304(0.031)    &0.300(0.068)	&0.301(0.041)	&0.296(0.029)		 \\ \hline
$d=0.2$   &0.199(0.045)	    &0.202(0.035)	&0.199(0.026)&0.190(0.047)	    &0.192(0.033)	&0.195(0.019)		 \\ \hline
$d=0.1$   &0.118(0.037)	    &0.117(0.028)	&0.116(0.018)   &0.099(0.025)	    &0.099(0.016)	&0.100(0.012)		 \\ \hline
$d=0$    &0.075(0.034) 	    &0.074(0.032)	&0.069(0.027)   &0.029(0.011) 	    &0.031(0.011)	&0.040(0.097)		\\ \hline \hline
\end{tabular}
\end{table}

\begin{table}[htbp!]
\caption{Average estimation errors $|\hat{r}-r_0|$ when $k_0$ is known for Example 4}
\centering
\begin{tabular}{ l| ccc| ccc}
\hline \hline \label{dist_r0}
$n$	  &\multicolumn{3}{c|}{$n=200$} &\multicolumn{3}{c}{$n=1000$}\\ \hline
$p$	         &$20$  &$40$   &$100$   &$20$  &$40$   &$100$   \\ \hline
$d=1$   	&0.043	    &0.048	&0.050		&0.020	&0.019	&0.019	 \\ \hline
$d=0.7$   &0.057	&0.061	&0.047		&0.020	&0.018	&0.016	 \\ \hline
$d=0.3$   &0.099	&0.086	&0.086		&0.035	&0.033	&0.033	 \\ \hline
$d=0.2$   &0.097	&0.114	&0.090		&0.052	&0.042	&0.041	 \\ \hline
$d=0.1$   &0.180	&0.148	&0.161		&0.092	&0.079	&0.084	 \\ \hline
$d=0$    &0.278 	&0.190	&0.187		&0.174	&0.192	&0.180	\\ \hline \hline
\end{tabular}
\end{table}

\section{Real Example}
We applied the proposed
approach to the daily returns of 123 stocks from January 2,
2002 to July 11, 2008. These stocks were selected among those included in the S\&P 500 and traded every day during the period. The returns were calculated in percentages based on daily closing prices.
This data set was analyzed by \cite{lam2012, chang2015, liu2016}.
The sample size $n$ is $1642$ and
 the dimension $p$ is $123$. We use $h_0=1$ in this analysis.

\cite{lam2012} and \cite{chang2015} used a factor model (with no switching) to
analyze the data. The estimated number of factors is 2.
\cite{liu2016} used a Markov switching factor model on the same data set
and found that there are two regimes, with one factor
in each regime. Here we analyze the data using a 2-regime threshold factor model.
We consider the lag cross-sectional standard deviation of $\by_{t-\ell}$ and the lag of squared S\&P 500 return $r_{t-\ell}$ as
the candidates for the threshold variable ($\ell=1,\ldots, 8)$. Since an overestimated number of factors still can identify the threshold variable, $\hat{k}=2$ is tentatively used when calculating $E$ defined in (\ref{E}). Table \ref{EE}
shows the value of $E$ for each candidate with $t_0=n/2$.
In the following we use the
cross-sectional standard deviation of $\by_{t-6}$
as the threshold variable, since it minimizes $E$.

\begin{table}[htbp!]
\footnotesize
\caption{$E$ divided by $10^5$ for all threshold variable candidates for real data analysis}
\centering
\begin{tabular}{l| cccccccc}
\hline \hline \label{EE}
Lag	     & $z_{t-1}$  &$z_{t-2}$ &$z_{t-3}$   &$z_{t-4}$  &$z_{t-5}$  &$z_{t-6}$ &$z_{t-7}$ &$z_{t-8}$\\ \hline
Lag of cross-section standard deviation   	
&2.097	&2.051	 &2.060	&1.984	&2.091 &1.975  &2.067 &1.993\\ \hline
Lag of squared S\&P return    			
&2.069	&2.053	 &2.069	&2.164	&2.043	&2.069	&2.086	&2.055	
\\ \hline \hline
\end{tabular}
\end{table}

We use 10-th and 90-th percentiles of the threshold variable as $\eta_1$ and
$\eta_2$ to estimate the number of factors. The left and right panels in Figure 6 display the ratio of eigenvalues of $\hat{\bM}_1(\eta_1,\eta_2)$ and $\hat{\bM}_2(\eta_1, \eta_2)$, respectively, where both the ratios reach their minimum values at 1. It yields that $\hat{k}=1$.

The above results indicate that only one factor drives the 123 stocks, but
the factor loadings switch between regimes according to the
cross-sectional standard deviation 6 trading days before. Ignoring
switching structure as in \cite{lam2012}, it would appear that there are two
different factors. Introducing a threshold variable reduces the number of factor by 1.

Figure 7 plots $\hat{G}(r)$, showing a V-shape curve with a relatively flat
bottom. By minimizing $\hat{G}(r)$, we have that $\hat{r}=1.850$.

\begin{figure}[htpb!]
\centering
\includegraphics[width=2.25in]{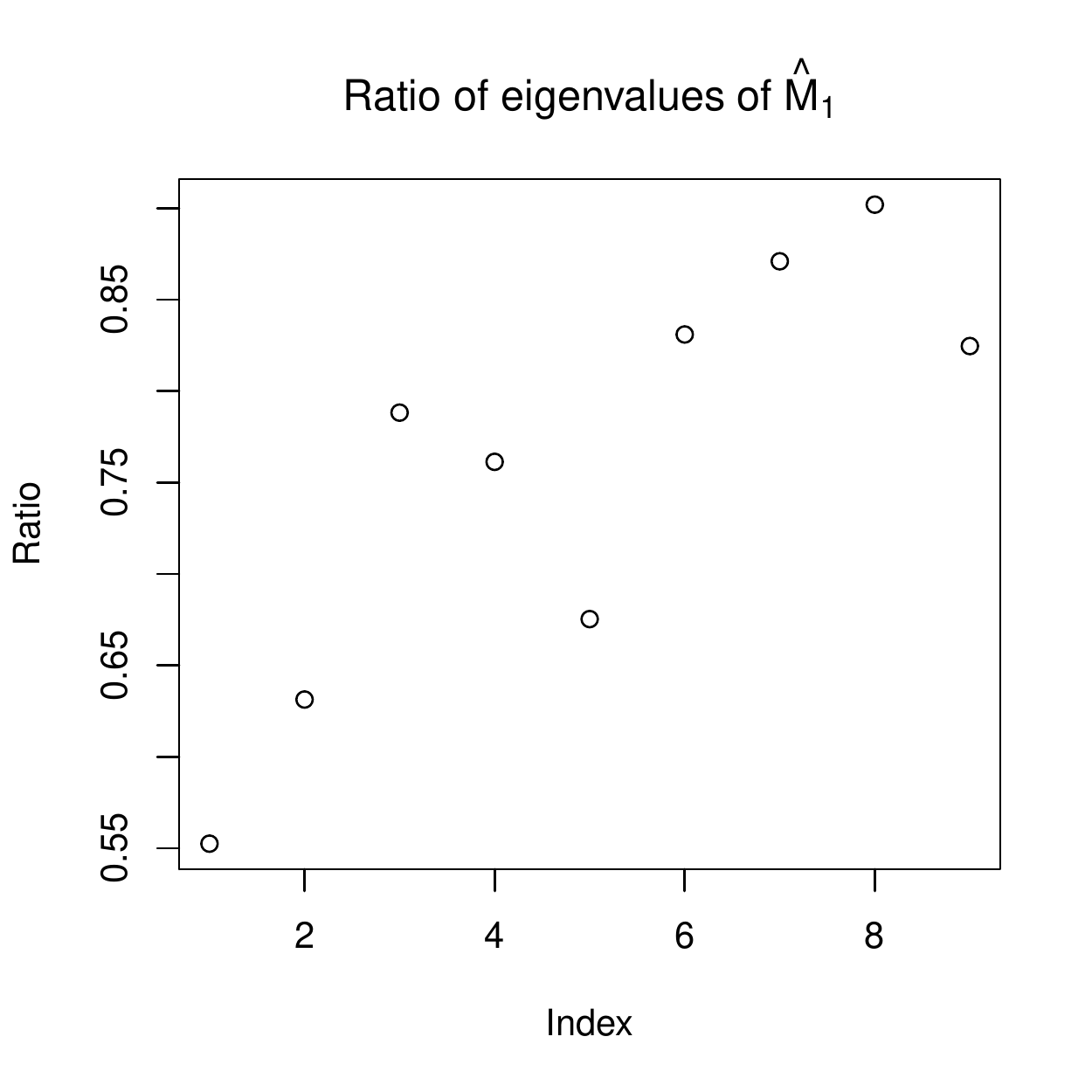}
\includegraphics[width=2.25in]{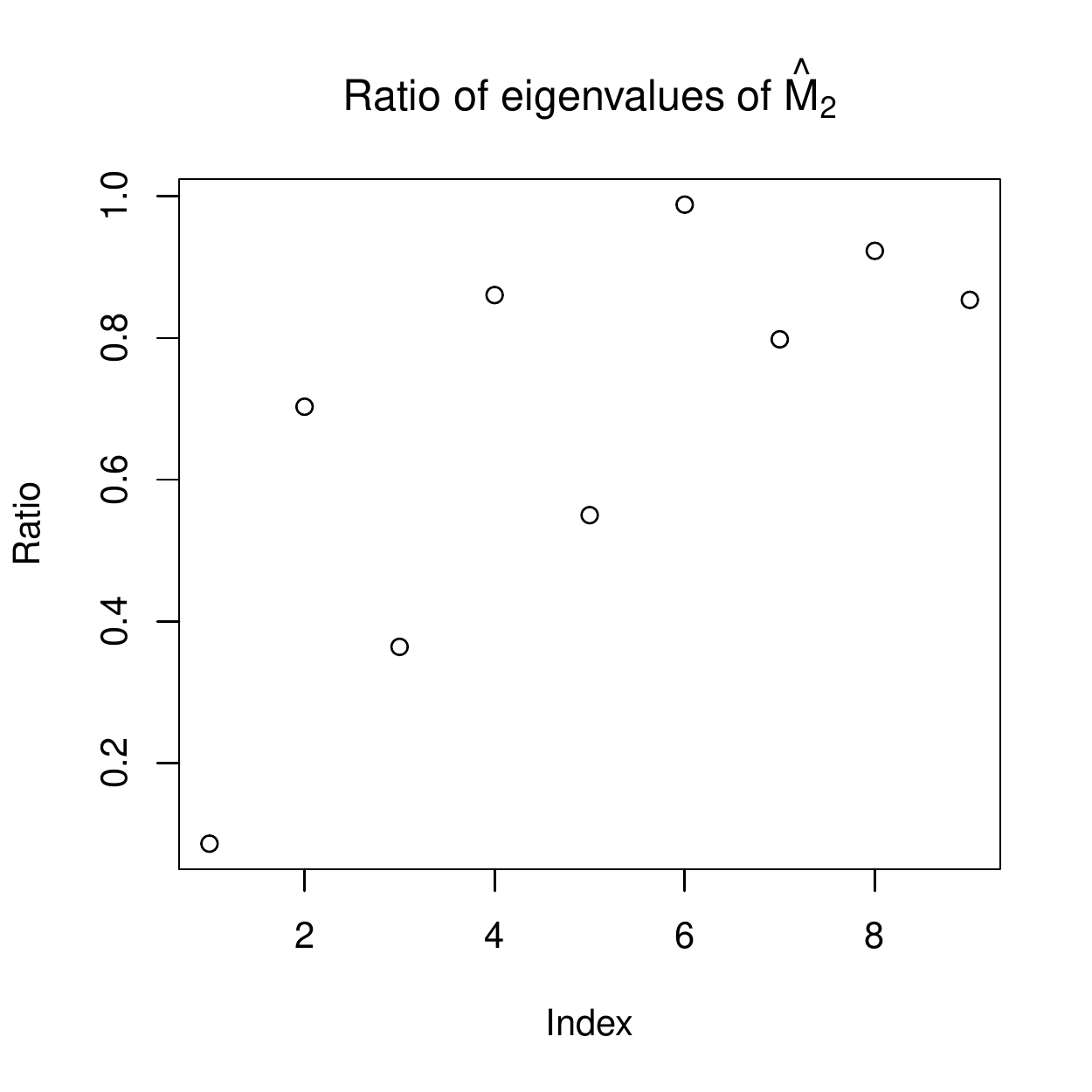}
\caption{Ratios of eigenvalues for real data analysis; left panel: ratio of eigenvalues of $\hat{\bM}_1(\eta_1,\eta_2)$; right panel: ratio of eigenvalues of $\hat{\bM}_2(\eta_1,\eta_2)$.}
\end{figure}

\begin{figure}[htpb!]
\centering
\includegraphics[width=4.75in]{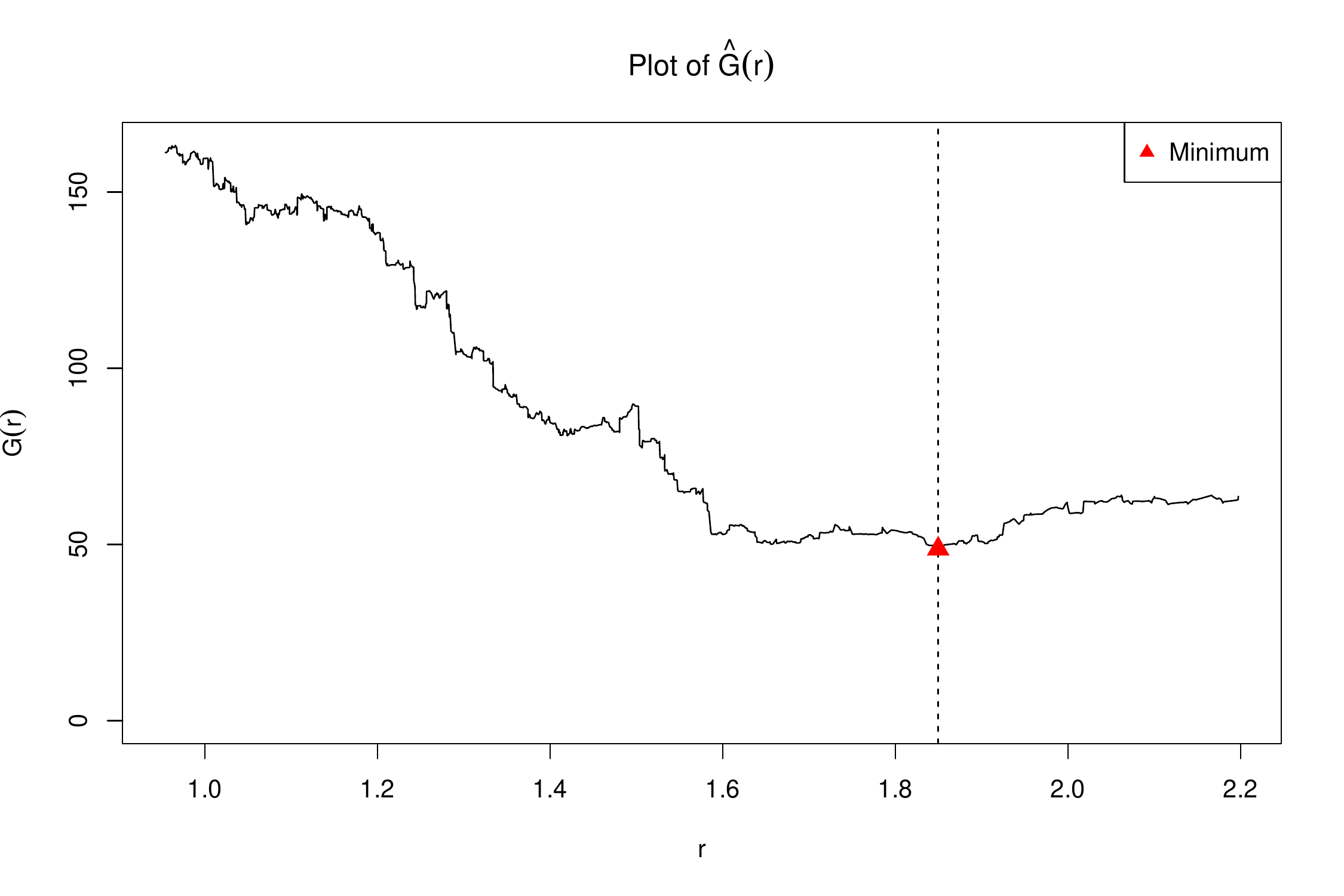}
\caption{Plot of $\hat{G}(r)$ for real data analysis.}
\end{figure}

\begin{figure}[htpb!]
\centering
\includegraphics[width=6in]{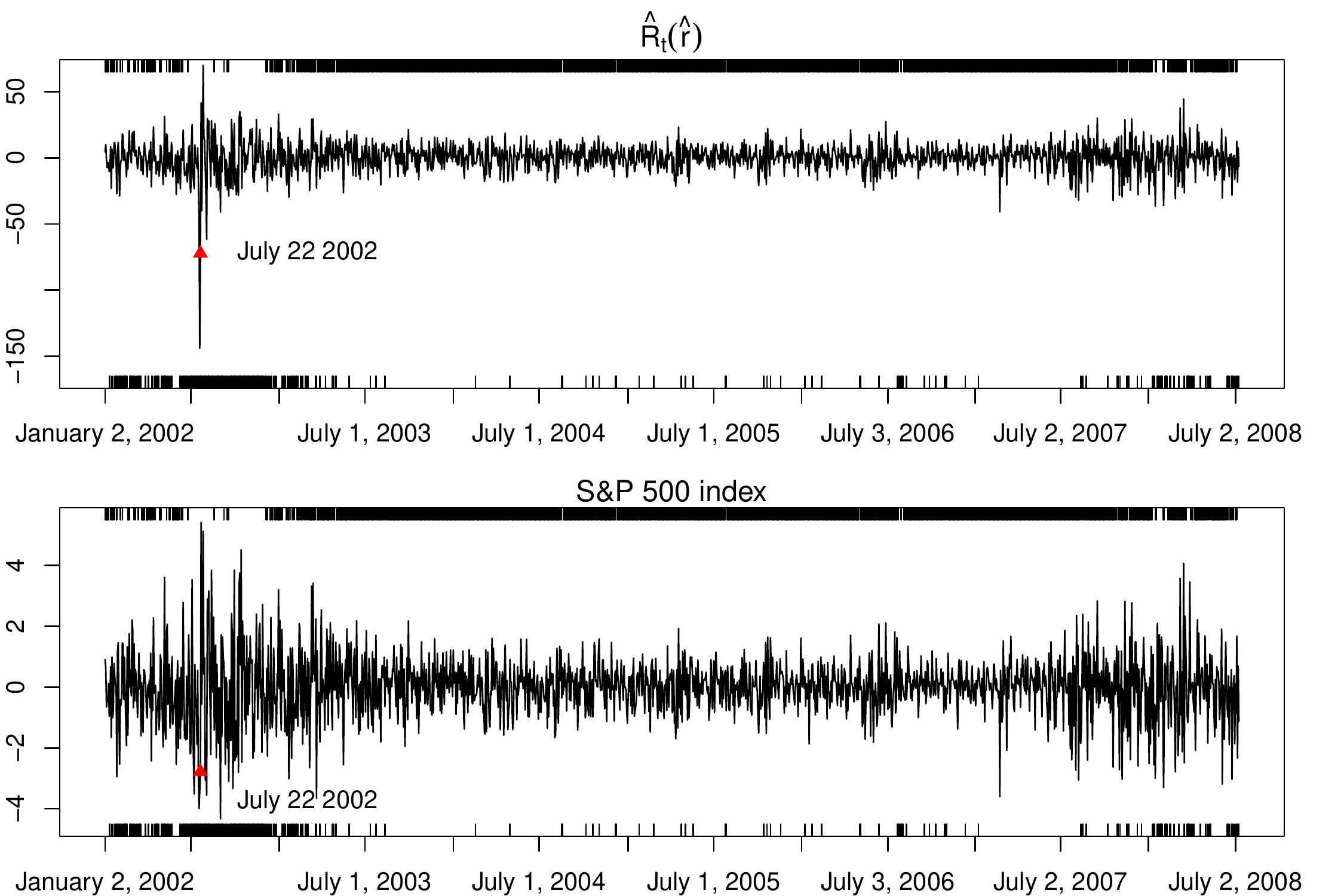}
\caption{Time series plots of $\hat{\bR}_{t}(\hat{r})$ (top panel) and the daily return of the S\&P 500 index (bottom panel) in the same period. Indicators of the estimated regimes of the observations ${I}_{t,i}(\hat{r})$ for $i=1,2$, are shown in the rug plots, on the top  for Regime 1 and at the bottom for Regime 2.}
\end{figure}

Note that
${\bR}_t$ is the common factor process with standardized loading matrix defined in (\ref{trans}), and can be estimated through (\ref{trans_est}). Figure 8 displays the time series plot of $\hat{\bR}_t(\hat{r})$ on the top panel and the daily returns of S\&P 500 index on the bottom panel. The estimated signal
$\hat{\bR}_{t}(\hat{r})$ in this period was closely correlated with
S\&P 500 index except for several days around July 22, 2002.
Figure 9 plots $\hat{\bR}_{t}(\hat{r})$ against daily returns of the S\&P 500 index,
and the correlation is 0.910.
Hence, this factor can be
regarded as a representation of market performance, which is in line with results in \cite{liu2016}.
\begin{figure}[htpb!]
\centering
\includegraphics[width=6in]{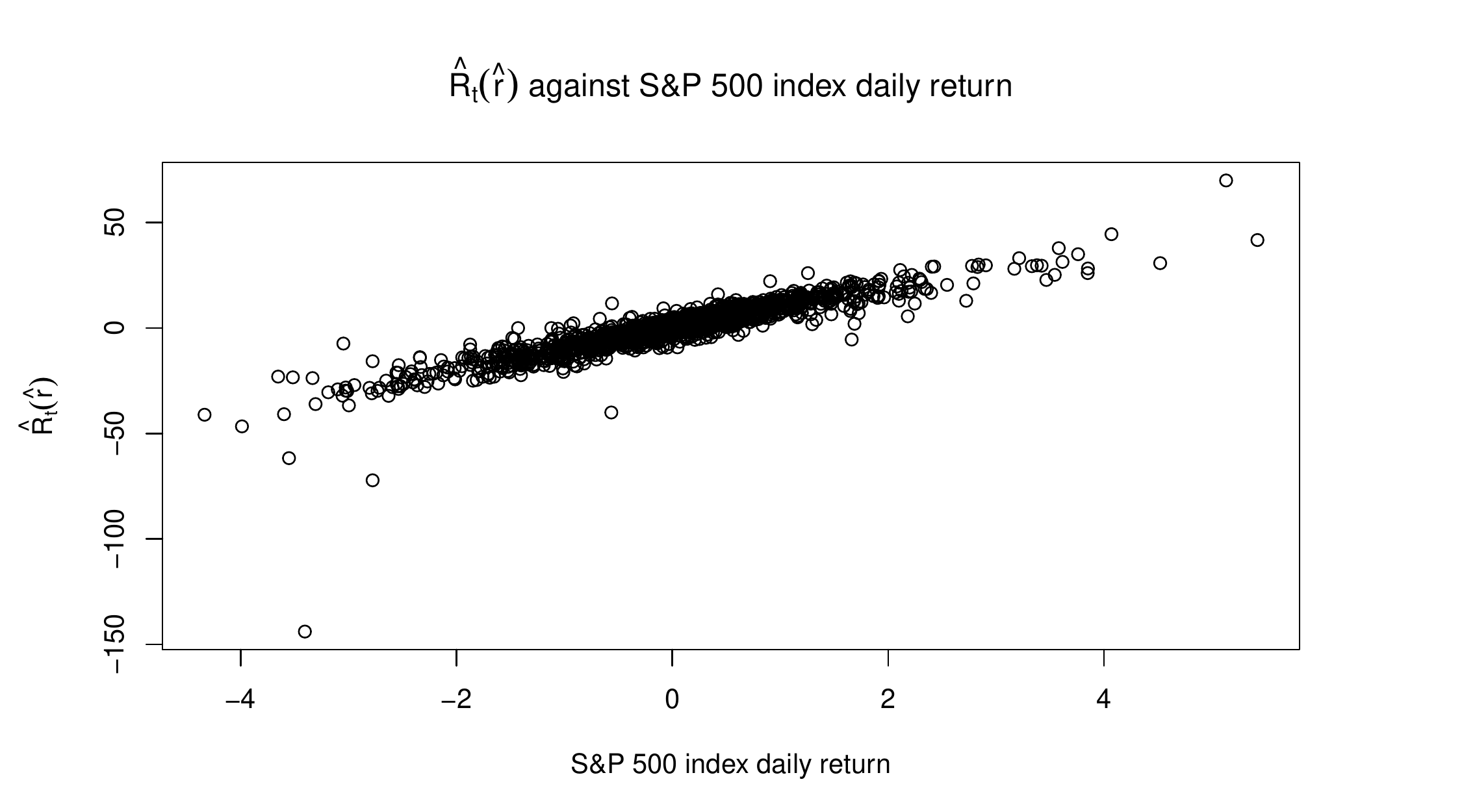}
\caption{Plot of $\hat{\bR}_{t}(\hat{r})$ against the daily return of the S\&P 500 index.}
\end{figure}

The estimated indicator functions for regimes $\{{I}_{t,i}(\hat{r}), i=1,2\}$, are
shown in the rug plots of both panels in Figure 8.
Regime 1 is plotted on the top and Regime 2 at the bottom.
When the market was volatile
in 2002, 2003, and 2008 due to the internet bubble, the invasion of Iraq, and the subprime crisis, respectively, the observations are more likely to belong to Regime 1; when the market was stable in 2004-2007,
 the observations tend to be in Regime 2.

 
{ The distance between the two estimated loading spaces defined in (\ref{distance_diff}) is 0.763. Since the loading spaces are one-dimensional (lines), the distance measures the absolute value of sine of the angle between the two lines.
  Although we do not have asymptotic distribution results for the estimated distance, the variation of estimated distance in Example 4 partially suggests that the two estimated spaces are quite different.}

To compare threshold factor models with regime-switching factor models \citep{liu2016}, we find the optimal path for the regime-switching factor model, calculate the in-sample residual sums of squares defined in (\ref{E}) for both models, and report in Table \ref{model_comp}. It shows that threshold factor models outperform the other.

\begin{table}
\begin{center}
\caption{In-sample residual sum of squares of different models for real data analysis}
\begin{tabular}{l| c}\hline \hline \label{model_comp}
Model	&Residual sum of squares\\ \hline
Threshold factor models	&606256.9\\ \hline
Regime-switching factor models&713742.6 \\ \hline \hline
\end{tabular}
\end{center}
\end{table}

\vspace{0.2in}

\bibliographystyle{apalike}
\bibliography{reference1}

\newpage

\section*{Appendix A.1 Regularity Conditions}

\noindent \textbf{Condition 1.}  Let $\calF_i^j$ be the $\sigma$-field generated by $\{(\bx_t, z_{t}): i \leq t \leq j\}$.
The joint process $(\bx_t,z_{t})$ is
$\alpha$-mixing with mixing coefficients satisfying
\[
\sum_{t=1}^{\infty} \alpha(t)^{1-2/\gamma}<\infty,
\]
for some $\gamma>2$, where
$\alpha(t)=\sup_{i} \sup_{A \in \calF_{-\infty}^i, B \in \calF_{i+t}^{\infty}} |P(A \cap B) -P(A)P(B)|$.

\noindent \textbf{Condition 2.} For any $j=1,\ldots, k_0$, $t=1, \ldots, n$, $E(|x_{t,j}|^{4\gamma})< \sigma_x^{4\gamma}$, where $x_{t,j}$ is the $j$-th element of $\bx_t$, $\sigma_x$ is a positive constant, and $\gamma$ is given in Condition 1.

\noindent \textbf{Condition 3.} $\{\bve_{t,1}\}$ and $\{\bve_{t,2}\}$ are two independent white noise processes. $\{ \bve_{t,1}, \bve_{t,2}\}$ and $\bx_t$ are uncorrelated with $z_{t}$ given ${\cal F}_{-\infty}^{t-1}$. Each element of $\bSigma_{t,i}$ remains bounded by $\sigma_{\epsilon}^2$ as $p$ increases to infinity, for $i=1,2$, and $t=1,\ldots, n$.

\noindent \textbf{Condition 4.} For $i=1,2$, there exists a constant $\delta_i \in [0,1]$ such that $\|\bA_i\|_2^2  \asymp \|\bA_i\|^2_{\min}\asymp p^{1-\delta_i}$, as $p$ goes to infinity.

\noindent \textbf{Condition 5.} For any $r_1, r_2 \in [\eta_1,\eta_2]$, there exist two positive integers $h_1,h_2 \in [1, h_0]$ such that
$\bSigma_{x,i,j}(h_i,r_1,r_2)$ is full rank,
and $\|\bSigma_{x,i,j}(h_i, r_1,r_2)\|_{\min}$ is uniformly bounded above 0, $j=1,2$.

\noindent \textbf{Condition 6.} $\bM_i(r_1,r_2)$ admits $k_0$ distinct positive eigenvalues, for $r_1,r_2 \in [\eta_1,\eta_2]$, $i=1,2$.

\medskip
\noindent{\bf Remark 15.} We do not impose the stationarity assumption on the latent process $\bx_t$, but we require it to
satisfy the generalized mixing condition in Condition 1 so the sample
version of $\bSigma_{y,i,j}(h,r_1,r_2)$ converges to its
population version for $i,j=1,2$.

\noindent{\bf Remark 16.} We allow $z_t$ to be a function of the
lag variables of $\bx_{t}$ or $\by_{t}$
or other observable processes that may also be dependent with
$\bx_{t}$ and $\by_{t}$. Therefore, Condition 3 also requires the noise at time $t$
is uncorrelated with $\bx_t$ given past information, in particular, given
$z_{t}$.

\noindent{\bf Remark 17.} Condition 5 ensures the existence of the 'helping effect' discussed in \cite{liu2016}. When $\bSigma_{x,i,1}(h_i, r_1,r_2)$ and
$\bSigma_{x,i,2}(h_i, r_1,r_2)$ are full rank, both regimes are relevant to $\bM_i(r_1,r_2)$, but the one corresponding to switching to the stronger regime carries more information about $\bx_t$
and improves the estimation results.

\noindent
{\bf Remark 18.}
Condition 6 makes $\bQ_i(r_1,r_2)$ and regime $i$ uniquely defined and
identifiable, where $\bQ_i(r_1,r_2)= (\bq_{i,1}(r_1,r_2), \ldots, \bq_{i,k_0}(r_1,r_2))$
collects the $k_0$ orthonormal
eigenvectors of $\bM_i(r_1,r_2)$ corresponding to the $k_0$ positive eigenvalues
$\lambda_{i,1}(r_1,r_2) >\ldots > \lambda_{i,k_0}(r_1,r_2)$.

\medskip

For $c_1 < c_2 <c_3 <c_4$, define
\[
\bSigma_x(h,c_1,c_2,c_3,c_4)=\frac{1}{n-h} \sum_{t=1}^{n-h} E\left[ \bx_t \bx_{t+h}' I(c_1 < z_{t} < c_2, c_3 < z_{t+h} < c_4)\right],
\]
where $c_i$ can be a constant, $-\infty$, or $+\infty$ for $1 \leq i \leq 4$.

The following additional conditions are needed for the asymptotic
results of $\hat{r}$ in (\ref{threshold}).


\noindent \textbf{Condition 7.}
Assume $r_0\in (\eta_1,\eta_2)$. $\{z_t\}$ is a strictly stationary process. The marginal probability of $z_t$ is continuous, $P(z_t\leq \eta_1)>0$ and $P(z_t\leq \eta_2)>0$. The conditional density of $z_{t+h}$ given $z_t$ is continuous, and there exists two positive constants $\tau_1$ and $\tau_2$ such that $\tau_1 \leq f(z_{t+h}\mid z_t)\leq \tau_2$ for any $z_{t+h}$ in $[\eta_1,\eta_2]$ and $z_t\in \mathbb{R}$.
The conditional probability of $z_{t+h}$ given $z_t$ satisfies that $P(z_{t+h}<\eta_1 \mid z_t)>0$ and $P(z_{t+h}>\eta_2 \mid z_t)>0$ for any $z_t \in (-\infty,\eta_1)$ and $(\eta_2,+\infty)$ and $h=1,\ldots h_0$.

\noindent{\bf Condition 8.} For any $r \in (\eta_1, r_0)$, there exists an integer $h_1^*\in [1,h_0]$ such that $\bSigma_x(h_1^*,r,r_0, -\infty,r)$ and $\bSigma_x(h_1^*,r,r_0,r_0,+\infty)$ are full rank. For any $r \in (r_0, \eta_2)$, there exists an integer $h_2^*\in [1,h_0]$ such that $\bSigma_x(h_2^*, r_0,r,r,+\infty)$ and $\bSigma_x(h_2^*,r_0,r,-\infty, r_0)$ are full rank. The minimum singular values of these four matrices mentioned are all uniformly
bounded above $\rho_0|r-r_0|>0$, where $\rho_0$ is a positive constant.



\noindent {\bf Condition 9.}  There exists a positive constant $d$ such that ${\cal D}({\cal M}(\bQ_1), {\cal M}(\bQ_2)) > d$  as $p$ goes to infinity.

\medskip
\noindent{\bf Remark 19.} Conditions 7 guarantees the consistency of the estimators for loadings spaces, when only data with $\{z_{t} \leq \eta_1\}$ and $\{z_{t} \geq \eta_2\}$ are used.  Condition 8 makes sure that  the cross moment matrices of $\by_t$ when $r$ is used as threshold value and these when $r_0$ is used as threshold value are differentiable. For example, if $k_0=1$, $x_t$ is stationary, independent of $z_t$, and has nonzero serial correlation at leads $h_1^*$ and $h_2^*$, then
Condition 8 is satisfied. Condition 9 is an identification condition that
ensures that the two regimes are uniformly sufficiently different. 

\medskip
\noindent {\bf Condition 10.}  When $\hat{k}>k_0$, there exists a positive constant $c_0$ such that ${\cal D}({\cal M}(\bQ_1), {\cal M}({\bQ}_2^*)) >c_0$ and ${\cal D}({\cal M}(\bQ_1^*), {\cal M}({\bQ}_2)) >c_0$, for any $p \times (\hat{k}-k_0)$ matrix $\bS_i$ such that $\dim({\cal M}(\bS_i)\cap {\cal M}(\bQ_i))=0$ for $i=1,2$, where $\bQ_i^*=(\bQ_i, \bS_i)$ is a $p \times \hat{k}$ matrix.

\medskip
\noindent{\bf Remark 20.} If the number of factors is overestimated, Condition 10 guarantees that the two augmented loadings spaces ${\cal M}(\bQ_1^*)$ and ${\cal M}(\bQ_2^*)$ can still be differentiated. With this condition, Theorem 4 shows that the estimator for the threshold value is consistent.

\section*{Appendix A.2 Lemmas and Proofs}
We mainly focus on the mathematical proofs for Regime 1 when $r>r_0$, $\epsilon>0$. The results for Regime 2, $r \leq r_0$ or $\epsilon \leq 0$ are included, but most of proofs are omitted since they are quite similar. We use $C$s and $C_i$s to denote the generic uniformly positive constants which only depend on the parameters. $C$s and $C_i$s may be different constants in different cases.

 Define
\[
I_t(h,c_1,c_2,c_3,c_4)=I(c_1<z_t<c_2,c_3<z_{t+h}<c_4),
\]
\[
\hat{\bSigma}_{x}(h, c_1,c_2,c_3,c_4)=\frac{1}{n-h} \sum_{t=1}^{n-h}\bx_t \bx_{t+h}' I_{t}(h,c_1,c_2,c_3,c_4),
\]
where $c_1<c_2<c_3<c_4$ can be real numbers, $-\infty$, or $+\infty$.
\begin{lemma}\label{eps}
Under Conditions 1, 2 and 7, when $n>2h_0$, for any $h\in[1, h_0]$, it holds that
\begin{align*}
{\rm E} \|\hat{\bSigma}_x(h, c_1,c_2,c_3,c_4)-\bSigma_x(h, c_1,c_2,c_3,c_4)\|_2^2\leq \frac{4(3h+4\alpha)\Delta_{c_1,c_2}\Delta_{c_3,c_4}k_0^2\sigma_x^4}{n},\\
\|\bSigma_x(h, c_1,c_2,c_3,c_4)\|_2\leq \Delta_{c_1,c_2} \Delta_{c_3,c_4}k_0 \sigma_x^2,
\end{align*}
where $\alpha=\sum_{t=1}^{\infty}\alpha(t)^{1-2/\gamma}$, and $c_1<c_2<c_3<c_4$ can be real numbers in $(\eta_1, \eta_2)$, $-\infty$, or $+\infty$. $\Delta_{c_1,c_2}=\tau_2|c_2-c_1|$ if $c_1$ and $c_2$ are both real numbers, $\Delta_{c_1,c_2}=1$ if at least one of them is $-\infty$ or $+\infty$.
\end{lemma}
\noindent{\it Proof:} Let $a_{q,\ell}$ and $\hat{a}_{q,\ell}$ be the $(q,\ell)$-th entry in $\bSigma_x(h, c_1,c_2,c_3,c_4)$ and $\hat{\bSigma}_x(h,c_1,c_2,c_3,c_4)$, for $q,\ell=1,\ldots, k_0$. Under Conditions 1, 2 and 7, by Jensen's inequality and Proposition 2.5 in \cite{fan2003}, we have
\begin{eqnarray*}
\lefteqn{{\rm E}|\hat{a}_{q,\ell}-a_{q,\ell}|^2
={\rm E} \left\{ \frac{1}{n-h} \sum_{t=1}^{n-h} \Big[x_{t,q}x_{t+h,\ell}I_t(h,c_1,c_2,c_3,c_4)-E(x_{t,q} x_{t+h,\ell} I_t(h,c_1,c_2,c_3,c_4))\Big] \right\}^2 }\\
&\leq& \frac{1}{(n-h)^2} \sum_{|t_1-t_2|\leq h} {\rm E} [x_{t_1,q}x_{t_1+h,\ell}I_{t_1}(h,c_1,c_2,c_3,c_4)-E(x_{t_1,q} x_{t_1+h,\ell} I_{t_1}(h,c_1,c_2,c_3,c_4))]\\
&&\cdot [x_{t_2,q}x_{t_2+h,\ell}I_{t_2}(h,c_1,c_2,c_3,c_4)-E(x_{t_2,q} x_{t_2+h,\ell} I_{t_2}(h,c_1,c_2,c_3,c_4))]\\
&&+\frac{1}{(n-h)^2} \sum_{|t_1-t_2|> h} {\rm E} [x_{t_1,q}x_{t_1+h,\ell}I_{t_1}(h,c_1,c_2,c_3,c_4)-E(x_{t_1,q} x_{t_1+h,\ell} I_{t_1}(h,c_1,c_2,c_3,c_4))]\\
&&\cdot [x_{t_2,q}x_{t_2+h,\ell}I_{t_2}(h,c_1,c_2,c_3,c_4)-E(x_{t_2,q} x_{t_2+h,\ell} I_{t_2}(h,c_1,c_2,c_3,c_4))]\\
&\leq&\frac{[(2h+1)n-h^2-h]\Delta_{c_1,c_2}\Delta_{c_3,c_4}\sigma_x^4}{(n-h)^2}+\frac{8(n-h)\Delta_{c_1,c_2}\Delta_{c_3,c_4}\sigma_x^4}{(n-h)^2}\sum_{u=1}^{n-2h-1}\alpha(u)^{1-2/\gamma}\\
&\leq&\frac{4[(2h+1)n-h^2-h]\Delta_{c_1,c_2}\Delta_{c_3,c_4}\sigma_x^4}{n^2}+\frac{16\alpha\Delta_{c_1,c_2}\Delta_{c_3,c_4}\sigma_x^4}{n}\\
&\leq&\frac{4(3h+4\alpha)\Delta_{c_1,c_2}\Delta_{c_3,c_4}\sigma_x^4}{n}.
\end{eqnarray*}
By the properties of Frobenius norm and $L$-2 norm,
\begin{eqnarray*}
\lefteqn{{\rm E} \|\hat{\bSigma}_x(h, c_1,c_2,c_3,c_4)- \bSigma_x(h, c_1,c_2,c_3,c_4)\|_2^2}\\
& \leq& {\rm E} \|\hat{\bSigma}_x(h, c_1,c_2,c_3,c_4) -\bSigma_x(h, c_1,c_2,c_3,c_4)\|_F^2\\
&= &\sum_{q=1}^{k_0}\sum_{\ell=1}^{k_0} {\rm E}|\hat{a}_{q,\ell}-a_{q,\ell}|^2 \leq  \frac{4(3h+4\alpha)\Delta_{c_1,c_2}\Delta_{c_3,c_4}k_0^2\sigma_x^4}{n}.
\end{eqnarray*}
Similar, we can show that
\begin{eqnarray*}
\lefteqn{|a_{q,\ell}| =\Big|\frac{1}{n-h} \sum_{t=1}^{n-h} E[x_{t,q} x_{t+h,\ell} I_t(h,c_1,c_2,c_3,c_4)]\Big|}\\
& \leq& \frac{1}{n-h}\sum_{t=1}^{n-h} {\rm E}(|x_{t,q}x_{t+h,\ell}|) \cdot {\rm E}[I_t(h,c_1,c_2,c_3,c_4)] \leq \Delta_{c_1,c_2}\Delta_{c_3,c_4} \sigma_x^2.
\end{eqnarray*}
Thus, $\|\bSigma_x(h,c_1,c_2,c_3,c_4)\|_2 \leq \|\bSigma_x(h,c_1,c_2,c_3,c_4)\|_F\leq \Delta_{c_1,c_2} \Delta_{c_3,c_4} k_0 \sigma_x^2$.
\endp

\begin{lemma} Under Conditions 1-4 and 7, when $n>2h_0$, for any $h\in [1,h_0]$ and $\epsilon \in (\eta_1-r_0,\eta_2-r_0)$, it holds that
\begin{align*}
{\rm E}\|\hat{\bSigma}_{y,i,j}(h,r_0+\epsilon)-\bSigma_{y,i,j}(h,r_0+\epsilon)\|_2^2 \leq 440(3h+4\alpha)a_2^4k_0^2\tau_0^2\delta_{\eta}^2\sigma_0^4 p^2n^{-1}, \mbox{ for } i,j=1,2,
\end{align*}
where $\tau_0=\max\{\tau_2,1\}$, $\delta_{\eta}=\max\{\eta_2-\eta_1, 1\}$, $\sigma_0=\max\{\sigma_x,\sigma_e\}$, and $a_2$ is a positive constant such that $\|\bA_i\|_2 \leq a_2 p^{1/2-\delta_i/2}$ for $i=1,2$.
\end{lemma}
\noindent{\it Proof:} When $\epsilon>0$, if we use $r_0+\epsilon$ as the threshold value, data are classified into 2 subsets, $S_1=\{t: I(z_t<r_0+\epsilon)\}$ and $S_2=\{t: I(z_t \geq r_0+\epsilon)\}$. Observation in $S_2$ are all from Regime 2 and correctly classified, but those in $S_1$ are mixed. In this case,
\begin{eqnarray*}
\lefteqn{\hat{\bSigma}_{y,1,1}(h,r_0+\epsilon)-\bSigma_{y,1,1}(h,r_0+\epsilon)}\\
&=&\left[ \bA_1 \left( \hat{\bSigma}_{x}(h, -\infty, r_0, -\infty, r_0) -{\bSigma}_{x}(h, -\infty, r_0, -\infty, r_0) \right) \bA_1' \right. \\
&&+ \bA_1 \left( \hat{\bSigma}_{x}(h, -\infty, r_0,r_0, r_0+\epsilon) -{\bSigma}_{x}(h, -\infty, r_0,  r_0+\epsilon) \right) \bA_2'  \\
&&+ \bA_2 \left( \hat{\bSigma}_{x}(h, r_0, r_0+\epsilon, -\infty, r_0) -{\bSigma}_{x}(h, r_0, r_0+\epsilon, -\infty, r_0) \right) \bA_1' \\
&&\left. + \bA_2 \left( \hat{\bSigma}_{x}(h, r_0, r_0+\epsilon,r_0,  r_0+\epsilon) -{\bSigma}_{x}(h, r_0, r_0+\epsilon, r_0, r_0+\epsilon) \right) \bA_2' \right] \\
&&+ \frac{1}{n-h}\sum_{t=1}^{n-h} \left[ \left(\bA_1 \bx_t \bve_{t+h,1}'+\bve_{t,1} \bx_{t+h}' \bA_1'+ \bve_{t,1} \bve_{t+h,1}'\right)I_{t,1}(r_0)I_{t+h,1}(r_0) \right.\\
&&+ \left(\bA_1 \bx_t \bve_{t+h,2}'+\bve_{t,1} \bx_{t+h}' \bA_2'+ \bve_{t,1} \bve_{t+h,2}'\right)I_{t,1}(r_0)I(r_0 \leq z_{t+h} <r+\epsilon)\\
&&+   \left( \bA_2 \bx_t \bve_{t+h,1}'+\bve_{t,2} \bx_{t+h}' \bA_1'+ \bve_{t,2} \bve_{t+h,1}'\right)I(r_0 \leq z_t <r_0+\epsilon)I_{t+h,1}(r_0)\\
&&+ \left. \left(\bA_2 \bx_t \bve_{t+h,2}'+\bve_{t,2} \bx_{t+h}' \bA_2'+ \bve_{t,2} \bve_{t+h,2}'\right)I(r_0 \leq z_t <r+\epsilon)I(r_0 \leq z_{t+h} <r+\epsilon) \right] \\
&=&I_1+I_2+I_3+I_4+I_5.
\end{eqnarray*}
Condition 4 implies that we can find two positive constants $a_1<1<a_2$ such that $a_1p^{1/2-\delta_i/2}\leq \|\bA_i\|_2 \leq a_2p^{1/2-\delta_i/2}$, for $i=1,2$. By Lemma 1, we have
\begin{eqnarray*}
\lefteqn{{\rm E}\| I_1\|_2^2
 \leq  4\|\bA_1\|_2^2 \cdot {\rm E} \|\hat{\bSigma}_{x}(h, -\infty, r_0, -\infty,r_0)-\bSigma_{x}(h,-\infty, r_0, -\infty, r_0)\|_2^2\cdot \|\bA_1\|_2^2}\\
&  & + 4\|\bA_1\|_2^2  \cdot {\rm E} \|\hat{\bSigma}_{x}(h, -\infty, r_0,r_0,r_0+\epsilon)-\bSigma_{x}(h,-\infty, r_0,  r_0, r+\epsilon)\|_2^2 \cdot  \|\bA_2\|_2^2 \\
&& +  4\|\bA_2\|_2^2 \cdot  {\rm E} \|\hat{\bSigma}_{x}(h, r_0, r_0+\epsilon, -\infty,r_0)-\bSigma_{x}(h,r_0, r_0+\epsilon, -\infty, r_0)\|_2^2 \cdot \|\bA_1\|_2^2 \\
&& +  4\|\bA_2\|_2^2\cdot  {\rm E} \|\hat{\bSigma}_{x}(h, r_0, r_0+\epsilon,r_0, r_0+\epsilon)-\bSigma_{x}(h,r_0, r_0+\epsilon, r_0, r_0+\epsilon)\|_2^2 \cdot \|\bA_2\|_2^2 \\
& \leq & 16(3h+4\alpha)a_2^4k_0^2\sigma_x^4(p^{2-2\delta_1}+2\tau_2 \epsilon p^{2-\delta_1-\delta_2}+\tau_2^2 \epsilon^2 p^{2-2\delta_2})n^{-1}.
\end{eqnarray*}
For the interaction terms of common component and noise in $I_2$, under Condition 3, if $n>2h_0$, we can show that
\begin{eqnarray*}
\lefteqn{{\rm E}\Big\| \frac{1}{n-h}\sum_{t=1}^{n-h}\bA_1 \bx_t \bve_{t+h,1}'I_{t,1}(r_0)I_{t+h,1}(r_0)\Big\|_2^2}\\
&\leq &\frac{a_2^2p^{1-\delta_1}}{(n-h)^2} {\rm E}\Big\|\sum_{t=1}^{n-h} \bx_t \bve_{t+h,1}'I_{t,1}(r_0)I_{t+h,1}(r_0) \Big\|_F^2\\
&\leq&\frac{a_2^2p^{1-\delta_1}}{(n-h)^2}\sum_{t=1}^{n-h}\sum_{m=1}^{k_0}\sum_{\ell=1}^p {\rm E}(x_{t,m}^2 \epsilon_{t+h,1,\ell}^2)
\leq 2a_2^2k_0\sigma_x^2\sigma_{\epsilon}^2p^{2-\delta_1}n^{-1},
\end{eqnarray*}
and
\begin{eqnarray*}
\lefteqn{{\rm E}\Big\| \frac{1}{n-h}\sum_{t=1}^{n-h}\bve_{t,1}\bx_{t+h}'\bA_1'I_{t,1}(r_0)I_{t+h,1}(r_0)\Big\|_2^2}\\
&\leq &\frac{a_2^2p^{1-\delta_1}}{(n-h)^2} {\rm E}\Big\|\sum_{t=1}^{n-h} \bve_{t,1} \bx_{t+h}'I_{t,1}(r_0)I_{t+h,1}(r_0) \Big\|_F^2\\
&\leq&\frac{a_2^2p^{1-\delta_1}}{(n-h)^2}\sum_{t=1}^{n-h}\sum_{m=1}^{p}\sum_{\ell=1}^{k_0} {\rm E}(\epsilon_{t,1,m}^2 x_{t+h,\ell}^2)
\leq 2a_2^2k_0 \sigma_x^2\sigma_{\epsilon}^2 p^{2-\delta_1}n^{-1},
\end{eqnarray*}
where $\epsilon_{t,i,m}$ is the $m$-th entry in $\bve_{t,i}$ for $m=1,\ldots,p$ and $i=1,2$.

For noise term in $I_2$,
\begin{eqnarray*}
{\rm E} \big\| \frac{1}{n-h} \sum_{t=1}^{n-h} \bve_{t,1}\bve_{t+h,1}'I_{t,1}(r_0)I_{t+h,1}(r_0)\big\|_2^2\leq \frac{1}{(n-h)^2} \sum_{t=1}^{n-h}\sum_{m=1}^p \sum_{\ell=1}^p {\rm E} (\epsilon_{t,1,m}^2 \epsilon_{t+h,1,\ell}^2)
\leq 2\sigma_{\epsilon}^4p^2n^{-1}.
\end{eqnarray*}
Hence,
\begin{eqnarray*}
\lefteqn{{\rm E}\|I_2\|_2^2\leq 3 {\rm E}\Big\| \frac{1}{n-h}\sum_{t=1}^{n-h}\bA_1 \bx_t \bve_{t+h,1}'I_{t,1}(r_0)I_{t+h,1}(r_0)\Big\|_2^2}\\
&&+3 {\rm E}\Big\| \frac{1}{n-h}\sum_{t=1}^{n-h}\bve_{t,1}\bx_{t+h}'\bA_1'I_{t,1}(r_0)I_{t+h,1}(r_0)\Big\|_2^2+ 3{\rm E} \big\| \frac{1}{n-h} \sum_{t=1}^{n-h} \bve_{t,1}\bve_{t+h,1}'I_{t,1}(r_0)I_{t+h,1}(r_0)\big\|_2^2\\
&\leq &12a_2^2k_0 \sigma_x^2\sigma_{\epsilon}^2 p^{2-\delta_1}n^{-1}+6\sigma_{\epsilon}^4p^2n^{-1}.
\end{eqnarray*}
Similarly, we can show that ${\rm E}\|I_i\|_2^2\leq 12 a_2^2k_0 \sigma_x^2\sigma_{\epsilon}^2 p^{2-\delta_1}n^{-1}+6\sigma_{\epsilon}^4p^2n^{-1}$, for $i=3,4,5$. Let $\delta_{\eta}=\max\{\eta_2-\eta_1, 1\}$, $\tau_0=\max\{\tau_2,1\}$, $\sigma_0=\max\{\sigma_x,\sigma_e\}$, and we have
\begin{eqnarray*}
\lefteqn{{\rm E}\|\hat{\bSigma}_{y,1,1}(h,r_0+\epsilon)-\bSigma_{y,1,1}(h,r_0+\epsilon)\|_2^2}\\
&\leq& 5{\rm E}\|I_1\|_2^2+5{\rm E}\|I_2\|_2^2+5{\rm E}\|I_3\|_2^2+5{\rm E}\|I_4\|_2^2+5{\rm E}\|I_5\|_2^2\\
&\leq& 80(3h+4\alpha)a_2^4k_0^2(p^{2-2\delta_1}+2\tau_2 \epsilon p^{2-\delta_1-\delta_2}+\tau_2^2 \epsilon^2 p^{2-2\delta_2})n^{-1}\\
&&+240a_2^2 k_0\sigma_x^2\sigma_{\epsilon}^2 p^{2-\delta_1}n^{-1}+120\sigma_{\epsilon}^4p^2n^{-1}\\
&\leq&440(3h+4\alpha)a_2^4k_0^2\tau_0^2\delta_{\eta}^2\sigma_0^4 p^2n^{-1}.
\end{eqnarray*}

For the other cases when $\epsilon \leq 0$ and $i=1,2$, the proof is tedious but similar.
\endp

\begin{lemma}
Under Conditions 1-4 and 7, when $n>2h_0$, for any $\epsilon \in (\eta_1-r_0,\eta_2-r_0)$, it holds that
\begin{eqnarray*}
\|\bSigma_{y,1,1}(h,r_0+\epsilon)\|_2 \leq \left\{
\begin{array}{ll}
a_2^2k_0\sigma_x^2 p^{1-\delta_1} & \epsilon \leq 0,\\
a_2^2k_0\sigma_x^2(p^{1-\delta_1}+ 2\epsilon p^{1-\delta_1/2-\delta_{2}/2}+\epsilon^2 p^{1-\delta_2})	&\epsilon>0,
\end{array}
\right.
\end{eqnarray*}
\begin{eqnarray*}
\|\bSigma_{y,2,2}(h,r_0+\epsilon)\|_2 \leq \left\{
\begin{array}{ll}
a_2^2k_0\sigma_x^2(p^{1-\delta_2}+ 2\epsilon p^{1-\delta_1/2-\delta_{2}/2}+\epsilon^2 p^{1-\delta_1})& \epsilon <0,\\
a_2^2k_0\sigma_x^2 p^{1-\delta_2} 	 &\epsilon \geq 0,
\end{array}
\right.
\end{eqnarray*}
when $i,j \in \{1,2\}$ and $i \neq j$,
\begin{eqnarray*}
\|\bSigma_{y,i,j}(h,r_0+\epsilon)\|_2 \leq \left\{
\begin{array}{ll}
a_2^2k_0 \sigma_x^2(p^{1-\delta_1/2-\delta_2/2} +\epsilon p^{1-\delta_1}) & \epsilon < 0,\\
a_2^2k_0 \sigma_x^2p^{1-\delta_1/2-\delta_2/2} 	& \epsilon=0,\\
a_2^2k_0 \sigma_x^2(p^{1-\delta_1/2-\delta_2/2} +\epsilon p^{1-\delta_2})	&\epsilon>0.
\end{array}
\right.
\end{eqnarray*}
\end{lemma}
\noindent{\it Proof:} We consider $S_1$ and $S_2$ as in the proof of Lemma 2.

By the definition of $\bSigma_{y,i,j}(h,r)$ and Lemma 1, when $\epsilon>0$, we have
\begin{eqnarray*}
\lefteqn{\big\|\bSigma_{y,1,1}(h,r_0+\epsilon)\big\|_2=\Big\|\frac{1}{n-h}\sum_{t=1}^{n-h} E\left[ \by_t \by_{t+h}' I( z_{t}<r_0+\epsilon, z_{t+h}<r_0+\epsilon)\right]\Big\|_2} \nonumber\\
&\leq&  \Big\| \bA_1 \bSigma_{x}(h,-\infty,r_0,-\infty,r_0) \bA_1' \Big\|_2 + \Big\| \bA_1 \bSigma_{x}(h,-\infty,r_0,r_0, r_0+\epsilon) \bA_2' \Big\|_2\\
&& +  \Big\| \bA_2 \bSigma_{x}(h,r_0,r_0+\epsilon,-\infty,r_0) \bA_1' \Big\|_2 + \Big\| \bA_2 \bSigma_{x}(h,r_0,r_0+\epsilon,r_0, r_0+\epsilon) \bA_2' \Big\|_2\\
& \leq & \| \bA_1\|_2^2\,  \| \bSigma_{x}(h,-\infty,r_0,-\infty,r_0) \|_2+ \|\bA_1\|_2\, \big\|\bSigma_{x}(h,-\infty,r_0,r_0, r_0+\epsilon) \big\|_2\, \|\bA_2\|_2 \\
&&+\| \bA_2\|_2\,  \| \bSigma_{x}(h,r_0,r_0+\epsilon,-\infty,r_0) \|_2 \, \|\bA_1\|_2+ \|\bA_2\|_2^2\, \big\|\bSigma_{x}(h,r_0,r_0+\epsilon,r_0, r_0+\epsilon) \big\|_2 \\
&=& a_2^2k_0\sigma_x^2(p^{1-\delta_1}+ 2\epsilon p^{1-\delta_1/2-\delta_{2}/2}+\epsilon^2 p^{1-\delta_2}),
\end{eqnarray*}
and
\begin{eqnarray*}
\lefteqn{\big\|\bSigma_{y,1,2}(h,r_0+\epsilon)\big\|_2=\Big\|\frac{1}{n-h}\sum_{t=1}^{n-h} E\left[ \by_t \by_{t+h}' I( z_{t}<r_0+\epsilon, z_{t+h} \geq r_0+\epsilon)\right]\Big\|_2}\\
&\leq &\Big\| \bA_1 \bSigma_{x}(h,-\infty, r_0,r_0+\epsilon, +\infty) \bA_2' \Big\|_2+\Big\| \bA_2 \bSigma_{x}(h,r_0, r_0+\epsilon,r_0+\epsilon, +\infty) \bA_2' \Big\|_2\\
& \leq & \| \bA_1\|_2\,  \| \bSigma_{x}(h,-\infty, r_0,r_0+\epsilon, +\infty) \|_2\, \|\bA_2\|_2+ \|\bA_2\|_2^2\, \big\|\bSigma_{x}(h,r_0, r_0+\epsilon, r_0+\epsilon, +\infty) \big\|_2 \\
&=&a_2^2k_0 \sigma_x^2(p^{1-\delta_1/2-\delta_2/2} +\epsilon p^{1-\delta_2}).
\end{eqnarray*}
Similarly, we can show the results for $\epsilon \leq 0$ and other equations.
\endp

\begin{lemma} Under Conditions 1-4 and 7, when $n>2h_0$, for any $\epsilon \in (\eta_1-r_0,\eta_2-r_0)$, it holds that
\begin{eqnarray*}
\|\bB_1(\eta_1,\eta_2)' \bSigma_{y,1,1}(h,r_0+\epsilon)\|_2 \left\{
\begin{array}{ll}
=0 & \epsilon \leq 0,\\
\leq a_2^2 k_0\sigma_x^2 (\epsilon p^{1-\delta_1/2-\delta_2/2}+\epsilon^2 p^{1-\delta_2})	&\epsilon>0,
\end{array}
\right.
\end{eqnarray*}

\begin{eqnarray*}
\|\bB_1(\eta_1,\eta_2)' \bSigma_{y,1,2}(h,r_0+\epsilon)\|_2\left\{
\begin{array}{ll}
=0 & \epsilon \leq 0,\\
\leq a_2^2 k_0 \sigma_x^2\epsilon p^{1-\delta_2}	&\epsilon>0,
\end{array}
\right.
\end{eqnarray*}

\begin{eqnarray*}
 \|\bB_2(\eta_1,\eta_2)' \bSigma_{y,2,1}(h,r_0+\epsilon)\|_2 \left\{
\begin{array}{ll}
\leq a_2^2 k_0 \sigma_x^2\epsilon p^{1-\delta_1}& \epsilon <0,\\
=0 &\epsilon \geq 0,
\end{array}
\right.
\end{eqnarray*}

\begin{eqnarray*}
 \|\bB_2(\eta_1,\eta_2)' \bSigma_{y,2,2}(h,r_0+\epsilon)\|_2 \left\{
\begin{array}{ll}
\leq a_2^2 k_0 \sigma_x^2(\epsilon p^{1-\delta_1/2-\delta_{2}/2}+\epsilon^2 p^{1-\delta_1}) & \epsilon <0,\\
=0 &\epsilon \geq 0.
\end{array}
\right.
\end{eqnarray*}
\end{lemma}
\noindent{\it Proof:} Since $r_0\in(\eta_1,\eta_2)$, by the definition, we have $\cM(\bB_i)=\cM(\bB_i(\eta_1,\eta_2))$. Hence, there exists a $p\times (p-k_0)$ orthogonal matrix $\bV_i$ such that $\bB_i=\bB_i(\eta_1,\eta_2)\bV_i$ for $i=1,2$.

When $\epsilon>0$, by Lemma 1 we have
\begin{eqnarray*}
\lefteqn{\big\|\bB_1(\eta_1,\eta_2)'\bSigma_{y,1,1}(h,r_0+\epsilon)\big\|_2= \Bigg\|\frac{1}{n-h}\sum_{t=1}^{n-h}\bB_1' {\rm E}\left[\by_t \by_{t+h}' I(z_t < r_0+\epsilon, z_{t+h} <r_0+\epsilon) \right] \Bigg\|}\nonumber \\
&=&\Bigg\| \frac{1}{n-h} \sum_{t=1}^{n-h}{\rm E}\left[\bB_1'\left(\bA_1 \bx_t \by_{t+h}' I(z_t < r_0) + \bA_2 \bx_t \by_{t+h}' I(r_0\leq z_t < r_0+\epsilon,)\right)I(z_{t+h}<r_0+\epsilon)\right] \Bigg\|_2 \\
&=&\big\| \bB_1' \bA_2 \bSigma_{x}(h,r_0,r_0+\epsilon,-\infty, r_0) \bA_1' +\bB_1' \bA_2 \bSigma_{x}(h,r_0,r_0+\epsilon, r_0,r_0+\epsilon) \bA_2' \big\|_2\\
&\leq & \|\bB_1\|_2 \, \|\bA_2\|_2 \, \|\bSigma_{x}(h,r_0, r_0+\epsilon,-\infty, r_0)\|_2 \, \|\bA_1\|_2+ \|\bB_1\|_2\, \|\bA_2\|_2^2\, \|\bSigma_{x}(h,r_0, r_0+\epsilon,r_0,r_0+\epsilon) \|  \\
&\leq& a_2^2 k_0\sigma_x^2(\epsilon p^{1-\delta_1/2-\delta_2/2}+\epsilon^2 p^{1-\delta_2}),
\end{eqnarray*}
and
\begin{eqnarray*}
\lefteqn{\big\| \bB_1(\eta_1,\eta_2)'\bSigma_{y,1,2}(h,r_0+\epsilon) \big\|_2=\Bigg\| \frac{1}{n-h}\sum_{t=1}^{n-h}\bB_1'{\rm E}\left[\by_t \by_{t+h}' I(z_t < r_0+\epsilon, z_{t+h}\geq r_0+\epsilon) \right] \Bigg\|_2}\nonumber \\
&=&\big\| \bB_1' \bA_2 \bSigma_{x}(h,r_0,r_0+\epsilon,r_0+\epsilon,+\infty) \bA_2' \big\|_2 \leq \|\bB_1\|_2 \, \|\bA_2\|_2^2 \, \| \bSigma_x(h,r_0,r_0+\epsilon, r_0+\epsilon, +\infty)\|_2 \\
&\leq&a_2^2k_0\sigma_x^2 \epsilon p^{1-\delta_2}.
\end{eqnarray*}
The proof for the case $\epsilon <0$ is similar.
\endp

\begin{lemma} Under Conditions 1-9, if $n>2h_0$, for any $\epsilon\in(\eta_1-r_0,\eta_2-r_0)$,  with true $k_0$, when $p$ is large enough, we have $G(r_0)=0$ and
\begin{eqnarray*}
{G}(r_0+\epsilon)  \geq \left\{\begin{array}{ll}
a_1^2 d^2 \rho_0^2 \epsilon^2 p^{2-\delta_1-\delta_{\min}}/2	&\epsilon <0,\\
a_1^2 d^2 \rho_0^2 \epsilon^2 p^{2-\delta_2-\delta_{\min}}/2	& \epsilon>0,
\end{array}\right.
\end{eqnarray*}
where $a_1$ satisfies that $\|\bA_i\|_2 \geq a_1p^{1/2-\delta_i/2}$ for $i=1,2$.
\end{lemma}
\noindent{\it Proof:} Note that
\begin{eqnarray*}
\lefteqn{\tr \left[ \bQ_2'  \left( \begin{array}{cc}
\bQ_1 	&\bB_1
\end{array}
\right)  \left(\begin{array}{c}
\bQ_1'\\
\bB_1'
\end{array} \right)
\bQ_2 \right] =\tr( \bQ_2'\bQ_1\bQ_1' \bQ_2) +\tr(\bQ_2' \bB_1 \bB_1' \bQ_2)}\\
&=&k_0\{1-  \left[{\cal D}({\cal M}(\bQ_2), {\cal M}(\bQ_1))\right]^2\} +\tr(\bQ_2' \bB_1 \bB_1' \bQ_2).
\end{eqnarray*}
On the other hand,
\begin{eqnarray*}
\tr \left[ \bQ_2'  \left( \begin{array}{cc}
\bQ_1 	&\bB_1	
\end{array}
\right)  \left(\begin{array}{c}
\bQ_1'\\
\bB_1'
\end{array} \right)
\bQ_2 \right] =\tr( \bQ_2'\bQ_2) =k_0.
\end{eqnarray*}
Hence $\tr((\bQ_2' \bB_1 \bB_1' \bQ_2)=k_0[{\cal D}({\cal M}(\bQ_2), {\cal M}(\bQ_1))]^2$.
In Lemma 2 we know that under Condition 4, there exists two positive constants $a_1<a_2$ such that $a_1 p^{1/2-\delta_i/2} \leq \|\bA_i\|_2 \leq a_2 p^{1/2-\delta_i/2}$ for $i=1,2$. Condition 9 indicates that $\|\bB_1' \bQ_2\|_2^2 \geq \tr(\bQ_2' \bB_1 \bB_1' \bQ_2)/k_0 \geq d^2$, and Condition 4 implies that $\|\bB_1' \bA_2\|_2^2 \geq  a_1^2d^2 p^{1-\delta_2}$. On the other hand,  $\|\bB_1' \bQ_2\|_2^2 \leq \tr(\bQ_2' \bB_1 \bB_1' \bQ_2) \leq k_0d^2$, and we have $\|\bB_1' \bA_2\|_2^2 \leq  a_2^2d^2k_0 p^{1-\delta_2}$.

If $\delta_1< \delta_2$, by Theorem 9 in \cite{merikoski2004}, we have
\begin{eqnarray*}
\lefteqn{G(r_0+\epsilon) \geq \|\bB_1' \bSigma_{y,1,1}(h^*_1, r_0+\epsilon)\|_2^2}\\
&=& \big\|\bB_1' \bA_2 \left( \bSigma_{x}(h_1^*,r_0, r_0+\epsilon,-\infty, r_0) \bA_1' +\bSigma_{x}(h_1^*,r_0, r_0+\epsilon, r_0, r_0+\epsilon) \bA_2' \right) \big\|_2^2\\
&\geq& \| \bB_1' \bA_2\|_2^2 \cdot \|\bSigma_{x}(h_1^*,r_0,r_0+\epsilon, -\infty,r_0) \bA_1' +\bSigma_{x}(h_1^*,r_0, r_0+\epsilon, r_0, r_0+\epsilon) \bA_2'\|_{\min}^2 \\
&\geq& \| \bB_1' \bA_2\|_2^2 \cdot \|\bSigma_{x}(h_1^*,r_0,r_0+\epsilon, -\infty,r_0)\|_{\min}^2\|\bA_1'\|_{2}^2 \\
&&- \| \bB_1' \bA_2\|_2^2\|\bSigma_{x}(h_1^*,r_0, r_0+\epsilon, r_0, r_0+\epsilon)\|_2^2 \| \bA_2\|_{2}^2 \\
&\geq & a_1^2d^2\rho_0^2 \epsilon^2   p^{2-\delta_1-\delta_2}- a_2^2d^2 k_0\sigma_x^4\epsilon^4 p^{2-2\delta_2}.
\end{eqnarray*}
If $\delta_1 \geq \delta_2$,
\begin{eqnarray*}
\lefteqn{G(r_0+\epsilon) \geq \|\bB_1' \bSigma_{y,1,2}(h^*_1, r_0+\epsilon)\|_2^2}\\
&=& \big\|\bB_1' \bA_2 \left( \bSigma_{x}(h_2^*,r_0, r_0+\epsilon, r_0+\epsilon, +\infty) \bA_2' \right) \big\|_2^2\\
&\geq& \| \bB_1' \bA_2\|_2^2 \cdot \|\bSigma_{x}(h_2^*,r_0,r_0+\epsilon, r_0+\epsilon, +\infty) \bA_2'\|_{\min}^2 \\
&\geq & a_1^2d^2 \rho_0^2   \epsilon^2  p^{2-2\delta_2}.
\end{eqnarray*}
If $p$ is large enough, we have $G(r_0+\epsilon)\geq a_1^2 d^2 \rho_0^2 \epsilon^2 p^{2-\delta_2-\delta_{\min}}/2$ when $\epsilon>0$. By the definition and Lemma 4, we can prove conclusions for $\epsilon=0$ and $\epsilon<0$ in a similar fashion.
\endp

\begin{lemma} Under Conditions 1-7, if $p^{\delta_1/2+\delta_2/2} n^{-1/2}=o(1)$, with true $k_0$, as $n,p\to \infty$, it holds that
\[
{\rm E} \|\bB_i(\eta_1,\eta_2)-\bB_i(\eta_1,\eta_2)\|_2^2\leq Cp^{\delta_i+\delta_{\min}}n^{-1}, \mbox{ for } i=1,2.
\]
\end{lemma}
\noindent{\it Proof:} Let $Y_{t}=x_{t,q}x_{t+h,\ell} I_t(h,c_1,c_2,c_3,c_4)-{\rm E}[x_{t,q}f_{t+h,\ell}I_t(h,c_1,c_2,c_3,c_4)]$, where $x_{t,q}$ is the $q$-th entry in $\bx_{t}$, and $c_1,c_2,c_3,c_4$ are real numbers in $[\eta_1,\eta_2]$, $-\infty$, or $+\infty$. Condition 2 indicates that there exists a positive constant $\sigma_y$ such that ${\rm E} Y_t^{2\gamma}<\sigma_y^{2\gamma}$, and $\sigma_y <\Delta_{c_1,c_2}\Delta_{c_3,c_4}\sigma_x^2$. Thus, if $n>2h_0$, by Cauchy-Schwartz inequality, Jensen's inequality, Proposition 2.5 in \cite{fan2003} and Lemma 1,
\begin{eqnarray*}
\lefteqn{\frac{1}{(n-h)^4} {\rm E}\left( \sum_{t=1}^{n-h} Y_{t}^4\right)}\\
&\leq&\frac{16}{n^4}\sum_{t=1}^{n-h} {\rm E}(Y_{t}^4)+\frac{32\binom 4 1}{n^4}\sum_{t_1< t_2} {\rm E}(Y_{t_1}^3Y_{t_2})+\frac{32\binom 4 2}{n^4}\sum_{t_1< t_2} {\rm E}(Y_{t_1}^2Y_{t_2}^2)\\
&&+ \frac{16\binom 4 2 \binom 2 1}{n^4}\sum_{\substack{t_1\neq t_2,t_2\neq t_3\\ t_1 \neq t_3}} {\rm E} (Y_{t_1}^2Y_{t_2}Y_{t_3})+\frac{16\cdot 4!}{n^4} \sum_{t_1<t_2<t_3<t_4}{\rm E} (Y_{t_1}Y_{t_2}Y_{t_3}Y_{t_4})\\
&<&\frac{16\sigma_y^4}{n^3}+\frac{160\sigma_y^4}{n^2} + \frac{192\sigma_y^4}{n}+\frac{384}{n^4}\sum_{\substack{t_1<t_2<t_3<t_4 \\ t_3-t_2 \leq h}} {\rm E} (Y_{t_1}Y_{t_2}Y_{t_3}Y_{t_4})+\frac{384}{n^4} \sum_{\substack{t_1<t_2<t_3<t_4 \\ t_3-t_2>h}}{\rm E} (Y_{t_1}Y_{t_2}Y_{t_3}Y_{t_4})\\
&<& \frac{368 \sigma_y^4}{n}+\frac{384h(n-2h)^3 \sigma_y^4}{n^4}+\frac{384}{n^4}\sum_{\substack{t_1<t_2<t_3<t_4\\ t_3-t_2> h}} \cov(Y_{t_1}Y_{t_2}, Y_{t_3}Y_{t_4})+{\rm E}(Y_{t_1}Y_{t_2}){\rm E}(Y_{t_3}Y_{t_4})\\
&<&\frac{752h\sigma_y^4}{n}+\frac{768\sigma_y^4}{n^2}\sum_{u=1}^{n-2h}\alpha(u)^{1-2/\gamma}\\
&&+\frac{3072}{n^4} \left( \sum_{t_1=1}^{n-h} \sum_{t_2=1}^{n-h} |\cov(Y_{t_1},Y_{t-2})|\right) \cdot \left( \sum_{t_3=1}^{n-h} \sum_{t_4=1}^{n-h} |\cov(Y_{t_3},Y_{t-4})|\right)\\
&<&\frac{16(47h+48\alpha)\sigma_y^4}{n}+\frac{48\sigma_y^4}{n^2}\left(\sum_{u=1}^{n-2h} \alpha(u)^{1-2/\gamma} \right)^2\\
&\leq& \frac{16(47h+48\alpha+192\alpha^2)\sigma_y^4}{n}.
\end{eqnarray*}
It follows,
\begin{eqnarray}
{\rm E}\| \hat{\bSigma}_{x}(h,c_1,c_2,c_3,c_4)-\bSigma_{x}(h,c_1,c_2,c_3,c_4)\|_2^4\leq \Delta_{c_1,c_2}\Delta_{c_3,c_4}16(47h+48\alpha+192\alpha^2)k_0^4 \sigma_0^8n^{-1}.\label{d2}
\end{eqnarray}
For the interaction terms of the common component and noise, since they are independent, similar to proof in Lemma 2, we can obtain that
\begin{eqnarray}
\lefteqn{{\rm E}\Big\| \frac{1}{n-h}\sum_{t=1}^{n-h}\bA_i \bx_{t}\bve_{t+h,j}'I_t(h,c_1,c_2,c_3,c_4)\Big\|_2^4}\nonumber\\
&\leq& \frac{a_2^4p^{2-2\delta_{i}}}{(n-h)^4}{\rm E} \left( \sum_{m=1}^{k_0} \sum_{\ell=1}^{p}\sum_{t=1}^{n-h}  (x_{t,m}^2  \epsilon_{t+h,j,\ell}^2 I_t(h,c_1,c_2,c_3,c_4) \right)^2
\leq  4a_2^4k_0^2 \sigma_0^8 p^{4-2\delta_{i}}n^{-2},\label{inter}
\end{eqnarray}
and
\begin{eqnarray}
\lefteqn{{\rm E}\Big\| \frac{1}{n-h}\sum_{t=1}^{n-h}\bve_{t,i} \bx_t'\bA_j' I_t(h,c_1,c_2,c_3,c_4)\Big\|_2^4}\nonumber\\
&\leq& \frac{a_2^4p^{2-2\delta_{j}}}{(n-h)^4}{\rm E} \left( \sum_{m=1}^{k_0} \sum_{\ell=1}^{p}\sum_{t=1}^{n-h}  (\epsilon_{t,i,m}^2  x_{t+h,\ell}^2 I_t(h,c_1,c_2,c_3,c_4) \right)^2
\leq  4a_2^4k_0^2 \sigma_0^8 p^{4-2\delta_{j}}n^{-2}, \label{inter2}
\end{eqnarray}
for $i,j=1,2$. For the noise,
\begin{eqnarray}
\lefteqn{{\rm E} \Big\| \sum_{t=1}^{n-h} \bve_{t,i} \bve_{t+h,j}'(h,c_1,c_2,c_3,c_4)\Big\|_2^4}\nonumber\\
&\leq&\frac{1}{(n-h)^4}{ \rm E} \left(\sum_{m=1}^{p}\sum_{\ell=1}^{p}  \sum_{t=1}^{n-h}\epsilon_{t,i,m}^2\epsilon_{t+h,\ell}^2 I(c_1,c_2,c_3,c_4)\right)^2\leq 16 \sigma_0^8 p^4n^{-2}, \label{noi}
\end{eqnarray}
for $i,j=1,2$. Similar to the proof of Lemma 2, with (\ref{d2})-(\ref{noi}), it can be shown that there exists a positive constant $C$ such that
\begin{eqnarray*}
{\rm E} \|\hat{\bSigma}_{y,i,j}(h,\eta_1,\eta_2)-\bSigma_{y,i,j}(h,\eta_1,\eta_2)\|_2^4 \leq C_1p^{4-2\delta_i-2\delta_j}n^{-1}+C_2p^4n^{-2}.
\end{eqnarray*}
for $i,j=1,2$.

If $p^{\delta_{1}/2+\delta_{2}/2}n^{-1/2}=o(1)$, by Lemma 2, we have
\begin{eqnarray*}
\lefteqn{{\rm E}\|\bM_i(\eta_1,\eta_2)-\bM_i(\eta_1,\eta_2)\|_2^2}\\
&\leq&4h_0 \sum_{h=1}^{h_0} \sum_{j=1}^2 {\rm E}\|\hat{\bSigma}_{y,i,j}(h,\eta_1,\eta_2)-\bSigma_{y,i,j}(h,\eta_1,\eta_2)\|_2^4\\
&&+4h_0 \sum_{h=1}^{h_0}\sum_{j=1}^2 2 \|\bSigma_{y,i,j}(h,\eta_1,\eta_2)\|_2\cdot  {\rm E} \|\hat{\bSigma}_{y,i,j}(h,\eta_1,\eta_2)-\bSigma_{y,i,j}(h,\eta_1,\eta_2)\|_2^2\\
&\leq&Cp^{4-\delta_i-\delta_{\min}}n^{-1}.
\end{eqnarray*}
Together with (\ref{Mmin}), following the proof for Theorem 1 in \cite{lam2011}, we can reach the conclusion.
\endp

\begin{lemma} Under Conditions 1-9, if $n>2h_0$, for any $\epsilon\in(\eta_1-r_0,\eta_2-r_0)$, with true $k_0$,
\begin{eqnarray*}
\lefteqn{{\rm E}|\hat{G}(r_0+\epsilon)-G(r_0+\epsilon)|}\\
& \leq \left\{\begin{array}{ll}
C_1 p^2 n^{-1}+C_2\epsilon p^{2-\delta_1/2-\delta_{\min}/2}n^{-1/2}+ C_3\epsilon^2 p^{2-\delta_1 +\delta_2/2-\delta_{\min}/2} n^{-1/2}	&\epsilon <0,\\
C_1p^2n^{-1}								& \epsilon=0,\\
C_1p^2 n^{-1}+C_2\epsilon p^{2-\delta_2/2-\delta_{\min}/2}n^{-1/2}+ C_3 \epsilon^2 p^{2+\delta_1/2 -\delta_2-\delta_{\min}/2} n^{-1/2}	&\epsilon>0.
\end{array}\right.
\end{eqnarray*}
\end{lemma}

\noindent{\it Proof:} Since $r_0 \in (\eta_1,\eta_2)$, it follows by the definition
\[
{\cal M}(\bB_i)={\cal M}(\bB_i(\eta_1,\eta_2)).
\]
Then there exists a $(p-k_0) \times (p-k_0)$ orthogonal matrix $\bV_i$ such that $\bB_i=\bB_i(\eta_1,\eta_2)\bV_i$ for $i=1,2$. Hence,
\[
G(r)= \sum_{i=1}^2\| {\bV'\bB_i(\eta_1,\eta_2)}' \bM_i(r) \bB_i(\eta_1,\eta_2)\bV\|_2= \sum_{i=1}^2\| {\bB_i(\eta_1,\eta_2)}' \bM_i(r) \bB_i(\eta_1,\eta_2)\|_2.
\]
By the definition of $\hat{G}(r)$ we have,
\begin{eqnarray}
\lefteqn{|\hat{G}(r)-G(r)|}\nonumber \\
 &\leq & \sum_{h=1}^{h_0}\sum_{i=1}^2\sum_{j=1}^2  \Big\| \hat{\bB}_i(\eta_1,\eta_2)'\hat{\bSigma}_{y,i,j}(h,r) \hat{\bSigma}_{y,i,j}(h,r) '\hat{\bB}_i(\eta_1,\eta_2) \nonumber\\
 && - \bB_i(\eta_1,\eta_2)' \bSigma_{y,i,j}(h,r) \bSigma_{y,i,j}(h,r)'\bB_i(\eta_1,\eta_2)\Big\|_2 \nonumber\\
&\leq& \sum_{h=1}^{h_0} \sum_{i=1}^2  \sum_{j=1}^2 \left[ \big\|\hat{\bB}_i(\eta_1,\eta_2)' \hat{\bSigma}_{y,i,j}(h,r) - \bB_i(\eta_1,\eta_2)' \bSigma_{y,i,j}(h,r)\big\|_2^2 \right.\nonumber \\
&&+ \left. 2 \big\| \bB_i(\eta_1,\eta_2)'\bSigma_{y,i,j}(h,r)\big\|_2 \cdot \big\| \hat{\bB}_i(\eta_1,\eta_2)' \hat{\bSigma}_{y,i,j}(h,r)-\bB_i(\eta_1,\eta_2)'\bSigma_{y,i,j}(h,r)\big\|_2 \right] \nonumber\\
&\leq & \sum_{h=1}^{h_0}\sum_{i=1}^2  \sum_{j=1}^2 \left[ \left( \| \hat{\bB}_i(\eta_1,\eta_2)\|_2 \cdot \big\| \hat{\bSigma}_{y,i,j}(h,r)  -{\bSigma}_{y,i,j}(h,r)\big\|_2\right.\right. \nonumber\\
&& +\left. \left. \|  \hat{\bB}_i(\eta_1,\eta_2)- \bB_i(\eta_1,\eta_2)\big\|_2 \cdot \| {\bSigma}_{y,i,j}(h,r) \|_2   \right)^2 \right. \nonumber\\
&&+  2 \| \bB_i(\eta_1,\eta_2)'\bSigma_{y,i,j}(h,r)\|_2 \left( \| \hat{\bB}_i(\eta_1,\eta_2)\|_2 \|\hat{\bSigma}_{y,i,j}(h,r)- {\bSigma}_{y,i,j}(h,r) \|_2 \right. \nonumber \\
&& \left.+ \|\hat{\bB}_i(\eta_1,\eta_2)- \bB_i(\eta_1,\eta_2) \|_2\| {\bSigma}_{y,i,j}(h,r)   \|_2 \right)  \Bigg] \nonumber\\
&=&\sum_{i=1}^2 \sum_{j=1}^2 \left[L_{i,j,1}(r)+L_{i,j,2}(r)\right].
\label{G}
\end{eqnarray}

When $\epsilon>0$, by Lemmas 1-6, we have
\begin{align*}
&{\rm E}(L_{1,1,1}(r_0+\epsilon))\leq C_1 p^2 n^{-1}+ C_2 \epsilon^4 p^{2+\delta_1-2\delta_2+\delta_{\min}}n^{-1},\\
 &{\rm E}(L_{1,1,2}(r_0+\epsilon))\leq C_1\epsilon p^{2-\delta_1/2-\delta_2/2} n^{-1/2}+C_2\epsilon^2 p^{2-\delta_2}n^{-1/2}+C_3\epsilon^4 p^{2+\delta_1/2-2\delta_2+\delta_{\min}/2}n^{-1/2},\\
&{\rm E}(L_{1,2,1}(r_0+\epsilon))\leq C_1p^2 n^{-1}+C_2\epsilon^2 p^{2+\delta_1-2\delta_2+\delta_{\min}} n^{-1}\\
 &{\rm E}(L_{1,2,2}(r_0+\epsilon))\leq C_1\epsilon p^{2-\delta_2} n^{-1/2} +C_2\epsilon^2 p^{2+\delta_1/2-2\delta_2 +\delta_{\min}/2}n^{-1/2},\\
& {\rm E}(L_{2,1,1}(r_0+\epsilon))\leq C_1p^2 n^{-1}, \quad
L_{2,1,2}(r_0+\epsilon)=0,\\
&{\rm E}(L_{2,2,1}(r_0+\epsilon))\leq C_1 p^2 n^{-1}, \quad
L_{2,2,2}(r_0+\epsilon)=0.
 \end{align*}
When $\epsilon>0$, it follows,
\begin{align*}
{\rm E} |\hat{G}(r_0+\epsilon)-G(r_0+\epsilon)|
\leq C_1 p^2 n^{-1}+C_2 \epsilon p^{2-\delta_2/2-\delta_{\min}/2}n^{-1/2}+ C_3 \epsilon^2 p^{2+\delta_1/2 -\delta_2-\delta_{\min}/2} n^{-1/2}.
\end{align*}
\endp

\noindent{\bf Proof of Theorem 1.}
When $r_1 \leq r_0$ and $r_2 \geq r_0$, under Condition 4, similar to Lemmas 2 and 3, we can show that
\[
\|\bSigma_{y,1,1}(h,r_1,r_2)\|_2=\| \bA_1 \bSigma_{x,1,1}(h,r_1,r_2) \bA_1' \|_2\leq Cp^{1-\delta_1},
\]
\[
\|\bSigma_{y,1,2}(h,r_1,r_2)\|_2= \|\bA_1 \bSigma_{x,1,2}(h,r_1,r_2) \bA_2'\|_2 \leq Cp^{1-\delta_1/2-\delta_2/2},
\]
and
\[
{\rm E}\|\bSigma_{y,1,j}(h,r_1,r_2)-\hat{\bSigma}_{y,1,j}(h,r_1,r_2)\|_2^2\leq Cp^2n^{-1}, \mbox{ for } j=1,2.
\]
It follows
\begin{eqnarray}
\lefteqn{\|\hat{\bM}_1({r_1,r_2})-\bM_1({r_1,r_2})\|_2}\nonumber \\
& \leq & \sum_{h=1}^{h_0} \sum_{j=1}^2 \left(\|\hat{\bSigma}_{y,1,j}(h,{r_1},r_2) -\bSigma_{y,1,j}(h,{r_1},r_2)\|_2^2\right. \nonumber \\
&&
+ \left.2 \|\bSigma_{y,1,j}(h,{r}_1,r_2)\|_2 \cdot  \| \hat{\bSigma}_{y,1,j}(h,{r}_1,r_2)-\bSigma_{y,1,j}(h,{r}_1,r_2)\|_2\right)\nonumber\\
&=&O_p(p^{2-\delta_1/2-\delta_{\min}/2}n^{-1/2}).\label{Mmin}
\end{eqnarray}
Under Condition 5, with Theorem 9 in \cite{merikoski2004}, we have
\begin{eqnarray*}
\lefteqn{\|\bM_1(r_1,r_2)\|_{\min}\geq \sum_{h=1}^{h_0} \sum_{j=1}^2  \big\|\bA_1 \bSigma_{x,1,j} (h,r_1,r_2)\bA_j' \bA_j \bSigma_{x,1,j}(h,r_1,r_2) \bA_1' \big\|_{\min}}\\
&\geq&  \sum_{j=1}^2  \|\bA_1\|_{\min}^2 \|\bSigma_{x,1,j}(h_1,r_1,r_2) \|_{\min}^2 \|\bA_j\|_{\min}^2
=O(p^{2-\delta_1-\delta_{\min}}).
\end{eqnarray*}
Following the proof of Theorem 1 in \cite{liu2016}, we can obtain that $\cD(\cM(\hat{\bQ}_1(r_1,r_2)), \cM({\bQ}_1))=O_p(p^{\delta_1/2+\delta_{\min}/2}n^{-1/2})$.
\endp

\noindent{\bf Proof of Proposition 1. } Similar to the proof of Lemma 5, we can see that when $r \neq r_0$, $G(r)>0$.\endp 

\noindent{\bf Proof of Theorem 2.}
Since $G(r)\geq 0$ and $G(r_0)=0$, if $p^{\delta_{1}/2+\delta_{2}/2}n^{-1/2}=o(1)$, let $g_0= a_1^2 d^2 \rho_0^2 p^{2-\delta_{2}-\delta_{\min}}/4$, then we have
\begin{eqnarray*}
\lefteqn{P(\hat{r}>r_0+\epsilon) =P[\hat{G}(r_0)> \hat{G}(\hat{r}),\hat{r}>r_0+\epsilon]}\\
&= &P\Big[\hat{G}(r_0) -G(r_0)> \hat{G}(\hat{r})-G(\hat{r})+G(\hat{r}),\hat{r}>r_0+\epsilon\Big]\\
&= &P\Big[\hat{G}(r_0) -G(r_0)+G(\hat{r})- \hat{G}(\hat{r})+\frac{3}{2}g_0 \epsilon^2- G(\hat{r})>\frac{3}{2}g_0\epsilon^2,\hat{r}>r_0+\epsilon\Big]\\
&\leq & P\Big[ \hat{G}(r_0)-G(r_0) >\frac{1}{2}g_0\epsilon^2  \Big] + P\Big[ \hat{G}(\hat{r})-G(\hat{r})>\frac{1}{2}g_0\epsilon^2,\hat{r}>r_0+\epsilon \Big]\\
&&+P\Big[ \frac{3}{2}g_0 \epsilon^2-G(\hat{r})>\frac{1}{2}g_0\epsilon^2, \hat{r}>r_0+\epsilon \Big] \\
&\leq& P\Big[ \big| \hat{G}(r_0)-G(r_0)\big|>\frac{1}{2}g_0\epsilon^2 \Big] + P\Big[ \big|\hat{G}(\hat{r})-G(\hat{r})\big|>\frac{1}{2}g_0\epsilon^2,\hat{r}>r_0+\epsilon \Big]\\
&&+P\Big[G(\hat{r})<g_0 \epsilon^2, \hat{r}>r_0+\epsilon \Big] \\
&\leq& \frac{2C_1p^2}{g_0\epsilon^2n}+\frac{2C_1d p^{2-\delta_{2}/2-\delta_{\min}/2}}{g_0 \epsilon n^{1/2}}\\
&\leq&\frac{Cp^{\delta_{2}/2+\delta_{\min}/2} }{\epsilon n^{1/2}}.
\end{eqnarray*}
\endp

\noindent{\bf Proof of Theorem 3.} When $\hat{r}>r_0$, for any $\epsilon>0$, Lemmas 2-4 imply that
\begin{eqnarray}
\lefteqn{\|\hat{\bM}_1({r_0+\epsilon})-\bM_1({r_0+\epsilon}) \|_2} \nonumber\\
 &\leq & \sum_{h=1}^{h_0}\sum_{j=1}^2 \left( \| \hat{\bSigma}_{y,1,j}(h,{r_0+\epsilon}) -\bSigma_{y,1,j}(h,{r_0+\epsilon})\|_2^2 \right.\nonumber \\
 &&\left. +2 \|\bSigma_{y,1,j}(h,r_0+\epsilon)\|_2 \cdot \|\hat{\bSigma}_{y,1,j}(h,{r_0+\epsilon})-\bSigma_{y,1,j}(h,{r_0+\epsilon})\|_2 \right) \nonumber\\
&=&O_p(p^2 n^{-1})+O_p(p^{2-\delta_1/2-\delta_{\min}/2}n^{-1/2})+O_p(\epsilon p^{2-\delta_2/2-\delta_{\min}/2} n^{-1/2}) \nonumber\\
&=&O_p(p^{2-\delta_1/2-\delta_{\min}/2}n^{-1/2})+O_p(\epsilon p^{2-\delta_2/2-\delta_{\min}/2} n^{-1/2}). \label{Ms}
\end{eqnarray}

Under Conditions 4 and 8, it follows from Lemma \ref{eps},
\begin{eqnarray*}
\lefteqn{\|\bSigma_{y,1,1}(h,r_0+\epsilon) -\bSigma_{y,1,1}(h,r_0)\|_2}\\
& = &\Big\| \frac{1}{n-h}\sum_{t=1}^{n-h} E\{\by_t \by_{t+h}' [I(z_t< r_0+\epsilon, z_{t+h} <r_0+\epsilon)- I( z_t < r_0, z_{t+h} <r_0)] \}\Big\|_2\\
&=&\Big\| \frac{1}{n-h}\sum_{t=1}^{n-h} E\{\by_t \by_{t+h}' [I(z_t<r_0+\epsilon,r_0 \leq z_{t+h}<r_0+\epsilon)+ I( r_0\leq z_t < r_0+\epsilon, z_{t+h} < r_0)] \}\Big\|_2\\
&=&\Big\| \bA_1\bSigma_x(h, -\infty, r_0,r_0, r_0+\epsilon )\bA_2'+ \bA_2\bSigma_x(h, r_0, r_0+\epsilon,r_0, r_0+\epsilon )\bA_2'\\
&& + \bA_2\bSigma_x(h,r_0,r_0+\epsilon, r_0, +\infty)\bA_1' \Big\|_2\\
&\leq& O(\epsilon p^{1-\delta_1/2-\delta_2/2})+ O(\epsilon^2 p^{1-\delta_2}),
\end{eqnarray*}
and
\begin{eqnarray*}
\lefteqn{\|\bSigma_{y,1,2}(h,r_0+\epsilon) -\bSigma_{y,1,2}(h,r_0)\|_2}\\
& = &\Big\| \frac{1}{n-h}\sum_{t=1}^{n-h} E\{\by_t \by_{t+h}' [I(z_t< r_0+\epsilon, z_{t+h} \geq r_0+\epsilon)- I( z_t < r_0, z_{t+h} \geq r_0)] \}\Big\|_2\\
&=&\Big\| \frac{1}{n-h}\sum_{t=1}^{n-h} E\{\by_t \by_{t+h}' [I( r_0\leq z_t < r_0+\epsilon, z_{t+h} \geq r_0+\epsilon)- I(z_t<r_0,r_0 \leq z_{t+h}<r_0+\epsilon)] \}\Big\|_2\\
&=&\Big\| \bA_2\bSigma_x(h, r_0, r_0+\epsilon, r_0+\epsilon,+\infty)\bA_2'- \bA_1\bSigma_x(h, -\infty, r_0, r_0, r_0+\epsilon )\bA_2' \Big\|_2\\
&\leq& O(\epsilon p^{1-\delta_2/2-\delta_{\min}/2}).
\end{eqnarray*}
Hence,
\begin{eqnarray*}
\lefteqn{\|\bM_1({r_0+\epsilon}) -\bM_1\|_2}\\
&\leq &  \sum_{h=1}^{h_0} \sum_{j=1}^2 \big\| \bSigma_{y,1,j}(h,r_0+\epsilon)\bSigma_{y,1,j}(h,r_0+\epsilon)' -\bSigma_{y,1,j}(h,r_0)\bSigma_{y,1,j}(h,r_0)' \big\|_2\\
&\leq &  \sum_{h=1}^{h_0}\sum_{j=1}^2  \left(\| \bSigma_{y,1,j}(h,{r}_0+\epsilon) - \bSigma_{y,1,j}(h,r_0)\|_2^2 \right.\\
&&\left.+ 2\|\bSigma_{y,1,j}(h,r_0)\|_2\cdot \|\bSigma_{y,1,j}(h,r_0+\epsilon)-\bSigma_{y,1,j}(h,{r_0}) \|_2\right)\\
&\leq& O(\epsilon^2 p^{2-\delta_2-\delta_{\min}})+ O(\epsilon p^{2-\delta_1/2-\delta_2/2-\delta_{\min}}).
\end{eqnarray*}
If $p^{\delta_1/2+\delta_{2}/2}n^{-1/2}=o(1)$, together with (\ref{Ms}), we have
\begin{eqnarray*}
\lefteqn{\| \hat{\bM}_1(r_0+\epsilon) -\bM_1\|_2} \nonumber\\
& \leq & \| \hat{\bM}_1(r_0+\epsilon) -{\bM}_1(r_0+\epsilon) \|_2 + \|\bM_1(r_0+\epsilon) -\bM_1\|_2\\
&=&O_p(p^{2-\delta_1/2-\delta_{\min}/2}n^{-1/2})+O(\epsilon p^{2-\delta_1/2-\delta_2/2-\delta_{\min}})+O(\epsilon^2 p^{2-\delta_2-\delta_{\min}}).
\end{eqnarray*}

Theorem 2 tells us if $\hat{r}>r_0$, $|\hat{r}-r_0|=O_p(p^{\delta_2/2+\delta_{\min}/2} n^{-1/2})$. Therefore,
\begin{eqnarray}\label{M-hatM}
\| \hat{\bM}_1(\hat{r}) -\bM_1\|_2 =O_p(p^{2-\delta_1/2-\delta_{\min}/2} n^{-1/2}), \mbox{ if } \hat{r}>r_0.
\end{eqnarray}
If $\delta_1 <\delta_2$, under Condition 5, by Theorems 6 and 9 in \cite{merikoski2004},
\begin{eqnarray*}
\lefteqn{\|\bM_1\|_{\min} \geq \|\bSigma_{y,1,1}(h_1,r_0)\|_{\min}^2
 \geq \big\| \bA_1\bSigma_{x,1,1}(h_1,r_0) \bA_1' \|_{\min}^2} \\
&\geq&\|\bA_2\|^4_{\min} \| \bSigma_{x,1,1}(h_1,r_0)\|_{\min}^2
\geq C p^{2-2\delta_2}.
\end{eqnarray*}
If $\delta_1 \geq \delta_2$,
\begin{eqnarray*}
\lefteqn{\|\bM_1\|_{\min} \geq \|\bSigma_{y,1,2}(h_1,r_0)\|_{\min}^2
 \geq \big\| \bA_1\bSigma_{x,1,2}(h_1,r_0) \bA_2' \|_{\min}^2}\\
&\geq&\|\bA_1\|_{\min}^2 \| \bSigma_{x,1,2}(h_1,r_0)\|_{\min}^2 \|\bA_2\|_{\min}^2 \geq C p^{2-\delta_1-\delta_2}.
\end{eqnarray*}
Hence, $\|\bM_1\|_{\min}\geq C p^{2-\delta_1-\delta_{\min}}$. Following the proof of Theorem 2 in \cite{liu2016}, together with (\ref{M-hatM}) we obtain the results. Similarly we can prove that $\cD( \cM(\hat{\bQ}_1(\hat{r}), \cM(\bQ_1))=O_p(p^{\delta_1/2+\delta_{\min}/2} n^{-1/2})$ when $\hat{r}< r_0$.
\endp

\begin{lemma}
Let ${\bB}_i^*$ be a $p \times (p-\hat{k})$ orthogonal matrix such that ${\cal M}(\bB_i^*) \in {\cal M}(\bB_i)$ for $i=1,2$. Under Conditions 1-4 and 10, for any ${\bB}_i^*$,
\begin{eqnarray*}
\|{\bB}_1^{*'} \bSigma_{y,1,1}(h,r_0+\epsilon)\|_2= \left\{
\begin{array}{ll}
0	&\epsilon \leq 0,\\
O(\epsilon p^{1-\delta_1/2-\delta_{2}/2})+O(\epsilon^2 p^{1-\delta_2})	&\epsilon>0,
\end{array}
\right.\\
\|{\bB}_1^{*'} \bSigma_{y,1,2}(h,r_0+\epsilon)\|_2= \left\{
\begin{array}{ll}
0	&\epsilon \leq 0,\\
O(\epsilon p^{1-\delta_{2}})	&\epsilon>0,
\end{array}
\right.\\
\|{\bB}_2^{*'} \bSigma_{y,2,1}(h,r_0+\epsilon)\|_2= \left\{
\begin{array}{ll}
O(\epsilon p^{1-\delta_{1}})	&\epsilon>0,\\
0	&\epsilon \leq 0,
\end{array}
\right.\\
\|{\bB}_2^{*'} \bSigma_{y,2,2}(h,r_0+\epsilon)\|_2= \left\{
\begin{array}{ll}
O(\epsilon p^{1-\delta_1/2-\delta_{2}/2})+O(\epsilon^2 p^{1-\delta_1})	&\epsilon < 0,\\
0	&\epsilon \geq 0.
\end{array}
\right.
\end{eqnarray*}
\end{lemma}
\noindent{\it Proof:} Note that for ${\bB}_i^{*}$ such that ${\bB}_i^{*'}\bA_i=\mathbf{0}$, following the proof of Lemma 4, we can reach the conclusion.
\endp

\noindent{\bf Proof of Corollary 1:} Similar to proof of Theorem 1 in \cite{liu2016}.\endp

\noindent{\bf Proof of Theorem 4:} Under Conditions 1-9, if $p^{\delta_1/2+ \delta_2/2}n^{-1/2}=o(1)$, similar to the proof of Theorem 1, we obtain that
\[
\big\|\hat{\bB}_i(\eta_1,\eta_2)-\bB_i(\eta_1,\eta_2) \big\|_2 =O_p(p^{\delta_i/2 +\delta_{\min}/2} n^{-1/2}).
\]
Since
\begin{eqnarray*}
\hat{\bB}_i(\eta_1,\eta_2)=\left( \begin{array}{llll} \hat{\bq}_{i,k_0+1}(\eta_1,\eta_2), &\ldots, &\hat{\bq}_{i,\hat{k}}(\eta_1,\eta_2), &\hat{\bB}_{i,\hat{k}}(\eta_1,\eta_2)
\end{array}\right),
\end{eqnarray*}
we have
\[
\big\|\hat{\bB}_{i,\hat{k}}(\eta_1,\eta_2)-\bB_{i,\hat{k}}(\eta_1,\eta_2) \big\|_2\leq \big\|\hat{\bB}_i(\eta_1,\eta_2)-\bB_i(\eta_1,\eta_2) \big\|_2 =O_p(p^{\delta_i/2 +\delta_{\min}/2} n^{-1/2}), \mbox{ for } i=1,2.
\]
With Lemma 7, similar to the proof of Theorem 2, we can complete the proof. \endp

\noindent{\bf Proof of Theorem 5:} Similar to proof of Theorem 3.\endp

\end{document}